\documentclass[onecolumn,a4paper,journal,draftclsnofoot,10pt]{IEEEtran}
\usepackage{qcfd_styles}
\usepackage{qcfd_macros}
\begin{document}
\title{Boundary Treatment for Variational Quantum Simulations of Partial Differential Equations on Quantum Computers}
\date{\today}
\author{
\IEEEauthorblockN{
    \href{mailto:paul.over@tuhh.de}{Paul~Over}$^{a,*}$\orcidlink{0000-0001-7436-5254},
    \footnotetext{\hspace{-0.5cm}\rule{2cm}{0.2mm}\\ ${}^*$ Corresponding author.\\ E-mail address: \href{mailto:paul.over@tuhh.de}{paul.over@tuhh.de}}
    \and
    \href{mailto:sergio.bengoechea@tuhh.de}{Sergio~Bengoechea}$^{a}$\orcidlink{0009-0001-8205-5878}, 
    \and
    \href{mailto:thomas.rung@tuhh.de}{Thomas~Rung}$^{a}$\orcidlink{0000-0002-3454-1804},
    \and
    \href{mailto:francesco.clerici94@gmail.com}{Francesco~Clerici}$^{b}$\orcidlink{0009-0002-8153-8111},
    \and 
    \href{mailto:l.scandurra@engys.com}{Leonardo~Scandurra}$^{b}$\orcidlink{0000-0003-3075-2919},
    \and 
    \href{mailto:e.devilliers@engys.com}{Eugene~de~Villiers}$^{c}$\orcidlink{0000-0002-0182-3637} and  
    \and 
    \href{mailto:dieter.jaksch@uni-hamburg.de}{Dieter~Jaksch}$^{d,e}$\orcidlink{0000-0002-9704-3941}
}

\IEEEauthorblockA{
  \small{$^{a}$ Institute for Fluid Dynamics and Ship Theory, Hamburg University of Technology, 21073 Hamburg, Germany;\\
  $^{b}$ ENGYS Srl, Via del Follatoio, 12, 34148 Trieste TS, Italy;  $^{c}$ ENGYS Ltd., London SW18 3SX, United Kingdom;\\
  $^{d}$ Institute for Quantum Physics, Universit{\"a}t Hamburg, Luruper Chaussee 149, 22761 Hamburg, Germany;\\
  $^{e}$ Clarendon Laboratory, University of Oxford, Parks Road, Oxford OX1 3PU, United Kingdom;
}}}

\maketitle

\begin{abstract}
The paper presents a variational quantum algorithm to solve initial-boundary value problems described by second-order partial differential equations. The approach uses hybrid classical/quantum hardware that is well suited for quantum computers of the current noisy intermediate-scale quantum era. The partial differential equation is initially translated into an optimal control problem with a modular control-to-state operator (ansatz). The objective function and its derivatives required by the optimizer can efficiently be evaluated on a quantum computer by measuring an ancilla qubit, while the optimization procedure employs classical hardware. The focal aspect of the study is the treatment of boundary conditions, which is tailored to the properties of the quantum hardware using a correction technique. For this purpose, the boundary conditions and the discretized terms of the partial differential equation are decomposed into a sequence of unitary operations and subsequently compiled into quantum gates. The accuracy and gate complexity of the approach are assessed for second-order partial differential equations by classically emulating the quantum hardware. 
The examples include steady and unsteady diffusive transport equations for a scalar property in combination with various Dirichlet, Neumann, or Robin conditions. The results of this flexible approach display a robust behavior and a strong predictive accuracy in combination with a remarkable \textit{polylog} complexity scaling in the number of qubits of the involved quantum circuits. Remaining challenges refer to adaptive ansatz strategies that speed up the optimization procedure.
\end{abstract} 
\clearpage
\section{Introduction}\label{sec:Introduction}

Partial Differential Equations (\textsc{PDE}) are ubiquitous for modeling problems in science and engineering. Areas of application include structural and fluid mechanics, electrodynamics, thermodynamics, and quantum physics. Today, sophisticated and dedicated numerical methods are well-established and widely available in all of those areas. However, the increasing demands to resolve wider ranges of spatial and temporal scales with great precision make these simulations expensive and energy-consuming. Furthermore, chip sizes for classical Central Processing Units (\textsc{CPU}s) are expected to converge over the next decade \cite{Burg2021} due to the limits of reducing the transistor's size further, putting an end to Moore's Law \cite{Khan2018}. 
In regard to increasing computing power, Quantum Computers (\textsc{QC}s) promise to address some future hardware challenges. Two categories for \textsc{QC}-supported \textsc{PDE} solutions have been proposed. They are either based on the direct encoding of the \textsc{PDE} solution in a large quantum circuit or on a Variational Quantum Algorithm (\textsc{VQA}), which evaluates shallow circuits in combination with classical optimization methods \cite{Suau2021}. 

For the first category, Quantum Linear Solvers (\textsc{QLS}) are applied to solve the algebraic equation systems derived from the discretization of linear \textsc{PDE}s. In a pioneering work \cite{Harrow2009}, Harrow, Hassidim, and Lloyd presented the \textsc{HH}L quantum algorithm to compute the solution of a Linear System of Equations (\textsc{LSE}). 
The \textsc{HHL} algorithm promises an exponential speedup, provided that the state preparation and information readout can efficiently be performed on the \textsc{QC}, which is by no means self-evident. The algorithm works best for non-stiff \textsc{LSE} but scales worse than classical methods for stiff problems. In \cite{Montanaro2016}, a modification of the \textsc{HHL} algorithm is applied to solve a \textsc{LSE} originating from a Finite-Element Method while the temporal evolution of a linear \textsc{PDE} is approximated in \cite{Berry2014}. Computational Fluid Dynamic (\textsc{CFD}) applications are reported in \cite{Steijl2018,Gaitan2020,Oz2021,Childs2020,Childs2020a,Cao2013,Chen2021}.
Steijl and Barakos \cite{Steijl2018} used a vortex-in-cell \textsc{CFD} method and employed a quantum Fourier solver for the Poisson equation. In \cite{Gaitan2020} and \cite{Oz2021} the \textsc{PDE} system is spatially discretized into a system of coupled Ordinary Differential Equations (\textsc{ODE}s) and approximated via a Quantum Amplitude Estimation Algorithm (\textsc{QAEA}) proposed by Kacewicz  \cite{Kacewicz2006}. In \cite{Childs2020} and \cite{Childs2020a}, the \textsc{PDE} solution is spatially approximated with truncated Fourier or Chebyshev series, in which the unknown coefficients are determined by a \textsc{QLS}. In the study of Cao et al. \cite{Cao2013}, the \textsc{HHL} is used to solve the Poisson equation within a predictor-corrector method to solve the incompressible Navier-Stokes equation. 
The \textsc{QLS} approaches mentioned above mostly require a large number of qubits devoted to error correction, and the number of operations associated with the \textsc{HHL} exceeds the capabilities of current \textsc{QC} hardware. Hence, \textsc{QLS} methods are more suitable for future fault-tolerant quantum machines and less suitable for \textsc{QC}s of the current Noise Intermediate-Scale Quantum (\textsc{NISQ}) generation \cite{Suau2021}. 

\smallskip 
The \textsc{VQA} is an alternative category to solve \textsc{PDE} problems and is dedicated to state-of-the-art \textsc{NISQ} devices~\cite{Peruzzo2014}. The noise-tolerant procedure is based on the efficient evaluation of a cost function by the \textsc{QC}, while the classical hardware optimizes the parameters of the ansatz circuit that encodes the solution on the \textsc{QC}. As a consequence, the \textsc{PDE} to solve must be reformulated as an optimization problem. The main advantage of this approach is the use of a small number of gate operations, making it more robust to decoherence effects of the quantum registers that compose the circuits. 

\textsc{VQA}s are used in many applications such as the identification of ground states \cite{Peruzzo2014}, or solving \textsc{LSE} \cite{Cerezo2021}. 
Recent applications to fluid dynamics and related \textsc{PDE}s are, for example, reported in \cite{Demirdjian2022,BravoPrieto2019,Kyriienko2021,Sato2021,Leong2022,Leong2023,Jaksch2023a}. Burgers' advection-diffusion equation is studied in \cite{Demirdjian2022} where the discretized sparse matrix problem is decomposed in a finite series of unitary operators and solved by the variational quantum solver proposed in~\cite{BravoPrieto2019}. In \cite{Kyriienko2021}, quantum neural networks are trained with a hybrid \textsc{VQA} to solve systems of nonlinear differential equations that model a supersonic nozzle flow. The work of Sato et al. \cite{Sato2021} is of particular relevance for this study and explores the ability of \textsc{VQA}s to solve the Poisson equation by minimizing the potential energy of an elliptic \textsc{PDE}. To this end, the dynamics of the Poisson equation can be linearly decomposed in a series of parameterized shallow quantum circuits. The idea of energy minimization from Sato et al. \cite{Sato2021} is extended to time-dependent problems in Leong et al. \cite{Leong2022,Leong2023}. 

The present effort aims to combine the quantum framework described in Lubasch et al. \cite{Lubasch2020} with the work of Sato et al. \cite{Sato2021}. We focus on a flexible treatment of engineering boundary conditions on \textsc{QC}s and deploy a Hadamard-test-based structure to gain quantum advantage \cite{Guseynov2023}. The related single-qubit measurements reduce expensive sampling operations to evaluate the results encoded in a quantum state and thus the associated loss of efficiency \cite{Berry2014,Chen2021,Kyriienko2021}.
Accurate, flexible, and robust treatment of diverse boundary conditions is crucial for engineering applications. Despite the importance, a rigorous algorithmic implementation in the context of \textsc{QC} is not trivial, as demonstrated by the variety of specific approaches proposed in the literature. The works of Cao et al. \cite{Cao2013} and Childs et al. \cite{Childs2020} are restricted to Dirichlet conditions, while a related study of Childs et al. \cite{Childs2020a} mirrors the solution into the symmetric and antisymmetric sub-spaces for Neumann and Dirichlet cases, respectively. 
The mixed boundary conditions for the Burgers' equation suggested in \cite{Oz2021} can only be enforced on classical hardware. 
The algorithm suggested in \cite{Costa2019} imposes Dirichlet or Neumann conditions on the discrete wave equation prior to the derivation of the Hamiltonian operator, i.e., the solutions in the Hilbert space (trial functions) need to satisfy the boundary conditions. Suau et al. \cite{Suau2021} present a modification of this approach, where only Dirichlet boundaries are included. 
For \textsc{VQA} applications, the boundary conditions in \cite{Kyriienko2021} are included either as part of the objective function or as an additional constraint on the ansatz function. 
The Dirichlet and Neumann boundary operators employed by Sato et al. \cite{Sato2021} do not form unitary operators. This either requires implementation on classical hardware due to the necessary non-unitary corrections or measuring the whole quantum register, which in turn reduces the efficiency. 

The paper suggests combining a classical CFD ghost-point technique with a unitary decomposition of the boundary operators by modifying the route proposed in \cite{Sato2021}. The decomposed boundary operators are subsequently compiled into quantum gates and finally included in the cost function using a deferred correction strategy. The proposed approach permits the full-quantum implementation of arbitrary Dirichlet, Neumann, and Robin conditions without additional constraints on the trial or ansatz functions. 

The remainder of the paper is structured as follows: Section \ref{sec:Mathematics} introduces the mathematical model, and the boundary treatment is described in Sec. \ref{sec:Discretization}. Subsequently Secs. \ref{sec:Strategy} and \ref{sec:Compilation} outline the computational model. The latter also addresses the \textsc{QC} implementation and the quantum circuits. Section \ref{sec:compilation:complexity} discusses means to reduce the gate complexity of the utilized quantum circuits. Section \ref{sec:Optimization} describes the employed optimization procedure in brief and Sec. \ref{sec:Applications} is devoted to numerical results. The applications are concerned with steady and transient heat conduction. Particular emphasis is given to temperature distributions obtained with various types of boundary conditions, which are compared to results of classical Finite-Differences (\textsc{FD}). Final conclusions and future directions are outlined in Sec. \ref{sec:Conclusion}. 
The presented quantum framework is emulated with \textsc{IBM}'s \textsc{Qiskit} environment \cite{Qiskit}. 
Within the publication, vectors and tensors are defined with reference to Cartesian coordinates. Note that binary representations follow the \textit{little-endian} convention.

\section{Mathematical Model}  \label{sec:Mathematics}

The considered \textsc{PDE} is first introduced and discretized in time. Subsequently, the \textsc{PDE} is cast into an optimization problem, and the objective function is prepared for being evaluated on \textsc{QC} hardware before the spatial discretization is outlined. 

\subsection{Governing Equation} \label{sec:GoverningEq}
The presented approach is restricted to unsteady, spatially \textsc{1D} problems.
An extension to two- or three spatial dimensions is straightforward \cite{Lynch1964}. For brevity, we use a {normalized} spatial coordinate $x \in [0,1]$ and a time domain $t \in [0,T]$. 
The problems considered herein are described by an unsteady reaction/diffusion \textsc{PDE} for an unknown field or state variable $y(x,t)$, viz. 
\begin{align}
   \frac{\partial y }{\partial t}-\nu \frac{\partial^2 y}{\partial x^2} -\zeta y \, p   = f \quad \text{in } \O_\text{T} : =  (0,1) \times (0,T] \, . 
   \label{eq:Generic_PDE-TR}
\end{align}
In \Eq{eq:Generic_PDE-TR}, the notation ${\partial^2 y}/{\partial x^2}$ describes the Laplace operator acting on $y$. 
Moreover, the right-hand side term $f(x)$ denotes a time-independent source, $p(x,t)$ is a time-dependent potential that is 
independent of $y$, $\zeta$ is a simple (real) coefficient, and $\nu$ refers to the inherently positive (real) diffusivity. 
In contrast to previous works published in  \cite{Cao2013,Sato2021,Leong2022,Leong2023,Arrazola2019}, the proposed framework to solve \Eq{eq:Generic_PDE-TR} applies to arbitrary combinations of Dirichlet and Neumann conditions. The details of the boundary treatment will be outlined in Sec. \ref{sec:Discretization}. 

\subsection{Temporal Discretization}
The time horizon $T$ is discretized into $N_\text{t}$ equidistant time instants $t^l$, where $l \in \left[0,N_\text{t}\right]$ and $\Delta t$ marks a constant time step. 
In line with traditional finite approximation strategies,  time derivatives are approximated by implicit backward \textsc{FD}, e.g., an implicit first-order Euler scheme or a second-order three-time-level scheme
\begin{align}
\begin{split}
     \frac{\partial y(x,t^l)}{\partial t} &= \frac{ y(x,t^l) -y(x,t^{l-1}) }{\Delta t} + \mathcal{O}(\Delta t) \, ,\\
 \frac{\partial y(x,t^l)}{\partial t} &= \frac{3 y(x,t^l) -4y(x,t^{l-1}) +y(x,t^{l-2}) }{2\Delta t} + \mathcal{O}(\Delta t^2)
 \; .
\end{split}
\label{eq:Generic_time}
\end{align}
The examples included herein use the implicit Euler scheme. This results in the following modifications of the potential term $p\to p + 1/(\zeta \Delta t)$ and the source term $f \to f+ y(x, t^{l-1})/\Delta t$, which yields a spatial \textsc{PDE} at time $t^l$ in residual form 
\begin{equation}
 R(y(x,t^l)) = 
  - \left( \nu \frac{\partial^2 y}{\partial x^2} + \zeta y \, p   + f \right) = 0 \quad \text{in }   \O : =  (0,1)\, . 
   \label{eq:Generic_PDE-TR-2}
\end{equation}

\subsection{Optimal Control Problem} 
\label{sec:optimal_control}

A weak form of \Eq{eq:Generic_PDE-TR-2} follows from a weighted residual formulation and reads
\begin{align}
 \int_{\O}  z(x,t^l) \, R(y(x,t^l)) \,  dx  = 0 \, , 
  \quad \text{or} \quad  
\underbrace{-\int_{\O} z \Big( \nu \frac{\partial^2 y}{\partial x^2} + \zeta y \, p\Big) \, dx}_{a(z,y)} 
 - \underbrace{\int_{\O} z \, f \, dx}_{F(z)} = 0  \, ,
 \label{equ:weighted_res-2}
\end{align}
where $z(x,t^l)$ is a weighting function. For $z=y$, the solution to \Eq{equ:weighted_res-2} is equivalent to the solution of the variational problem \cite{Grossmann2007,Glowinski2015} characterized by minimizing an objective function $J$,
\begin{align}
    \min\limits_{y} J(y) \quad  \text{with} \quad J(y):= a(y,y) - 2F(y).
    \label{eq:Generic_Opt_Approx}
\end{align}
Mind that the equivalence between the solutions restricts the function space of $y$ to an elliptic $a(z,y)$ \cite{Grossmann2007} and that the computed solution is, of course, only an approximation to $y$.

The limited ability to represent arbitrary states $y$ on \textsc{QC} hardware, yields the \textit{partition of unity} constraint 
given by the $L^2$ scalar product as \cite{Troutman1996,Werner2005}
\begin{align}
    (y^*,y)_{L^2} = \int_{\O}  y^*(x,t^l) \, y(x,t^l) \,  dx =  1 \, ,
    \label{eq:partition_of_unity}
\end{align}
where the asterisk indicates the complex conjugate. Turning our attention to the variational framework \eqref{eq:Generic_Opt_Approx}, the function $y$ is made variable by the control $\upb^l \in \mathbb{R}^{c+1},  \upb^l: = \left(\lambda_0^l, \lamb_c^l\right)^\intercal$ ($c \in \mathbb{N}$), using the time-dependent ansatz 
\begin{align}
    y(x,t^l, \upb^l) = \lambda_{0}^{l} \, u(x, {t^l}, {\lamb_c^l}) \, , \quad \text{with} \quad 
    \big(u^*, u\big)_{L^2} = 1 \, 
        \label{eq:Generic_Ansatz}
\end{align} 
to enforce the normalization restriction \eqref{eq:partition_of_unity}. Suppressing the temporal index $(...)^l$ and substituting the ansatz~\eqref{eq:Generic_Ansatz} into the objective function, one obtains the (time-dependent) minimization problem 
\begin{align}
	\begin{split}
     \min \limits_{\upb} J(\upb)  \quad \text{with} \quad J(\upb) :&= a\big(y(\upb),y(\upb)\big) - 2F(y(\upb)),\\
			 &=  - \nu  \, \lambda_{0}^2 \int_{\O}  u(x, \lamb_{c})  \frac{\partial^2 }{\partial x^2}  u(x, \lamb_{c}) \, dx - \zeta \, \lambda_{0}^2  \int_{\O} u(x, \lamb_{c}) \, p(x) \, u(x, \lamb_{c}) \, dx \\
        &\quad- 2  \lambda_{0} \int_{\O}  u(x, \lamb_{c}) \, f(x) \, dx \, .
	\end{split}
\label{eq:Reduced_Opt}
\end{align}
For each time step, this minimization requires the derivatives of the objective function with respect to the control. The Gâteaux derivative with respect to the control $\upb$ provides a first-order optimality condition \cite{Troutman1996}
\begin{align}
     \frac{\partial J}{\partial \upb} \delta \upb &= {\frac{\partial J}{\partial y} \frac{\partial y}{\partial \upb} \delta\upb} =  \begin{pmatrix}
            \nabla_{\lambda_0} J \\ 
            \nabla_{\lamb_c} J 
        \end{pmatrix}  \cdot \begin{pmatrix}
            \delta \lambda_0\\
            \delta \lamb_c
        \end{pmatrix} =  0 \quad \forall \, \delta\upb.
\end{align}
The partial derivatives of the objective function with respect to the control 
are not discussed in the continuous form here. Instead, a \textsc{QC}-compatible discretized form is developed in Sec. \ref{sec:Optimization}. 
 
\subsection{Spatial Discretization}
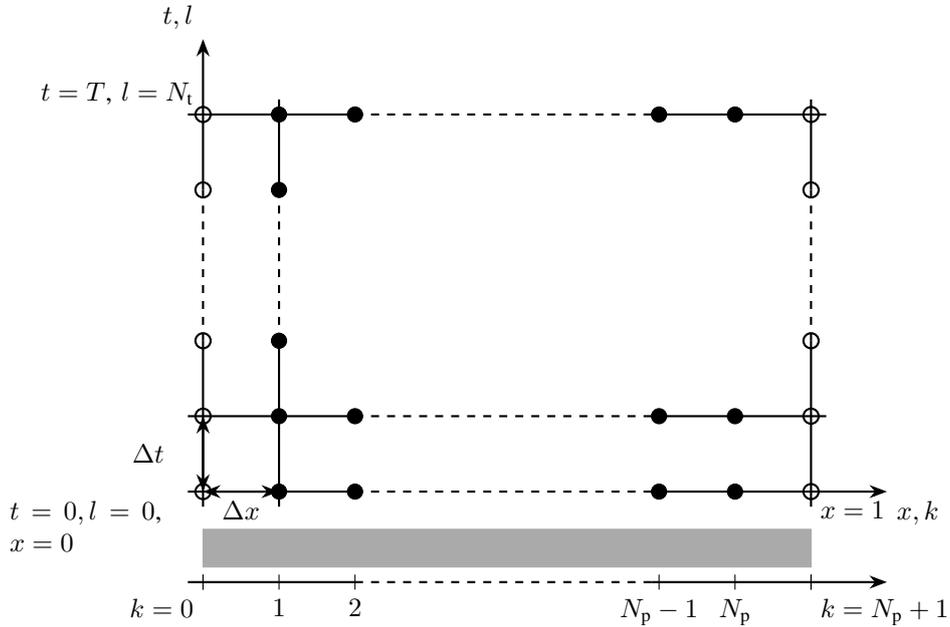
\begin{figure}[htbp]
    \centering
    \tikzsetnextfilename{Discretization}
\begin{tikzpicture}
        \draw (0,1,0) node [below left, text width=2.42cm] {$t=0,l=0$, $x=0$};
        
        \foreach \y in {1,2,6}{
            \draw[thick] (-0.2,\y) -- (2,\y);
            \draw[thick, dashed] (2,\y) -- (6,\y);
            \ifthenelse{\y=1}{
                \draw[thick, -Stealth] (6,\y) -- (9,\y)  node [below right] {$x,k$};
            }{
                \draw[thick] (6,\y) -- (8.2,\y);
            };
            \foreach \x in {1,2,6,7}{
        	   \draw[fill]  (\x,\y) circle [radius=0.1];
            };
        };
        
        \draw (8,1) -- (8,1) node [below right] {$x=1$};
        \draw[thick, Stealth-Stealth] (0,1) -- node [below] {$\Delta x$} (1,1);
        
        \foreach \x in {0,1,8}{
            \draw[thick] (\x,0.8) -- (\x,3);
            \draw[thick, dashed] (\x,3) -- (\x,5);
            \ifthenelse{\x=0}{
                \draw[thick, -Stealth] (\x,5.) -- (\x,7.)  node [above left] {$t,l$};
            }{
                \draw[thick] (\x,5.) -- (\x,6.2);
            };
            
            \ifthenelse{\x=1}{
                \foreach \y in {3,5}{
            	   \draw[fill,black] (\x,\y) circle [radius=0.1];
                };
            }{
                \foreach \y in {1,2,3,5,6}{
            	   \draw[thick,black] (\x,\y) circle [radius=0.1];
            };
            };
        };
        \draw (0,6) -- (0,6) node [above left] {$t=T$, $l=N_\text{t}$};
        \draw[thick, Stealth-Stealth] (0,1) -- node [left,text width=0.8cm] {$\Delta t$} (0,2);
       
        \draw [fill,gray] (0,0) rectangle (8,0.5);
        \draw (0,-0.1) -- (0,-0.3);
        \draw (8,-0.1) -- (8,-0.3);
        \draw [thick] (-0.2,-0.2) -- (2.,-0.2);
        \draw [thick, dashed] (2,-0.2) -- (6,-0.2);
        \draw [thick, -Stealth] (5.9,-0.2) -- (9,-0.2);		
        
        \foreach \x in {1,2}{
        	\draw (\x,-0.1) -- (\x,-0.3) node [below] {$\x$};
        };
        \draw (6,-0.1) -- (6,-0.3) node [below] {$N_\text{p}-1$};
        \draw (7,-0.1) -- (7,-0.3) node [below] {$N_\text{p}$};

        \draw (0,-0.1) -- (0,-0.3) node [below left] {$k = 0$};
        \draw (8,-0.1) -- (8,-0.3) node [below right] {$k = {N_\text{p}+1}$};
\end{tikzpicture}
    \caption{Illustration of the spatial-temporal discretization on an equidistant grid with 
    boundary/ghost nodes indicated by open symbols~(\protect\tikz \protect\draw[thick,black] (1ex,1ex)  circle [radius=0.1];) and interior nodes by closed symbols~(\protect\tikz \protect\draw[fill] (1ex,1ex)  circle [radius=0.1];).}
    \label{fig:discretization_encoding}
\end{figure}

The unit interval is discretized by $N_\text{p}+2$ equidistantly spaced points  $x_k$, with $k \in \left[0,N_\text{p}+1\right]$. As alluded to earlier,  the time $t\in [0,T]$ is discretized by $N_\text{t}+1$ equidistant time instants $t^l$, which leads to the structured grid depicted in \Fig{fig:discretization_encoding}. As indicated by the figure, we distinguish between interior points (closed circles) and ghost points (open circles), which serve to introduce boundary conditions. 
Similar to classical finite approximation schemes, variables located along the spatial boundaries can either be part of the unknowns or separated from the unknowns. In the latter case, the boundary conditions are represented by the influence of prescribed ghost point variables on the interior point approximations. In the case of Neumann or Robin conditions, this usually involves the additional discretization of the spatial derivative.   

The procedure is as follows: Boundary nodes are assigned to ghost nodes (\protect\tikz \protect\draw[thick,black] (1ex,1ex)  circle [radius=0.1];)  and the spatially discretized objective function \eqref{eq:Reduced_Opt} is evaluated for the interior nodes (\protect\tikz \protect\draw[fill,black] (1ex,1ex)  circle [radius=0.1];), with the contributions of the ghost nodes arising from discretized boundary conditions. Thus, the discrete ansatz function vector $\big(u(x_0,t^l), u(x_1,t^l), \dots, u(x_{N_\text{p}+1},t^l)\big)=\ub^l$ is represented using the orthogonal basis $\eb_k$, viz. 
\begin{align}
    \ub^l = u_0^l \eb_0 +  \underbrace{\sum\limits_{k=1}^{N_\text{p}} u_k^l \eb_k}_\text{register} + u_{N_\text{p}+1}^l \eb_{N_\text{p}+1} \, ,  
    \label{eq:discretization_register}
\end{align}
with the interior to be stored in a register of length $\log_2(N_\text{p})$ as shown in Sec.~\ref{sec:Strategy}. The derivatives of the Laplace operator required to evaluate the objective function (\refeq{eq:Reduced_Opt}) are approximated by
\begin{align}
\frac{\partial^2 u^l_k}{\partial x^2} =
\frac{u_{k+1}^l-2u_k^l+u_{k-1}^l}{\Delta x^2} +  \mathcal{O}(\Delta x^2) \, , 
\label{eq:derivatives_space}
\end{align}
which is second-order accurate for an equidistant grid. 

\section{Discrete Objective Function and  Boundary Treatment}  \label{sec:Discretization}
The objective function \eqref{eq:Reduced_Opt} consists of contributions from the Laplace operator $j_L$, the potential $j_P$, and the source term $j_S$. The integration over the interior domain 
in \Eq{eq:Reduced_Opt} is fractioned into sub-integrals around the centrally spaced interior nodes, which are approximated by a  second-order accurate midpoint rule, i.e.,
\begin{equation}
     J^{l} = \int_\Omega (j_L+ j_P+j_S)^l \,  dx = {\sum_{k=1}^{N_\text{p}}} (j_{L_k} + j_{P_k}+ j_{S_k})^l \, \Delta x \, .
     \label{eq:Reduced_Opt-Disc}
\end{equation} 
Subsequently, the discretized boundary conditions are introduced. These are formulated as corrections to established periodic boundary conditions, which yield augmentations of the discrete objective function \eqref{eq:Reduced_Opt-Disc}. The approach is compatible for an evaluation on \textsc{QC} hardware and is discussed below using the left boundary ($k=0$) as an example. The procedure for the right boundary is analogous and is therefore not discussed in detail. The time index is again deliberately suppressed for the sake of clarity.

The potential and source term contributions $j_{P_k}$, $j_{S_k}$ to \Eq{eq:Reduced_Opt-Disc} are strictly local and therefore do not interact with the ghost point (boundary) values. The contribution of the discrete Laplace operator, however, is non-local (but homogeneous) and interacts with the boundary. At the first interior point ($k=1$), the contribution reads
\begin{equation}
j_{L_1} =-\left(\frac{\nu  \, \lambda_{0}^2}{\Delta x^2} \right) \; u_1 \bigg( u_0 - 2u_1 +u_2\bigg) \, , 
 \label{eq:Laplace-left}
\end{equation}
where $u_0$ is an unknown that must be closed by the boundary condition. As already mentioned, the strategy refers to corrections of periodic boundaries, for which the left ghost point receives the value from the right interior point and the right ghost point gets the value from the left interior point, i.e., $u_0 =  u_{N_\text{p}}$ and  $u_{N_\text{p}+1} =  u_{1}$. The periodic contribution of the left boundary to $J$ reads 
\begin{align}
    j_{L_1}^P   =-\left(
 \frac{\nu  \, \lambda_{0}^2}{\Delta x^2 }
  \right) \; u_1 \bigg( u_{N_\text{p}} - 2u_1 +u_2\bigg) \, .
  \label{eq:periodic-left}
\end{align}
On the other hand, a Dirichlet condition $D = u_0$ yields  
\begin{align}
    j_{L_1}^D  =-\left(\frac{\nu  \, \lambda_{0}^2}{\Delta x^2}\right) \; u_1 \bigg( D - 2u_1 +u_2\bigg) \, .
     \label{eq:dirichlet-left}
\end{align}
 Aiming to reconstruct the Dirichlet condition \eqref{eq:dirichlet-left} from the periodic approach \eqref{eq:periodic-left}, one arrives at 
\begin{align}
   j_{L_1}^D = j_{L_1}^P
  + u_1 \, \left(\frac{\nu \lambda_0^2}{\Delta x^2} \right) \left( 
   u_{N_\text{p}} - D
  \right)  = j_{L_1}^P \; \underbrace{+ u_1 u_{N_\text{p}} \, \left(\frac{\nu \lambda_0^2}{\Delta x^2}\right)}_{\text{new J$_1$-contrib.} \, j_{DN_1}} \;   \underbrace{-u_1 D \, \left(\frac{\nu \lambda_0^2}{\Delta x^2}\right)}_{\text{source-term modif.} \, f_1}  \, .
   \label{eq:defcoD-left}
\end{align}
In the case of Neumann conditions, the prescribed gradient $N= \partial u/\partial x\rvert_0$ first needs to be approximated. The exemplary use of simple first-order \textsc{FD} $N \approx (u_1-u_0)/(\Delta x)$ to substitute $u_0 = u_1 - N \Delta x $ in \Eq{eq:Laplace-left} yields 
\begin{align}
\begin{split}
    j_{L_1}^N   &=-\left(\frac{\nu  \, \lambda_{0}^2}{\Delta x^2}\right) \; u_1 \bigg( (u_1 -N \Delta x) - 2u_1 +u_2\bigg) \\
    &= 
  j_{L_1}^P
   \; \underbrace{+ u_1 u_{N_\text{p}} \, \left(\frac{\nu \lambda_0^2}{\Delta x^2}\right)}_{\text{new J$_1$-contrib.} \, j_{DN_1}} \; 
     \; \underbrace{- u_1 u_{1} \, \left(\frac{\nu \lambda_0^2}{\Delta x^2}\right)}_{\text{new J$_1$-contrib.} \, j_{N_1}} \; 
  \underbrace{+ u_1 \, N \left( \frac{\nu \lambda_0^2}{\Delta x} \right)}_{\text{source-term modif.} f_1}
  \, .
\end{split}
  \label{eq:defcoN-left}
\end{align}
Note that the contribution $j_{DN_1}$, which neutralizes the periodic term, occurs for any conditions. The technique can be combined with higher-order \textsc{FD} to approximate derivative expressions in Neumann or Robin conditions. 
To demonstrate this procedure, a central second-order \textsc{FD} approximates the left boundary at $x_0$, viz. $N \approx (u_{1}-u_{-1})/(2 \Delta x)$. The terms $u_{-1}$ and $u_{0}$ are recursively eliminated by combining the discrete Laplace operator of \Eq{eq:Laplace-left} for $x_{0}$ and $x_{1}$ and the second-order derivative approximation at $x_{0}$. In the case of second-order Neumann, the modifications result in the additional contribution of $ 0.5 u_1 f_0 \nu \lambda_0^2$ to the source term $f_1$. The $f_0$ indicates the extrapolation of the right-hand side $f$ in \Eq{eq:Generic_PDE-TR} to $x_0$. Analogously, the source term on the right boundary $f_{N_\text{p}}$ is modified by $0.5 u_{N_\text{p}} f_{N_\text{p+1}} \nu \lambda_0^2$.

At the right boundary, the corresponding correction terms read
 \begin{align}
   j_{L_{N_\text{p}}}^D 
= j_{L_{N_\text{p}}}^P
  + u_{N_\text{p}} \, \left(\frac{\nu \lambda_0^2}{\Delta x^2} \right) \left( 
   u_{1} - D
  \right) 
  = j_{L_{N_\text{p}}}^P 
   \; \underbrace{+ u_1 u_{N_\text{p}} \, \left(\frac{\nu \lambda_0^2}{\Delta x^2}\right)}_{\text{new J$_{N_\text{p}}$-contrib.} \, j_{DN_{N_\text{p}}}} \;   \underbrace{-u_{N_\text{p}} D \, \left(\frac{\nu \lambda_0^2}{\Delta x^2}\right)}_{\text{source-term modif.} \, f_{N_\text{p}}} \,
   \label{eq:defcoD-right}\\
   \text{ and } \qquad  j_{L_{N_\text{p}}}^N   = j_{L_{N_\text{p}}}^P
   \; \underbrace{+ u_1 u_{N_\text{p}} \, \left(\frac{\nu \lambda_0^2}{\Delta x^2}\right)}_{\text{new J$_{N_\text{p}}$-contrib.} \, j_{DN_{N_\text{p}}}} \; 
     \; \underbrace{- u_{N_\text{p}} u_{N_\text{p}} \, \left(\frac{\nu \lambda_0^2}{\Delta x^2}\right)}_{\text{new J$_{N_\text{p}}$-contrib.} \, j_{N_{N_\text{p}}}} \; 
  \underbrace{- u_{N_\text{p}} \, N \left( \frac{\nu \lambda_0^2}{\Delta x} \right)}_{\text{source-term modif.} f_{N_\text{p}}}
  \, .
  \label{eq:defcoN-right}
\end{align}

Again, the contribution $j_{DN_{N_\text{p}}}$, which neutralizes the periodic entries, occurs for both Neumann and Dirichlet conditions, and it is identical for both boundary locations ($j_{DN_{N_\text{p}}}=j_{DN_{1}}$). 
The actual correction implementation is given in \Eq{eq:Discrete_Opt_BC}. 
It employs three terms: (a) the periodic baseline term, (b) corrections based upon combinations of unknown interior points ($j_{DN_k}$, $j_{N_k}$), and (c) modifications of the source terms $j_{S_k}$ arising from known boundary condition values ($D$, $N$), viz. 
\begin{align}
\begin{split}
   J(\lambda_0, \lamb_c) =
   &\sum\limits_{k=1}^{N_\text{p}} \underbrace{ -\left(\frac{\nu \lambda_{0}^2}{\Delta x^2}\right) u_k(\lamb_{c}) \big(u_{k+1}(\lamb_{c})-2u_k(\lamb_{c})+u_{k-1}(\lamb_{c})\big) }_{j_{L_k}: \; \text{periodic \& interior term}} \Delta x \, \\
    & \underbrace{+\left(\frac{\nu \lambda_{0}^2}{\Delta x^2}\right) \bigg( u_1(\lamb_{c})u_{N_\text{p}}(\lamb_{c}) + u_{1}(\lamb_{c}) u_{N_\text{p}}(\lamb_{c}) \bigg)}_{ j_{DN_1} + j_{ DN_{N_\text{p}} }: \, \text{neutralizing left/right term (Dirichlet \& Neumann)} } \Delta x   \\
    & \underbrace{-\left(\frac{\nu \lambda_{0}^2}{\Delta x^2}\right)  u_1^2 (\lamb_{c})}_{j_{N_1}:  \, 
   \text{left Neumann cond.}} \Delta x 
   \underbrace{-\left(\frac{\nu \lambda_{0}^2}{\Delta x^2}\right)    u_{N_\text{p}}^2 (\lamb_{c}) }_{j_{N_{N_\text{p}}}: \,   
   \text{right Neumann cond.}} \Delta x \,\\
   + &\sum\limits_{k=1}^{N_\text{p}} \, \underbrace{ -\zeta \lambda_{0}^2 u_k(\lamb_{c}) p_k u_k(\lamb_{c})  }_{{j_{P_k}: \, \text{potential term}}}  \, \Delta x + \sum\limits_{k=1}^{N_\text{p}} \underbrace{- 2\lambda_{0} u_k(\lamb_{c}) \Tilde{f}_k } _{j_{S_k}: \, \text{source term}} \, \Delta x.
    \label{eq:Discrete_Opt_BC}
    \end{split}
\end{align}

This approach is applied {in the sense of the desired boundary conditions} for all time steps and can be adapted to combinations of Dirichlet, Neumann, {Robin} or periodic conditions. \Equ{eq:Discrete_Opt_BC} utilizes a modified source term $\tilde f_k$ which inheres the additional  sources  given in Eqns. (\ref{eq:defcoD-left}-\ref{eq:defcoN-right}). For the example of a Dirichlet conditions on the left boundary, cf. \Eq{eq:defcoD-left}, the source term 
modifies to $j_{S_1} = - 2 \lambda_0 u_1 \tilde f_1$, with $\tilde f_1 = f_1 + 0.5  D \nu \lambda_0/\Delta x^2 $.

\medskip
The following two sections outline the framework for evaluating \Eq{eq:Discrete_Opt_BC} on a QC. 
Sec. \ref{sec:Strategy} describes the \textsc{VQA} framework, which shares features with \cite{Lubasch2020}. Subsequently,  the quantum circuits used to calculate the different contributions ($j_{L_k}, j_{S_k}, j_{P_k}$) to \Eq{eq:Discrete_Opt_BC} on a QC are described in Sec. \ref{sec:Compilation}. The gate
complexity of these circuits is optimized in Sec. \ref{sec:compilation:complexity} before they will be used to approximate the derivative of the objective function with respect to the parameters in Sec. \ref{sec:Optimization}, which in turn is required to advance the optimization.

\medskip
Before we continue with the \textsc{QC} implementation of the different objective function contributions, we remark that the Neumann terms $ j_{N_1}$ and $j_{ N_{N_\text{p}}}$ could also be cast into potential term contributions derived from the reciprocal of the diffusive time scale, i.e., $\tilde p_1 = p_1 + \nu/(\zeta \Delta x^2)$ and $\tilde p_{N_\text{p}} = p_{N_\text{p}} + \nu/(\zeta \Delta x^2)$. 

\section{Variational Quantum Algorithm} \label{sec:Strategy}

As outlined above, a quantum register of $n =\log_2\left(N_\text{p}\right)$ qubits is required to evaluate the objective function on a \textsc{QC}, cf. \Eq{eq:discretization_register}. Each qubit possesses two distinguishable computational states $\ket{0}$ and $\ket{1}$, which form an orthonormal basis of a complex Hilbert space and are frequently identified with basis vectors $\ket{0} = \left(1,0\right)^\intercal$ and $\ket{1}= \left(0,1\right)^\intercal$ \cite{Nielsen2010}.
The $n$ qubit register spans a Hilbert space whose basis is formed by tensor products of the computational basis states of individual qubits, e.g., $\eb_{i=0} = \ket{0}^{\otimes n}=\ket{0}\otimes\ket{0}\otimes~\hdots~\otimes~\ket{0} $, where ${\otimes n}$ indicates an n-fold tensor product. Thus, the $i$-th element of the orthonormal basis can be expressed via the binary representation $\text{bin}(i)$ as $\ket{\text{bin}(i)} = \eb_i$, following the \textit{little-endian} convention where the least significant qubit corresponds to the right-most register position ($q_1$) and the most significant qubit corresponds to the left-most register position ($q_n$), i.e., $\ket{q_n~q_{n-1}~\hdots~q_1}$. Furthermore, the least significant qubit ($q_1$) is drawn as the uppermost qubit in the network figures and the lowermost qubit ($q_n$) indicates the most significant. This encoding enables the representation of the discrete vector $\ub$ by the amplitudes of the quantum register, viz.  $\ub = \sum\limits_{k=1}^{\text{N}_p} u_k \eb_{k-1} =  \sum\limits_{k=1}^{\text{N}_p}u_k \ket{\text{bin}\left(k-1\right)}= \ket{\ub}$, where the offset of the indices results from the technicality that the basis vectors $\eb_k$ start at $k=0$ but the interior unknowns $u_k$  begin with $k=1$. Mind that, though complex $u_k$ are generally allowed, we assume real solutions of \Eq{eq:Generic_PDE-TR}, hence $u_k^* =  u_k$, and that the time index is mostly suppressed in this section to improve the readability. 

\subsection{Variational Quantum Algorithm} \label{sec:VQA}
\begin{figure}[htbp]
    \centering
\tikzexternaldisable
\tikzstyle{decision} = [diamond, draw, text width=5em, text badly centered, node distance=3cm, inner sep=0pt]
\tikzstyle{decisionklein} = [diamond, draw, text width=3.5em, text badly centered, node distance=3cm, inner sep=0pt]
\tikzstyle{decisiongross} = [diamond, draw, text width=5em, text badly centered, node distance=3cm, inner sep=0pt]
\tikzstyle{block} = [rectangle, draw, text width=10em, text centered,  minimum height=4em]
\tikzstyle{blockklein} = [rectangle, draw, text width=6em, text centered,  minimum height=4em]
\tikzstyle{blockgross} = [rectangle, draw, text width=8em, text centered,  minimum height=4em]
\tikzstyle{blockmitte} = [rectangle, draw, text width=9em, text centered,  minimum height=4em]
\tikzstyle{blockmitteclean} = [rectangle, text width=8em, text centered,  minimum height=4em]

\tikzstyle{quantum} = [rectangle, draw, text width=9em, text centered,  minimum height=2em, fill=gray]
\tikzstyle{classical} = [rectangle, draw, text width=8em, text centered,  minimum height=2em]
\tikzstyle{exit} = [rectangle, draw, text width=5em, text centered,  minimum height=4em]
\tikzstyle{blockk} = [rectangle, draw=blue, text width=7em, text centered, rounded corners, minimum height=4em] 
\tikzstyle{cloud} = [draw, ellipse, node distance=4cm,minimum height=2em]
\tikzstyle{line} = [draw]
\tikzstyle{arrow} = [draw, -latex] 
\tikzstyle{stopblock} = [rectangle, minimum width=1cm, text width=3.5cm, minimum height=1cm,text centered]
\tikzstyle{startblock} = [rectangle, minimum width=1cm, text width=3.5cm, minimum height=1cm,text centered]
  
\begin{tikzpicture}[node distance = 1.5cm, scale=1]
    \node [blockmitte] (opt) {$\min\limits_{\upb^{l}}J^{l}\left(\upb^{l}\right)$};
    \node[startblock, left of= opt,xshift=-2cm](start){ \textbf{start}\;$\bullet$};
    \node[stopblock, right of=opt,xshift=+1.5cm](stop){$\bullet$\; \textbf{stop}};

    \node [blockklein, below of= opt,xshift=-2.5cm, node distance = 3.2cm](co1){Evaluate the Objective $J(\lambda_0, \lamb_c)$ \eqref{eq:Discrete_Opt_BC}}; 
    \node [decisionklein, below of=co1,yshift=+1cm] (co2){conv.?};
    \node[blockgross, below of= co2, yshift=-0.5cm] (co3){Update\\
    $\upb_{i+1} = \upb_{i} + \Delta\upb$};

    \node [quantum, right of=co1, xshift=3.5cm,yshift=-1.4cm](qc1){ Evaluate\\[3mm] $\sum\limits_{k=1}^{N_\text{p}}j_{L_k} \Delta x,$\\[1mm] $j_{DN_{1,N_\text{p}}}\Delta x, $\\[1mm]
    $j_{N_{1,N_\text{p}}}\Delta x, $\\[1mm]
    $\sum\limits_{k=1}^{N_\text{p}}j_{P_k}\Delta x, $\\[1mm]
     $\sum\limits_{k=1}^{N_\text{p}}j_{S_k}\Delta x$ \\[1mm]
     $\vdots$ 
     };

\node[draw,inner xsep=4mm, inner ysep=2.5mm, dashed,overlay, label=below : Optimization ,fit=(co1) (co2) (co3) (qc1) ] (Optimization_Box) {};
\node [classical, below of=co2, xshift=0cm, yshift=-2.5cm ](CDevice){\textbf{Classical Device}};
\node [quantum, right of= CDevice, xshift=3.5cm](QDevice){\textbf{Quantum Device}};

\path [arrow] ([xshift=-1.3cm]start.east) -- (opt.west);
\path [arrow] (opt.east) -- ([xshift=+1.3cm]stop.west);
\path [arrow] ([xshift=-0.1cm]opt.south) |- node [anchor=east] {} ([xshift=+1.3cm, yshift=-1cm]start.south) -| (co1.north);
\path [arrow] ([yshift=0.2cm]co1.east) -- node [anchor=south] {$\lamb_{{c}_i}$} ([yshift=1.6cm]qc1.west);
\path [arrow] ([yshift=1.2cm]qc1.west) --node [anchor=north] {$J_i$} ([yshift=-0.2cm]co1.east);
\path[arrow] (co1.south) -- (co2.north);
\path[arrow] (co2.east) -| node [anchor=north] {yes} ([xshift=+0.1cm]opt.south);
\path[arrow] (co2.south) --node [anchor=east] {no}  (co3.north);
\path[arrow] (co3.west) -| ++(-0.55,0) -- ++(0,2) -- ++(0,0.55)
|- node[xshift=+0.1cm, yshift=-2.0cm, text width=2.5cm,anchor=west] {$i+1$}(co1.west);
\end{tikzpicture}
\tikzexternalenable
    \caption{Sketch of the hybrid  \textsc{VQA} optimization (time dependencies {in the optimization step} are suppressed for the sake of clarity). 
    }
    \label{fig:optimizer}
\end{figure}
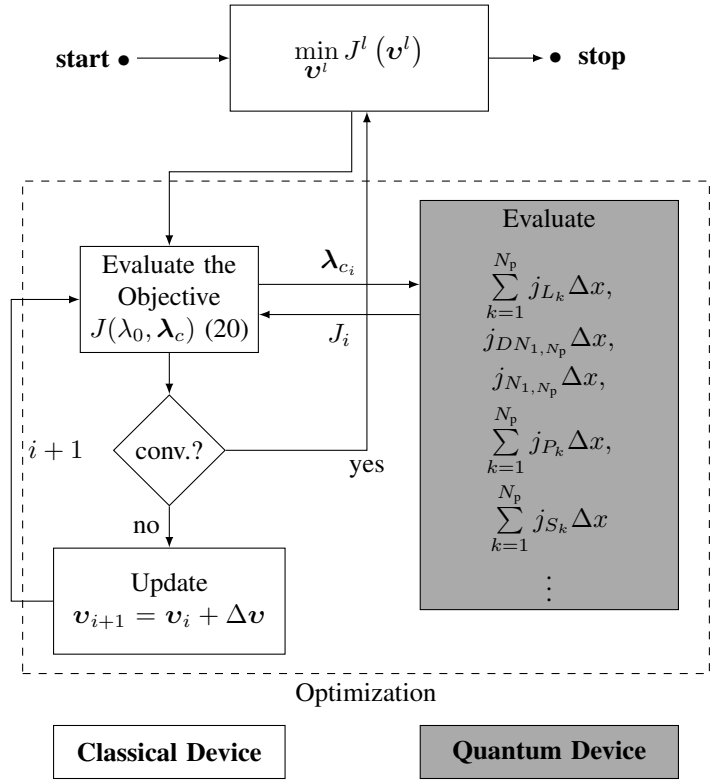

The \textsc{VQA} methodology aims to minimize an objective function $J^{l}$ given in \Eq{eq:Discrete_Opt_BC}, by optimizing the control parameters $\upb^{l}= (\lambda_0^{l}, \lamb_c^{l})^\intercal$ for every time step $l$. The hybrid framework utilizes both classical and quantum hardware, as displayed in \Fig{fig:optimizer}. 
The function $J^l$ is evaluated on the quantum device, as illustrated by the shaded box in Fig. \ref{fig:optimizer}, whereas the optimization process runs on the classical hardware that controls the process and iteratively adjusts the control $\upb$. For a particular set of ansatz parameters $\lamb_c$, the objective function is evaluated from dedicated, individual so-called Quantum-Nonlinear-Processing-Units (\textsc{QNPU}) for each group of terms in \Eq{eq:Discrete_Opt_BC}, i.e., for $j_{L_k},{j_{DN}, j_{N}}, j_{P_k}, j_{S_k}$, cf. Sec. \ref{sec:Compilation}. An update $\Delta\upb$ is determined by the optimizer as outlined in Sec. \ref{sec:Derivative1} before a new iteration starts.

\begin{figure}[htbp]
\centering
    \begin{subfigure}[t]{1.0\textwidth}
        \centering
        \input{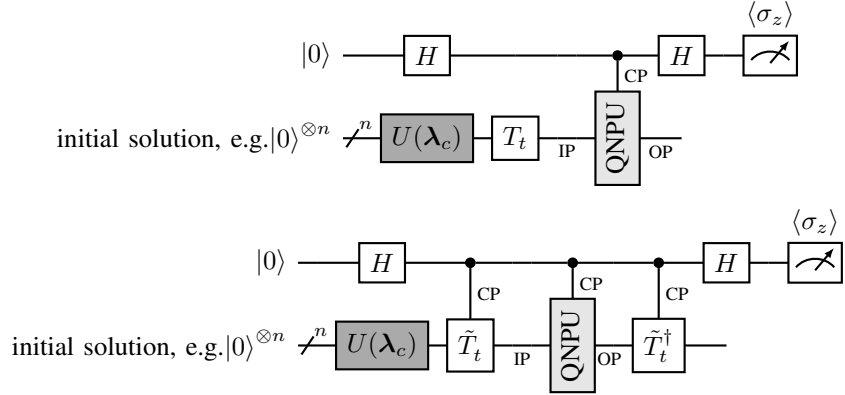}
        \caption{Sketch of two \textsc{VQ} circuits composed of the ansatz $U(\lamb_c)$, the $T_t$ or $\tilde T_t (\tilde T^\dagger_{t})$ transformation, a specific \textsc{QNPU}, and a measurement qubit (ancilla qubit) for indirect measurements. $H$ indicates a Hadamard gate, \textsc{CP} denotes a control port, \textsc{IP} marks the input port \& \textsc{OP} the output port, and $\langle \sigma_z \rangle$ provides the expectation value of the measurement operation, cf. \cite{Lubasch2020}. 
        }
        \label{fig:VQA_QNPU}
    \end{subfigure}
    \begin{subfigure}[t]{1.0\textwidth}
         \vspace*{2ex}
        \centering
        \input{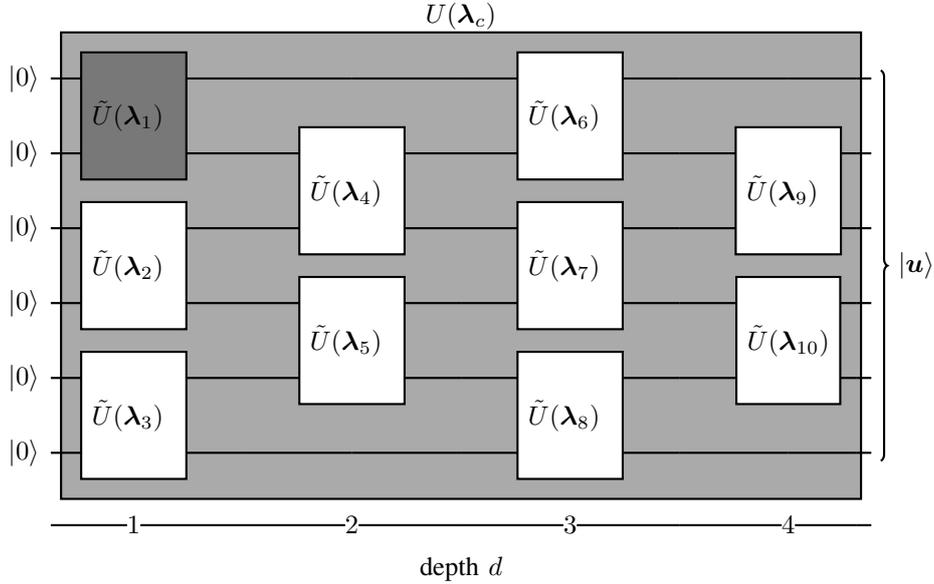}
        \caption{Example of an ansatz circuit $U(\lamb_c)$  composed from  $a=10$ fundamental gates $\Tilde{U}(\lamb_i)$ featuring a bricklayer topology,  parameterized by $\lamb_c = (\lamb_1, \lamb_2,\hdots, \lamb_a)^\intercal$ parameters to encode a test function $\ket{\ub}$ on $n = 6$ qubits, cf. \cite{Lubasch2020}. }
        \label{fig:VQA_Ansatz}
    \end{subfigure}\\
    \begin{subfigure}[b]{1.0\textwidth}
     \vspace*{2ex}
        \centering
        \input{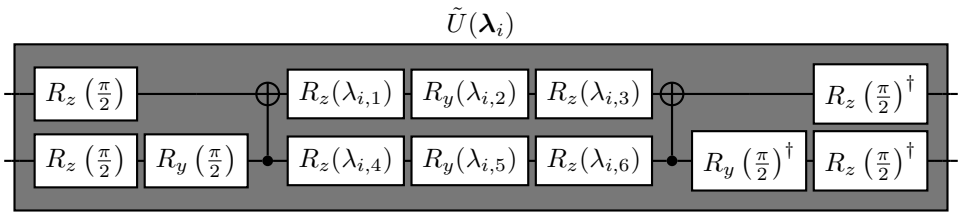}
        \caption{Two qubit \textsc{\it SO4} ansatz $\Tilde{U}(\lamb_i)$ 
        to create an optimal variational state \cite{Vatan2004} based on $\lamb_i = \left(\lambda_{i,1},\lambda_{i,2},\hdots,\lambda_{i,6}\right)^\intercal$ individual parameters. In the first layer ($d=1$) the ansatz has only three parameters in $\lamb_i$, the rotation gates corresponding to $\lambda_{i,4}$, $\lambda_{i,5}$, and $\lambda_{i,6}$ are identity gates.}
        \label{fig:quantum_ansatz_SO4}
    \end{subfigure}\\      
    \caption{Illustration of the employed nested \textsc{VQA} circuits.}
    \label{fig:VQA}
\end{figure}

Figure \ref{fig:VQA_QNPU} describes two {equivalent} options of the general layout of circuits on the quantum device. Grossly speaking, both options consist of a measuring  (aka. ancilla) qubit in the upper ``wire''  and calculation qubits in the lower ``wires''. In general, the qubits are initialized to the base state $\ket{0}$. The calculation qubits are then introduced to an ansatz circuit called $U$, highlighted in grey in \Fig{fig:VQA_QNPU}. The ansatz $U$ can be denoted by a $N_\text{p} \times N_\text{p}$ matrix composed of an embedded combination of two-qubit units, i.e., fundamental gates $\tilde U (4 \times 4)$, as illustrated in \Fig{fig:VQA_Ansatz} for six qubits. The particular combination employed in this research refers to a scalable depth bricklayer topology and shares some similarities with machine learning topologies. The benefit of the bricklayer topology over alternative topologies has been communicated in \cite{Nakaji2021}.

The employed fundamental gate is displayed in dark grey in Figs. \ref{fig:VQA_Ansatz}-\ref{fig:quantum_ansatz_SO4}. It has been suggested by Vatan and Williams \cite{Vatan2004} and is frequently labeled \textsc{\it SO4}. This gate operates in real and imaginary space, which is rather unusual for engineering applications that aim for real-valued results \cite{Sato2021,Leong2022}. Its application promises a richer ansatz function space that, in turn, supports the use of shallower depth $d$ in $U$.
As displayed in \Fig{fig:quantum_ansatz_SO4}, the fundamental gate consists of \textsc{R$_y$}, \textsc{R$_z$} and \textsc{CNOT} gates, where \textsc{R$_y$} and \textsc{R$_z$} are parameterized versions of classical Pauli $\sigma_y$ and $\sigma_z$ matrices \cite{Nielsen2010}, cf. App.~\ref{app:Fundamentals}, and the $(\dots)^\dagger$ symbol indicates the transposed of the complex conjugate. 

Turning back to \Fig{fig:VQA_QNPU}, the result of the ansatz circuit $U(\lamb_c)$ is the parameterized state $\ket{\ub(\lamb_c)}$. 
Subsequently, a particular transformation matrix $T_t$ or $\tilde T_t$ is introduced
in case of boundary contributions $j_{DN}$ and $j_{N}$ before the data enters a  boundary-specific \textsc{QNPU}. 
The transformation essentially modifies the order of qubits in the register $\ket{\ub}$ where necessary. As will be shown in Sec.~\ref{sec:compilation:complexity}, two implementations are conceivable, cf. \Fig{fig:VQA_QNPU}, by either using highly complex multi-controlled gates or by employing additional carry qubits. 

The prepared state ($T_t \ket{\ub}$ or $\tilde T_t \ket{\ub}$) inputs to the main \textsc{QNPU} gate, indicated in light grey in \Fig{fig:VQA_QNPU}, through the input port (\textsc{IP}) and outputs the specific objective function contributions for a group of terms ($j_{L_k},j_{DN}, j_{N}, j_{P_k}, j_{S_k}$) through and output port (\textsc{OP}). Finally, the register is reversed to its initial order after applying the boundary-specific \textsc{QNPU}. To this end, the  Hadamard test method inherently reverses the gates in the upper variant of \Fig{fig:VQA_QNPU}. In contrast, the implementation example at the bottom of \Fig{fig:VQA_QNPU} requires the explicit application of $\tilde T^\dagger_{t}$. 
Note that for periodic conditions, where neither $j_{DN}$ nor $j_{N}$ contributions occur, no transformation is needed and $T_t, \tilde T_t, \tilde T_t^\dagger$ simply degenerates to the identity matrix $I$ in line with the upper part of \Fig{fig:VQA_QNPU}. 
The use of the Hadamard test \cite{Endo2020} is motivated by previous studies \cite{Guseynov2023}, which indicated quantum advantage with this approach. Here, rather than assessing the individual contributions at each grid point $x_k$ \cite{Endo2020},  
the integral over all interior points is directly compiled, for example $\sum_k j_{L_k}$, and measured statistically on the upper ``wire'' by the expectation value denoted by $\langle \sigma_z \rangle$ \cite{Nielsen2010}. This is indicated by the upper ``wire'' of \Fig{fig:VQA_QNPU} that connects to the calculation ``wires'' through a control port (\textsc{CP}). 

Before we discuss the individual \textsc{QNPU}s in Sec.~\ref{sec:Compilation}, the measurement strategy implemented here should be described for the sake of completeness. We rely on the density (or autocorrelation) matrix $\rho_\text{ancilla}$ of the ancilla qubit to approximate the expectation $\langle \sigma_z \rangle$ at the measurement gauge in \Fig{fig:VQA_QNPU}. To eliminate stochastic influences, evaluating \textsc{QC}-calculated results is usually based on evaluating sufficiently large samples. However, if the \textsc{QC} is only emulated, as in the current study, the procedure can be simplified using the trace-out method~\cite{Nielsen2010}. 
First, the whole system state $\ket{\psib}$ (ancilla and calculation qubits) is computed by a state-vector simulation. As we are only interested in the information contained in the ancilla portion of the state $\ket{\psib}$, the remainder is traced out, and the density operator $\rho_\text{ancilla}= \text{Tr}_\text{QNPU}\{ \ket{\psib}\bra{\psib}\}$ of the ancilla remains. The state-vector simulation thus yields the same result as the measurement of the ancilla performed on real hardware, without noise and sampling errors \cite{Nielsen2010}. 

\section{Quantum-Nonlinear-Processing-Units}
\label{sec:Compilation}
For the introduction of our library of {Quantum-Nonlinear-Processing-Units} (\textsc{QNPU}s), we use a complex number notation. All \textsc{QNPU}s are highlighted in light grey in the respective figures, which links them to the \textsc{VQ} circuits depicted in \Fig{fig:VQA_QNPU}. They are adjustable to an arbitrary equidistant one-dimensional discretization and have been designed to keep the gate sequence shallow. Quantum gates are often associated with matrices or matrix operations. Therefore, the interested reader is referred to App.~\ref{app:MatrixGateConversion} for a brief introduction to converting gate sequences into matrices. 

\subsection{Periodic Laplace operator} \label{sec:Compilation:Laplace}

The Laplacian is given by the {\textsc{FD}} stencil in \Eq{eq:derivatives_space} and partly resembles the classical bit shift operator \cite{Nielsen2010,Vedral1996} due to shifted products, e.g., $u_k^{l*} u_{k+1}^l$, $u_k^{l*} u_{k-1}^l$. 
In contrast to other approaches \cite{Sato2021,Liu2021} the \textsc{QNPU} circuit employed herein is based on the adder circuit with a polynomial scaling of the gate complexity, cf. Sec.~\ref{sec:compilation:complexity}. 
Using $n=2$ qubits, a half-adder circuit can be used, which consists of one \textsc{CNOT} gate and an additional Toffoli gate. For $n > 2$, additional ($n-2$) carry qubits (auxiliary qubits) are required, which results in a full-adder 
circuit, cf. \cite{Parhami2010, Draper2000}. 
Assuming a periodic state, it can be shown \cite{Lubasch2020} that the general structure of the discrete Laplace operator is represented by  
\begin{align}
\label{eq:Laplace}
    \sum\limits_{k=1}^{N_\text{p}}u_k \ket{\text{bin}\left(k-1\right)} \to  \sum \limits_{k=1}^{N_\text{p}} u_{k+1}  \ket{\text{bin}\left(k-1\right)} \, , \qquad \text{with} \qquad {1= \sum_{k=1}^{N_\text{p}} u_k^* u_k}.
\end{align}

\begin{figure}[htbp]
    \centering
    \input{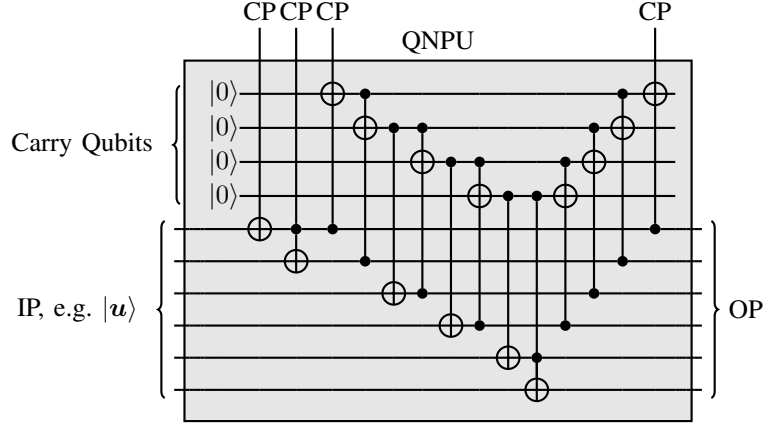}
    \caption{Exemplary quantum circuit of the discrete Laplacian in the Hadamard test configuration. The example refers to $n=6$ qubits and  $n-2=4$ carry bits. It requires $n-2=4$ \textsc{CNOT} and $2n-2=10$ Toffoli gates. For $n=2$ qubits, this circuit is exactly implementing a half-adder, cf. \cite{Lubasch2020}.
    }
    \label{fig:Laplace_compiled}
\end{figure}

Mind that an adder \textsc{QNPU} to compute a periodic Laplace operator has previously been reported by Lubasch et al. \cite{Lubasch2020}.  \Fig{fig:Laplace_compiled} depicts an example of the employed Laplace QNPU circuit 
for $n=6$ qubits and $n-2 = 4$ carry bits. The circuit involves $n-2=4$ \textsc{CNOT} gates and $2n-2=10$ Toffoli gates. 

\subsection{External Contributions (potential and source terms)} \label{sec:Compilation:Potential}

The computation of external contributions follows from two subsequent steps. First, the representation of the discrete linear source  $\tilde f_k^l$ and potential terms $p_k^l$  are assembled. This enables computing their contributions to the objective function {($j_S$,$j_P$)} in a second step. 

For simple, analytically described terms, e.g., $exp(x)$ or $sin(x)$,  Lubasch et al.  \cite{Lubasch2020} introduced corresponding gate sequences that allow such functions to be represented on a \textsc{QC}. Other suggestions intensively deploy Matrix-Product-State (\textsc{MPS}) techniques, \cite{Oseledets2012,GonzalezConde2023,Oseledets2010}. 
In our case, however, we first solve a prior optimization problem to represent arbitrary external contributions by quantum gates, taking the ansatz $U(\lamb)$ into account, viz. 
\begin{align}
     \bar{\lamb}_c =  \min\limits_{\lamb_c} \big(1 -\lambda_0\bra{\ub(\lamb_c)} \ket{\gb}\big) \text{ with } \frac{1}{\lambda_0^2}=\sum \limits_1^{N_\text{p}} g_k^{l*} g_k^l \, . 
    \label{eq:OptAnsatz}
\end{align}

In \Eq{eq:OptAnsatz} $\gb = \Tilde{\fb}$ or $\gb = \pb$ for the source and the potential term, respectively. The benefit of this approach is that it re-uses existing implementations and also keeps the depth of the resulting circuits shallow. 
The optimal parameters $\bar{\lamb}_c$ determined in this way result in the required gates $U(\bar{\lamb}_c) = P$ and $U(\bar{\lamb}_c) =  {F}$ to be used inside the \textsc{QNPU}s that compute the potential and the source term contributions to $J$ in \Eq{eq:Discrete_Opt_BC}.

\begin{figure}[htpb]
    \centering
    \begin{subfigure}[t]{0.48\textwidth}
        \centering
        \input{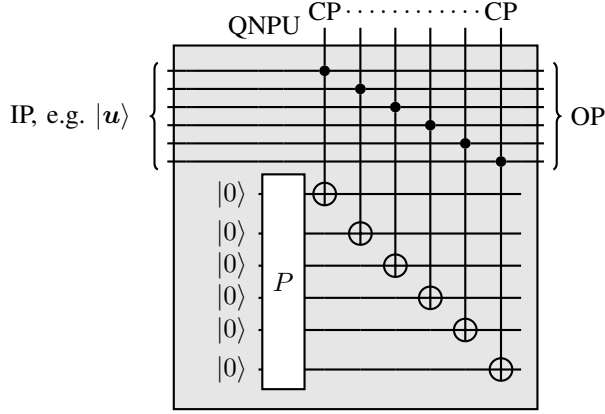}
        \caption{Exemplary quantum circuit for a discrete potential function $\pb=( p_1, p_2,p_3, \dots p_{N_\text{p}})^\intercal$ represented by the gate ``$P$''. }
        \label{fig:QNPU_Potential}
    \end{subfigure}
    \hspace{0.2cm}
    \begin{subfigure}[t]{0.48\textwidth}
        \centering
        \input{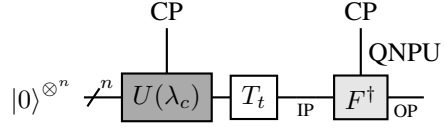}
        \caption{Evaluation of external source term $f$  by $\bra{\ub}\ket{\fb}$, with $T_t = I$.}
        \label{fig:overlapp}
    \end{subfigure}
    \caption{External contributions on the Hadamard test. The displayed example refers to $\text{N}_p=64$, $n=6$ qubits.}
    \label{fig:externals}
\end{figure}

The contribution of the potential term $p(x_k, t^l)$ to $J$ follows from $\sum\limits_{k=1}^{\text{N}_p} \left[ \lambda_0^2 u_k^*p_k u_k \right]^l $. The corresponding \textsc{QNPU} circuit is assembled by Toffoli gates according to \Fig{fig:QNPU_Potential}. Since the product $u_k^* p_k u_k$ is non-unitary, additional $n$ ancilla qubits occur in the circuit \cite{Sarma2023}. Similarly, the representation of $\tilde{\ket{\fb}} = U(\bar{\lamb}_c)\ket{0}$ is implemented into the source term \textsc{QNPU} (light grey) given in \Fig{fig:overlapp}. Note that thanks to the versatility of the \textsc{\it SO4} ansatz, the minimum depth $d=1$ is usually sufficient to reconstruct $\tilde \fb$ and $\pb$. 
Instead of the uncontrolled ansatz $U(\lamb)$ displayed in Fig. \ref{fig:VQA_QNPU}, the linearity of the source term requires that $U(\lamb)$  is also controlled. 

\subsection{Boundary Conditions}\label{sec:Compilation:BC}
With the intention of developing a quantum equivalent of the ghost point approach discussed in Sec. \ref{sec:Discretization}, we first elaborate on a matrix-based representation to facilitate the understanding of the quantum implementation. Subsequently, the novel \textsc{QNPU}s for the boundary treatment in variational quantum simulations are introduced. 
\medskip

The implementation of boundary conditions is based on the correction of the periodic contributions in the derivative matrix. To outline the transfer of \Eq{eq:Discrete_Opt_BC} into quantum-inspired or true quantum computations, we restrict ourselves to Dirichlet conditions imposed for a simple Laplace equation $-\nu \Delta u=0$ in the 2-qubit~case. 
Next, let us recover a slightly reduced version of \Eq{eq:Discrete_Opt_BC} for this problem in a matrix-based approach, assuming $\ub \in \mathbb{C}^{N_p}$ and $A \in \mathbb{R}^{N_p \text{x} N_p}$. For the sake of clarity, we also define $(\nu \lambda_0^2/\Delta x) = 1$, viz. 
\begin{align}
\begin{split}
    \ub^{\dagger} \cdot A \cdot \ub &= \begin{pmatrix} u_{1}^* & u_{2}^* & u_{3}^* & u_{4}^* \end{pmatrix} \cdot \begin{pmatrix} 2 & -1 & 0 & -1 \\ -1 & 2 & -1 & 0 \\ 0 & -1 & 2 & -1 \\ -1 & 0 & -1 & 2 \end{pmatrix} \cdot \begin{pmatrix} u_{1} \\ u_{2} \\ u_{3} \\ u_{4} \end{pmatrix} \\ 
    & = 2 u^{*}_1 u_1 - u_1^*u_2 - u_2^* u_1 + 2 u_2 u_2^* - u_2^*u_3 - u_3^*u_2 \\
    &\quad + 2 u_3 u_3^* - u_3^*u_4 - u_4^*u_{3} + 2 u_4^* u_4 - \underbrace{u_1^*u_4 - u_4^*u_1}_{\text{periodic boundary}}.
\end{split}
\label{eq:Expect_perio}
\end{align}

The sum on the right-hand side of \Eq{eq:Expect_perio} distinguishes between regular interior contributions and periodic boundary contributions. In line with \Eq{eq:Discrete_Opt_BC}, the periodic contributions of $-u_1^*u_4 - u_4^*u_1$ need to be eliminated by another term in the cost function, viz. $J = \sum\limits_{k=1}^4 j_{L_k} + j_{DN_1} + j_{DN_4}$. This is achieved by 
\begin{align}
\begin{split}
    J_{DN} = j_{DN_1} + j_{DN_4} = \ub^{\dagger} \cdot D \cdot \ub &= \begin{pmatrix} u_{1}^* & u_{2}^* & u_{3}^* & u_{4}^* \end{pmatrix} \cdot \begin{pmatrix}0 & 0 & 0 & 1 \\ 0 & 0 & 0 & 0 \\ 0 & 0 & 0 & 0 \\ 1 & 0 & 0 & 0 \end{pmatrix} \cdot \begin{pmatrix} u_{1} \\ u_{2} \\ u_{3} \\ u_{4} \end{pmatrix} \\ 
    & = \underbrace{u_1^*u_4 + u_4^*u_1}_{\text{boundary}}. 
\end{split}
\label{eq:Bound_J}
\end{align}

The matrix $D \in \mathbb{R}^{N_p \times N_p}$ needs to be unitary for its representation by quantum gates. This distinguishes the quantum-inspired approaches from true \textsc{QC} methods. A \textsc{QC}-compatible representation of $D$ follows from a modification with the basis of its null space removing the zero eigenvalues and intuitively matching the boundary contributions of \Eq{eq:Bound_J}. Accordingly, the unitary matrix $C \in \mathbb{R}^{N_p \times N_p}$ is  proposed as 
\begin{align}
\begin{split}
    J_{DN} = \ub^{\dagger} \cdot C \cdot \ub &= \begin{pmatrix} u_{1}^* & u_{2}^* & u_{3}^* & u_{4}^* \end{pmatrix} \cdot \begin{pmatrix}0 & 0 & 0 & 1 \\ 0 & 0 & -1 & 0 \\ 0 & 1 & 0 & 0 \\ 1 & 0 & 0 & 0 \end{pmatrix} \cdot \begin{pmatrix} u_{1} \\ u_{2} \\ u_{3} \\ u_{4} \end{pmatrix} \\ 
    & = \underbrace{ u_1^*u_4 -u_2^*u_3 + u_3^* u_2 + u_4^*u_1 }_{\text{if } \bm{u} \in \mathbb{R}^n}  = \underbrace{2u_1u_4}_{\text{boundary}} \, . 
\end{split}
\label{eq:Bound_J_unit}
\end{align}

The unitary matrix $C$ only inheres nonzero entries on the antidiagonal. The upper and lower corners must have the same sign to neutralize the periodic contributions of matrix $A$. The other terms need to change sign with respect to the main diagonal for these contributions to be canceled when the expectation value $\langle \sigma_z \rangle$ is measured. Note that for the summation of the interior antidiagonal elements, one obtains zero only if $\ub \in \mathbb{R}^n$. 

\begin{figure}[htpb]
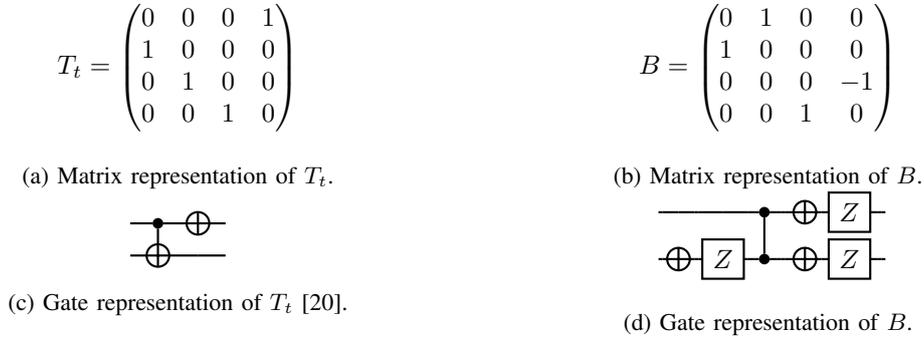

    \centering
    \begin{subfigure}[t]{0.48\textwidth}
        \centering
        \begin{equation*}
        T_t = \begin{pmatrix}0 & 0 & 0 & 1 \\ 1 & 0 & 0 & 0 \\ 0 & 1 & 0 & 0 \\ 0 & 0 & 1 & 0 \end{pmatrix}
        \end{equation*}
        \caption{Matrix representation of $T_t$.}
        \label{fig:matrix2gate:MatrixT}
    \end{subfigure}
    \begin{subfigure}[t]{0.48\textwidth}
        \centering
        \begin{equation*}
        B = \begin{pmatrix}0 & 1 & 0 & 0 \\ 1 & 0 & 0 & 0 \\ 0 & 0 & 0 & -1 \\ 0 & 0 & 1 & 0 \end{pmatrix}
        \end{equation*}
        \caption{Matrix representation of $B$.}
        \label{fig:matrix2gate:MatrixB}
    \end{subfigure}

    \begin{subfigure}[t]{0.48\textwidth}
        \centering
        \input{Figures/F_Transformation_n2}
        \caption{Gate representation of $T_t$ \cite{Sato2021}.}
        \label{fig:matrix2gate:shift}
    \end{subfigure}
    \begin{subfigure}[t]{0.48\textwidth}
        \centering
        \input{Figures/F_Dirichlet_n2}
        \caption{Gate representation of $B$.}
        \label{fig:matrix2gate:dir}
    \end{subfigure}
    \caption{Matrices and their corresponding circuits for 2-qubits.}
    \label{fig:matrix2gate}
\end{figure}

The remaining task is to design the circuits that implement the transformation described by the matrix $C$. To this end, the heuristic approach requires another decomposition, i.e., $C = T_t^{-1} \cdot B \cdot T_t$, where  $T_t$ relates to a permutation operator and $B$ to a boundary \textsc{QNPU}, which both directly translate into quantum gates, cf. \Fig{fig:matrix2gate}. 

In general, deriving the gate sequence corresponding to a given matrix is neither trivial nor unique. 
The application of heuristic approaches leads to the gate-level implementation for two qubits outlined in \Fig{fig:matrix2gate:shift} and \Fig{fig:matrix2gate:dir}. 

In contrast to the method of Sato et al. \cite{Sato2021} and the related studies of Leong et al. \cite{Leong2022}, the proposed boundary corrections are represented as unitary circuits and, therefore, as quantum gates. Further insight into the conversion between matrix and circuit representation is given in App.~\ref{app:MatrixGateConversion}. For the sake of completeness, the four qubit version of the circuits in \Fig{fig:matrix2gate:shift} and \Fig{fig:matrix2gate:dir} with the corresponding unitary matrix $C$ are also included in App.~\ref{app:BoundaryMatrixGate}. 

\begin{figure}[htbp]
    \centering
    \input{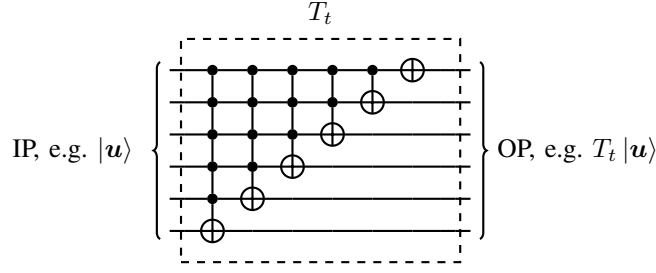}
    \caption{Six qubit example of the permutation  operator $T_t$ represented by a sequence of multi-controlled Toffoli gates cf. \cite{Sato2021}.}
    \label{fig:Transformation}
\end{figure}

\bigskip

\begin{figure}[htbp]
\centering
    \begin{subfigure}[t]{1.0\textwidth}
       \centering
        \input{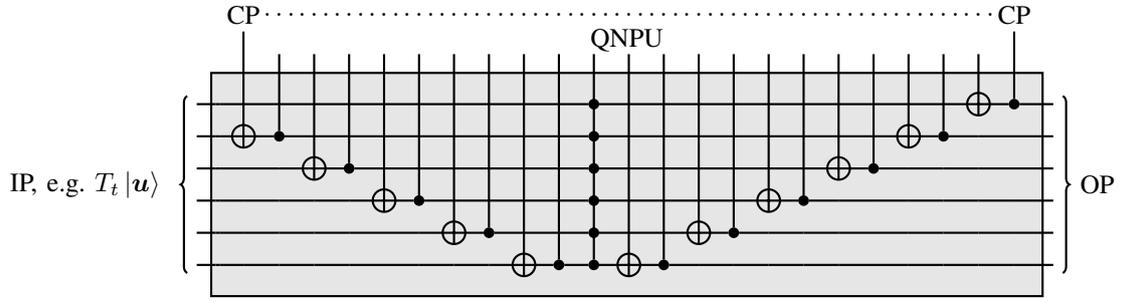}
        \caption{Compiled \textsc{QNPU} for the implementation of  $j_{DN}$ used to realize both Dirichlet and Neumann conditions, cf. \Eq{eq:Discrete_Opt_BC}.}
        \label{fig:QNPU_BC_Dirichlet}
    \end{subfigure}\\
    \vspace*{2ex}
    
    \begin{subfigure}[t]{1.0\textwidth}
        \centering
        \input{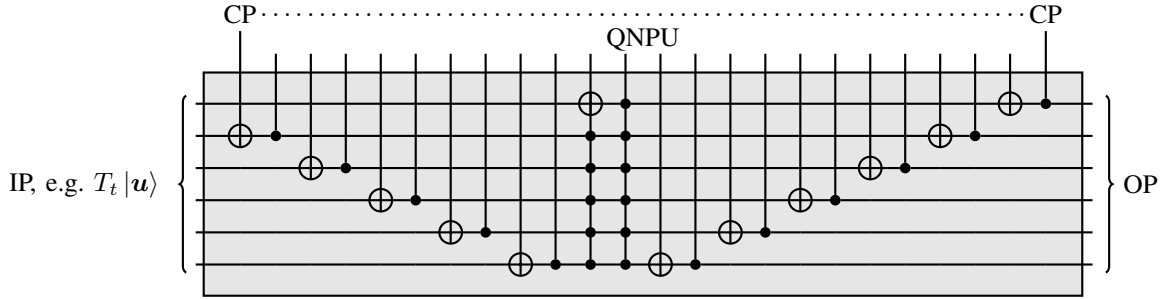}
        \caption{Compiled \textsc{QNPU} for the implementation of  $j_{N}$ used to realize Neumann boundary conditions, cf. \Eq{eq:Discrete_Opt_BC}.}  
        \label{fig:QNPU_BC_Neumann}
    \end{subfigure}\\
    
    \caption{Illustration of the novel boundary correction \textsc{QNPU}s for a $n=6$ qubits setup.}
    \label{fig:QNPU_BC}
\end{figure}

The extension to an $n$ qubit framework implies the multi-qubit implementation of the transformation $T_t$ with a sequence of $n-2$ multi-controlled \textsc{NOT} gates, cf. \Fig{fig:Transformation}. Similarly, the boundary corrections $j_{DN}$ and $j_{N}$ consist of $2n-1$ \textsc{CNOTS} and controlled-\textsc{Z} gates for an arbitrary number of qubits. Furthermore, one $n$ multi-qubit controlled-\textsc{Z} gate for both circuits and one $n$ multi-controlled \textsc{NOT} gate for $j_{DN}$ complete the circuits, as illustrated in \Fig{fig:QNPU_BC} for $n=6$ qubits. Both circuits shown in \Fig{fig:QNPU_BC_Dirichlet} and \Fig{fig:QNPU_BC_Neumann} use operations that only manipulate the dynamics at the boundaries. In the interior domain, the effect of the operator neutralizes for the expectation value $\langle \sigma_z \rangle$. 

\medskip

The resulting circuits for $T_t$, $j_{DN}$, and $j_N$ in \Fig{fig:Transformation} and \Fig{fig:QNPU_BC} 
inhere sequences of multi-controlled gates, which substantially deteriorate the gate complexity. In this regard, the most efficient implementation depends on the available \textsc{QC} hardware and its support for multi-controlled gates. Alternatively, additional carry qubits can be used to obtain more shallow circuits. This technique is illustrated in more detail in Sec. \ref{sec:compilation:complexity}.

\section{Gate Complexity} 
\label{sec:compilation:complexity}
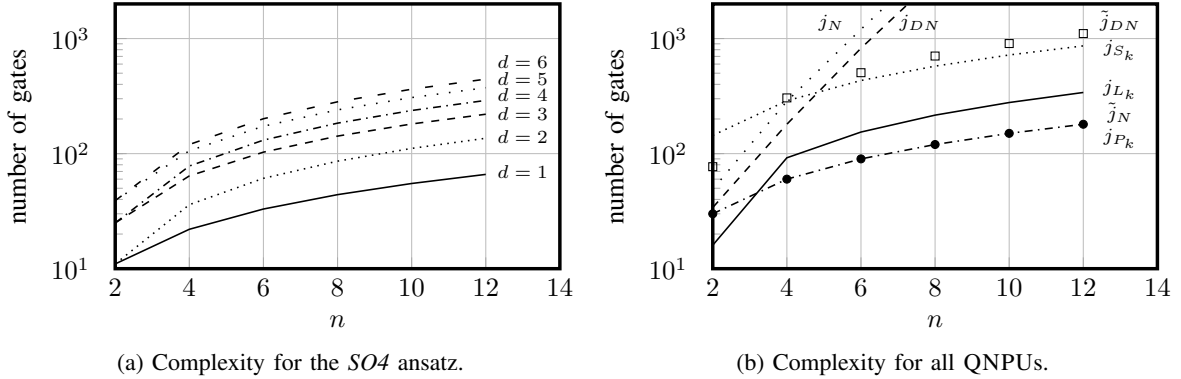
\begin{figure}[htbp]
    \begin{subfigure}[t]{0.49\textwidth}
       \centering
        \tikzsetnextfilename{Complexity-SO4}
\begin{tikzpicture}
\begin{semilogyaxis}[
tick pos=left,
xmajorgrids,
xlabel = $n$,
ylabel = number of gates,
xmin=2, xmax=14, 
ymin= 10,
ymax = 2000,
ymajorgrids,
]

\addplot [semithick,solid]
table [col sep = comma , row sep=crcr]{%
2.0,  11 \\  
4.0,  22 \\  
6.0,  33 \\  
8.0,  44 \\
10.0, 55 \\
12.0, 66 \\
};
\node[above] at (13,50) {\scriptsize{$d=1$}};

\addplot [semithick,dotted]
table [col sep = comma , row sep=crcr]{%
2.0,  11 \\  
4.0,  36 \\  
6.0,  61 \\  
8.0,  86 \\
10.0, 111 \\
12.0, 136 \\
};
\node[above] at (13,100) {\scriptsize{$d=2$}};


\addplot [semithick,dashed]
table [col sep = comma , row sep=crcr]{%
2.0,  25 \\  
4.0,  64 \\  
6.0,  103 \\  
8.0,  142 \\
10.0, 181 \\
12.0, 220 \\
};
\node[above] at (13,160) {\scriptsize{$d=3$}};

\addplot [semithick,dashdotted]
table [col sep = comma , row sep=crcr]{%
2.0,  25 \\  
4.0,  78 \\  
6.0,  131 \\  
8.0,  184 \\
10.0, 237 \\
12.0, 290 \\
};
\node[above] at (13,230) {\scriptsize{$d=4$}};

\addplot [semithick,loosely dotted]
table [col sep = comma , row sep=crcr]{%
2.0,  39 \\  
4.0,  106 \\  
6.0,  173 \\  
8.0,  240 \\
10.0, 307 \\
12.0, 374 \\
};
\node[above] at (13,320) {\scriptsize{$d=5$}};

\addplot [semithick,loosely dashed]
table [col sep = comma , row sep=crcr]{%
2.0,  39 \\  
4.0,  120 \\  
6.0,  201 \\  
8.0,  282 \\
10.0, 363 \\
12.0, 444 \\
};
\node[above] at (13,450) {\scriptsize{$d=6$}};

\end{semilogyaxis}
\end{tikzpicture}

        \caption{Complexity for the \textsc{\it SO4} ansatz.} 
        \label{fig:Results:SO4complex}
    \end{subfigure}
    \begin{subfigure}[t]{0.49\textwidth}
        \centering
        \tikzsetnextfilename{Complexity-QNPU}
\begin{tikzpicture}
\begin{semilogyaxis}[
tick pos=left,
xmajorgrids,
xlabel = $n$,
ylabel = number of gates,
xmin=2, xmax=14,
ymin=10,
ymax=2000,
ymajorgrids,
]

\addplot [semithick,black,solid]
table [col sep = comma , row sep=crcr]{%
2.0,  16 \\  
4.0,  92 \\  
6.0,  154 \\  
8.0,  216 \\
10.0, 278 \\
12.0, 340 \\
};
\node[above] at (13,240) {\scriptsize{$j_{L_k}$}};


\addplot [semithick,black,dotted]
table [col sep = comma , row sep=crcr]{%
2.0,  144 \\  
4.0,  288 \\  
6.0,  432 \\  
8.0,  576 \\
10.0, 720 \\
12.0, 864 \\
};
\node[above] at (13,530) {\scriptsize{$j_{S_k}$}};


\addplot [semithick,black,dashdotted]
table [col sep = comma , row sep=crcr]{%
2.0,  30 \\  
4.0,  60 \\  
6.0,  90 \\  
8.0,  120 \\
10.0, 150 \\
12.0, 180 \\
};
\node[above] at (13,90) {\scriptsize{$j_{P_k}$}};

\addplot [semithick,black,dashed]
table [col sep = comma , row sep=crcr]{
2.0,  34 \\   
4.0,  180 \\   
6.0,  824 \\   
8.0,  3326 \\  
10.0, 13316 \\  
12.0, 53258 \\  
};
\node[above] at (7.6,950) {\scriptsize{$j_{DN}$}};

\addplot [thin, mark={square},only marks,mark size=1.5pt]
  table [col sep = comma , row sep=crcr]{
 2.0, 77\\
 4.0,  305\\
 6.0,  505\\
 8.0,  705\\
 10.0, 905\\
 12.0, 1105\\
 };
 \node[above] at (13,950) {\scriptsize{$\tilde{j}_{DN}$}};

\addplot [semithick,black,loosely dotted]
table [col sep = comma , row sep=crcr]{%
2.0,  49 \\  
4.0,  285 \\  
6.0,  1203 \\  
8.0,  4857 \\
10.0, 19455 \\
12.0, 77829 \\
};
\node[above] at (5.2,950) {\scriptsize{$j_{N}$}};

\addplot [thin, mark={*},only marks,mark size=1.5pt]
table [col sep = comma , row sep=crcr]{%
2.0,  30 \\  
4.0,  60 \\  
6.0,  90 \\  
8.0,  120 \\
10.0, 150 \\
12.0, 180 \\
};
\node[above] at (13,150) {\scriptsize{$\tilde {j}_{N}$}};

\end{semilogyaxis}
\end{tikzpicture}
        \caption{Complexity for all \textsc{QNPU}s.} 
        \label{fig:Results:QNPUcomplex}
    \end{subfigure}
    \caption{Gate complexity of the employed \textsc{QNPU}s and ansatz gate compilations.}
    \label{fig:compilation:GateComplexity}
\end{figure}

To scrutinize the scalability of the algorithm, the number of single- (\textsc{R$_z$,R$_y$,U},...) and two-qubit (\textsc{CNOT},...) gates per qubit are outlined in this section. For this purpose, multi-qubit gates are decomposed using known identities, which are available, for example, in \textsc{Qiskit} \cite{Qiskit}. The scaling of the number of gates involved in the \textsc{\it SO4} ansatz with respect to the number of qubits $n$ and the depth $d$ is reported in \Fig{fig:Results:SO4complex}. It is observed that the number of gates scales linearly with $n$ and $d$, i.e., $\mathcal{O}(n \, d)$.

The graphs displayed in \Fig{fig:Results:QNPUcomplex} illustrate the polynomial relation between the number of gates and the number of qubits for the adder ($j_{L}$), the source ($j_{S}$), the potential ($j_{P}$) and the boundary circuits -- the latter using either multi-controlled gates ($j_{DN}, j_{N}$) described in Sec. \ref{sec:Compilation:BC} or the corresponding modifications described below ($\tilde j_{DN}, \tilde j_{N}$). 
 
Most \textsc{QNPU}s in \Fig{fig:Results:QNPUcomplex} depict a favorable complexity. The linear complexity of the adder circuit, $\mathcal{O}( n)$, is due to the very local reach of the \textsc{FD} stencil for approximating the second derivative. In this case, the resulting circuit conserves the local property. An analysis of the source and potential \textsc{QNPU}s also reveals a linear scaling of  $\mathcal{O}( n)$, which underlines the importance of an efficient state initialization for arbitrary generic functions \cite{Chen2021,Kyriienko2021,Creevey2023,Melnikov2023,Araujo2021}, cf. Sec. \ref{sec:Compilation:Potential}. The similarity to the scaling behavior examined for the \textsc{\it SO4} ansatz is obvious and indicates the strong influence of the prior optimization of $U(\lamb_c)$ to determine $\Tilde{\fb}$ or $\pb$, cf. \Eq{eq:OptAnsatz}. The related overhead in the number of gates with respect to the ansatz is due to the control operation of the ansatz in both \textsc{QNPU}s ($j_S$,$j_P$) and the additional \textsc{CNOT} gates following the $P$ gate for the potential \textsc{QNPU}, see \Fig{fig:QNPU_Potential}.  Note that for the values given in \Fig{fig:compilation:GateComplexity}, the depth of the ansatz for $j_{P}$ and $j_{S}$ is assigned to $d=1$. 

\begin{figure}[htbp]
    \centering
    \input{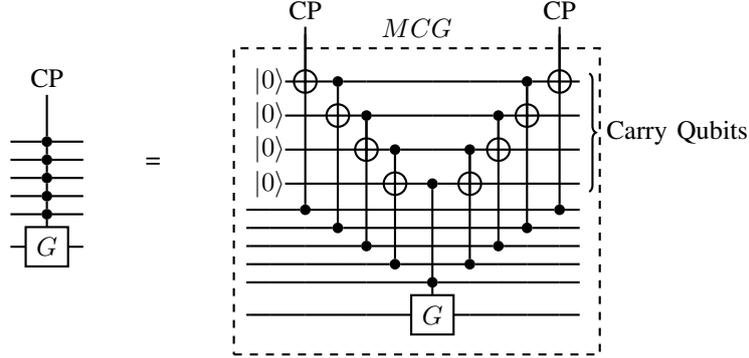}
    \caption{Equivalent networks using a deep qubit-saving implementation (left) and a shallow circuit depth approach (right) to realize a multi-controlled gate \textsc{G}.}
    \label{fig:multicontrol}
\end{figure}

The boundary \textsc{QNPU}s ($j_{DN}, j_{N}$), however, scale polynomially, on the maximum order of $\mathcal{O}(n^{5})$. This was also observed by other researchers when using multi-controlled quantum gates, for example, \cite{Sato2021,Leong2022,Leong2023}. The reason for that is the non-locality of the implementation and the multi-controlled feature, which yields interaction with more than just the nearest neighbor qubits. 
In practical applications, e.g., an implementation on \textsc{NISQ} hardware, there is a demand for circuit efficiency. To obtain more shallow circuits, additional carry qubits are used. They implement intermediate information associated with \textsc{AND} operations by a staggered series of Toffoli gates. The technique is illustrated for a multi-controlled-\textsc{G} gate (\textsc{MCG}) in \Fig{fig:multicontrol}. 

The example consists of $n=6$ qubits and requires $n-2=4$ carry qubits to transform the deep multi-controlled gate on the left of \Fig{fig:multicontrol} into $n+2$ Toffoli- and one controlled G-gate on the right of \Fig{fig:multicontrol}. Mind that after applying \textsc{G}, all carry bits need to be reversed to their initial state, i.e., $\ket{0}$. 

\begin{figure}
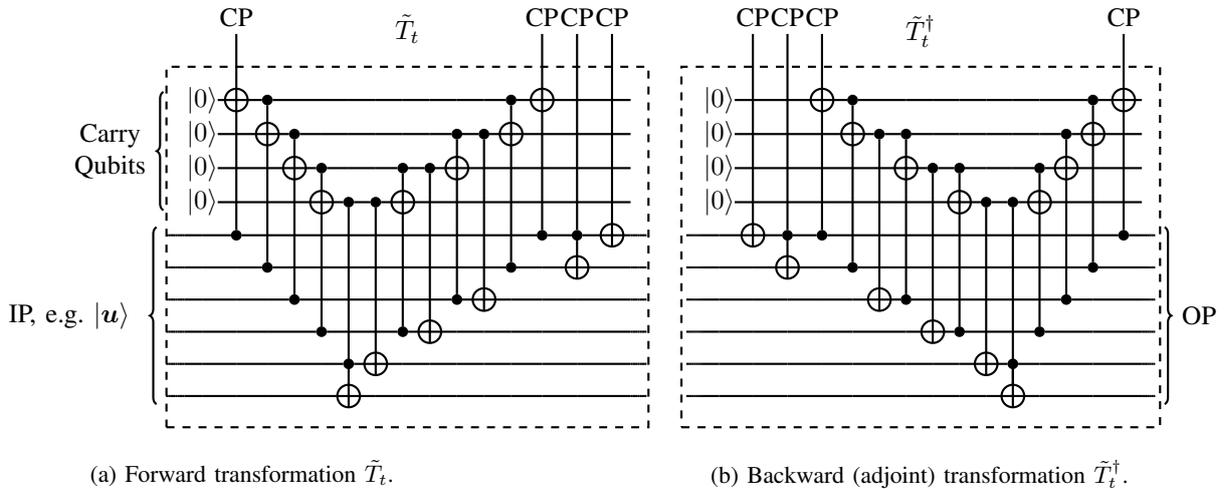

\begin{subfigure}[t]{0.4\textwidth}
      \include{Figures/F_ShiftUP}
      \caption{Forward transformation $\tilde T_t$.}
      \label{fig:new-Tt:up}
\end{subfigure}
\hspace{2.3cm}
\begin{subfigure}[t]{0.4\textwidth}
      \include{Figures/F_ShiftDOWN}
      \caption{Backward (adjoint) transformation $\tilde T^\dagger_{t}$.}
      \label{fig:new-Tt:down}
\end{subfigure}
\caption{Alternative implementations of the forward ($T_t$, $\tilde T_t$) and backward ($T_{t}^\dagger$, $\tilde T^\dagger_{t}$) transformation introduced in \Fig{fig:Transformation}.}
\label{fig:new-Tt}
\end{figure}

Applying this concept to the transformation circuit $T_t$ outlined in \Fig{fig:Transformation} allows to derive a shallower alternative $\Tilde{T}_t$ resulting in a permutation operator, cf. \Fig{fig:new-Tt:up}. 
This figure also provides the corresponding circuit for the inverted transformation  $\Tilde{T}^\dagger_{t}$  in \Fig{fig:new-Tt:down}. The latter is required for measuring the expectation value $\langle \sigma_z \rangle = \bra{0} U^\dagger T_t^\dagger \; [\text{QNPU}] \; T_t U \ket{0}$ as illustrated by the Hadamard-test based  structure in \Fig{fig:VQA_QNPU}. Both $\tilde T_t$ and $\tilde T^\dagger_{t}$ are controlled by an ancilla qubit,  which follows the same strategy as depicted for the \textsc{QNPU} assembly in \Fig{fig:VQA_QNPU}. 

Introducing the above-mentioned shallow circuit alternatives to the $j_{DN}$ circuit is straightforward and allows replacing the multi-controlled-\textsc{Z} gate by the usage of $n-2$ carry qubits. As already outlined at the end of Sec. \ref{sec:Discretization}, the \textsc{QNPU} related to the Neumann condition ($j_{N}$) shown in \Fig{fig:QNPU_BC_Neumann} can be cast in terms of the potential \textsc{QNPU}, depicted in \Fig{fig:QNPU_Potential}. A consistent application of shallow circuit (\textsc{MCG}) strategies yields the complexity graphs for $\tilde j_N$, $\tilde j_{DN}$ marked by symbols in \Fig{fig:compilation:GateComplexity}. Advantageous differences to the initial implementation
$j_N, j_{DN}$, given by lines in \Fig{fig:compilation:GateComplexity}, are clearly visible but, of course, come at the expense of additional $(n-2)$ carry qubits. The graph displays an efficient scaling below $\mathcal{O}(n^2)$. In terms of grid-point behavior, the complexity is of \textit{polylog} scaling. We again emphasize that the chosen reduction in complexity, i.e., the number of carry qubits used, depends on the capabilities of the \textsc{QC} hardware. 

\section{Optimization Strategy} \label{sec:Optimization}

A common denominator of all \textsc{VQA}s is the efficiency of the optimization process. As stated in \cite{Bittel2021}, this is non-trivial since the problem covered by \textsc{VQA} is theoretically \textsc{NP}-hard, meaning that the optimization is often characterized as non-convex and features multiple local minima and/or vanishing derivatives \cite{Leong2022,Sarma2023,Wiersema2023}. 
To update the controls ($\lambda_0, \lamb_c$) along the route outlined in \Fig{fig:optimizer}, we apply the sequence of a zero- and a first-order method. The combination aims at a larger sample size to address non-convex functions and higher accuracy in the vicinity of the minima and might be replaced by a more sophisticated strategy. The optimization starts with the zero-order approach and employs an evolutionary \textit{Particle-Swarm Optimization} (\textsc{PSO}) algorithm \cite{Kennedy1995} using $\mathfrak{p}$ particles. In the case of non-periodic BC, the \textsc{PSO} step typically involves $\mathcal{O}(1000)$ iterations, where around $\mathfrak{p} \approx 100\,n$ particles are employed. The best candidate initializes the control $\upb$ for the subsequent first-order classical \textit{Gradient Descent} (\textsc{GD}) method. The required derivative of the objective function with respect to the controls $\nabla_{\upb} J(\upb_{i})$ is obtained from the \textsc{QC} and follows the expressions given below, cf. \Eq{eq:quantum_derivative_0} and \Eq{eq:quantum_derivative_control}. Convergence is assessed from the absolute change of the derivative
\begin{align}
   \textsf{ch}( \nabla J^l_i) = | \nabla J^l_i- \nabla J^l_{i-1}|_{2} \quad \le 10^{-m}  \, , 
   \label{alg:convergence}
\end{align}
where $m\in[3,7]$ depending on the number of employed qubits, cf. Sec. \ref{sec:Applications}.  
The solution is advanced in time as described in \Fig{fig:discretization_encoding} and in the algorithm \ref{alg:VQA}. For time steps $l>1$, the optimal parameters of the previous step $\upb^{l-1}$ serve as an initial guess for the \textsc{PSO} to reduce the efforts, as proposed by \cite{Sato2021,Leong2022}.

\begin{algorithm}[htbp]
\caption{VQ Algorithm}
\label{alg:VQA}
    \begin{algorithmic}
    \Require $N_\text{t}, \tilde f, \texttt{maxit}_\text{PSO}, \texttt{tol}_\text{PSO}, \texttt{maxit}_\text{GD},\texttt{tol}_\text{GD} $
    \State $l \gets 1$
    \While{$l \leq N_\text{t}$}  \Comment{Time Evolution}
        \If{$l=1$}
            \State $\Tilde{f}^1 \gets \frac{y(x,0,\upb_{0}) }{\Delta t}+ \Tilde{f}^1$ \Comment{Initial setup of the external contribution}
            \State $\upb^1 \gets \left(\lambda_0^1,\lamb_c^1\right)^\intercal$  \Comment{Random initialization of $\lambda_0^1\in [0,1]$ and $\lamb_c^1 \in [0,4\pi)$ for all $\mathfrak{p}$ particles}
        \Else
            \State $\Tilde{f}^l \gets \frac{y(x, t^l, \upb_{l-1}) }{\Delta t}+ \Tilde{f}^l$ \Comment{Adjust the contributions}
            \State $\upb^l \gets \upb^{l-1}$  \Comment{Propagate previous final control $\upb^{l-1}$}
        \EndIf
        \State $\upb^{l} \gets$ \textsc{PSO}$(\upb^l)$  \Comment{Iterate until $\texttt{maxit}_\text{PSO}$ or $\texttt{tol}_\text{PSO}>\textsf{ch}(\nabla J_i^l)$ (cf. \Fig{fig:optimizer})} \label{alg:VQA:PSO}
        \State $\upb^{l} \gets$ \textsc{GD}$(\upb^{l})$  \Comment{Iterate until $\texttt{maxit}_\text{GD}$ or $\texttt{tol}_\text{GD}>\textsf{ch}(\nabla J_i^l)$ (cf. \Fig{fig:optimizer})}\label{alg:VQA:GRAD}
        \State $l \gets l+1$
    \EndWhile
    \end{algorithmic}
\end{algorithm}

The whole framework is implemented by an in-house toolbox, with \textsc{Qiskit} v.0.42.1 \cite{Qiskit} as a quantum backend. 

\subsection{Derivative Computation} \label{sec:Derivative1}
The determination of the descent direction is supported by the \textsc{QC}. To this end, the quantum circuits derived in Secs. \ref{sec:Compilation} and \ref{sec:Strategy} are re-used, as outlined below. The gradient of the objective function with respect to the scalar parameter $\lambda_0$ follows from \Eq{eq:Discrete_Opt_BC} and simply reads
\begin{align}
\begin{split}
\nabla_{\lambda_0} J(\lambda_0, \lamb_c) =
   &\sum\limits_{k=1}^{N_\text{p}} \underbrace{ -\left(\frac{2\nu \lambda_{0}}{\Delta x^2}\right) u_k(\lamb_{c}) \bigg(u_{k+1}(\lamb_{c})-2u_k(\lamb_{c})+u_{k-1}(\lamb_{c})\bigg) }_{\nabla_{\lambda_0}j_{L_k} } 
   \Delta x \, \\
    &\underbrace{+\left(\frac{2\nu \lambda_{0}}{\Delta x^2}\right) \bigg( u_1(\lamb_{c})u_{N_\text{p}}(\lamb_{c}) + u_1(\lamb_{c}) u_{N_\text{p}}(\lamb_{c})  \bigg)}_{\nabla_{\lambda_0} (j_{DN_1} + j_{ DN_{N_\text{p}}}) } \Delta x \,
     \underbrace{-\left(\frac{2\nu \lambda_{0}}{\Delta x^2}\right) \bigg(  u_1(\lamb_{c})^2 + u_{N_\text{p}}(\lamb_{c}) ^2 \bigg)}_{\nabla_{\lambda_0}(j_{N_1} +j_{N_{N_\text{p}}}) } \Delta x \, \\
   + &\sum\limits_{k=1}^{N_\text{p}} \, \underbrace{ - 2\zeta \lambda_{0} u_k(\lamb_{c}) p_k u_k(\lamb_{c})  }_{\nabla_{\lambda_0} j_{P_k} }  \, \Delta x + \sum\limits_{k=1}^{N_\text{p}} \underbrace{ - 2 u_k(\lamb_{c}) \Tilde{f}_k } _{\nabla_{\lambda_0} j_{S_k}  } \, \Delta x \, . 
   \label{eq:deriv0-a}
\end{split}
\end{align}
Again, the temporal index has been omitted for better clarity. Comparing \Eq{eq:deriv0-a} with \Eq{eq:Discrete_Opt_BC} reveals
\begin{equation}
\label{eq:quantum_derivative_0}
\nabla_{\lambda_0} J
= \frac{1}{\lambda_0} \bigg[  \sum\limits_{k=1}^{N_\text{p}} 
 \big( 2  j_{L_k}
+ 2 j_{P_k} 
+ j_{S_k} \big)  \; \Delta x \bigg]  
+ \frac{2}{\lambda_0} 
\left[ \bigg(
  j_{DN_1} + j_{DN_{N_\text{p}}}
+ j_{N_1} + j_{N_\text{p}}
 \bigg)
 \, \Delta x \right] \, ,
\end{equation}
which can be evaluated on the \textsc{QC}. The derivative of $J$  with respect to $\lamb_c$ can also be computed on the \textsc{QC} without deriving a new gate sequence. The derivative of the objective function with respect to the control vector $\lamb_c$ follows from 
\begin{align}
\begin{split}
\nabla_{\lamb_c} J (\lambda_0, \lamb_c) 
   =
   &\sum\limits_{k=1}^{N_\text{p}} \nabla_{\lamb_c} \big( 
j_{L_k} + j_{P_k} + j_{S_k} \big) \Delta x 
   +
   \big( \nabla_{\lamb_c} (j_{DN_1} +j_{DN_{N_\text{p}}}) 
   +
   \nabla_{\lamb_c}(j_{N_1} +j_{N_{N_\text{p}}}) \big)   \,\Delta x  \, .
   \label{eq:derivc-a}
\end{split}
\end{align}

The linear source term $\nabla_{\lamb_c} j_{S_k}$ is approximated with second-order accurate central differences that operate on the complete (linear) objective function contribution \cite{Nocedal2006}, viz.
\begin{align}
    \nabla_{\lamb_c}j_{S_k}   = -2\lambda_{0} 
 \big( \nabla_{\lamb_c} u_k \big)    
\Tilde{f}_k \approx 
  \frac{j_{S_k}(\lambda_0, \lamb_c+h\eb_c)-j_{S_k}(\lambda_0, \lamb_c-h\eb_c)}{2h} 
   + \mathcal{O}(h^2) \, .
   \label{eq:opt-source-FD}
\end{align}
Mind that \Eq{eq:opt-source-FD} has to be executed for all entries of the parameter vector $\lamb_c$, and the step size is typically assigned to a constant value $h$, depending on the number of control parameters. For the quadratic terms $j_L, j_P, j_{DN}, \text{ and } j_N$, an exact representation of the derivative is given by the \textit{Parameter Shift Rule}, cf.  \cite{Schuld2019,Banchi2021,Crooks2019}. The derivative follows using only two additional evaluations of existing quantum circuits, and the nonlinearity is implicitly considered. Applying these techniques yields an expression for the derivative of the objective function $J$ with respect to the control vector $\lamb_c$ that utilizes the existing circuits, viz.  
\begin{align}
\begin{split}
\nabla_{\lamb_c} J(\lambda_0, \lamb_c) =&  \sum\limits_{k=1}^{N_\text{p}} \frac{1}{2} \bigg(j_{L_k}\big(\lambda_0, \lamb_c+\frac{\pi}{2}\eb_c\big) - j_{L_k}\big(\lambda_0, \lamb_c-\frac{\pi}{2}\eb_c\big) \bigg) \Delta x\\
&+ \frac{1}{2} \bigg(j_{DN}(\lambda_0, \lamb_c+\frac{\pi}{2}\eb_c)- j_{DN}(\lambda_0, \lamb_c-\frac{\pi}{2}\eb_c) \bigg) \Delta x \\
&+  \frac{1}{2} \bigg(j_{N}(\lambda_0, \lamb_c+\frac{\pi}{2}\eb_c)-j_{N}(\lambda_0, \lamb_c-\frac{\pi}{2}\eb_c) \bigg) \Delta x\\
+&\sum\limits_{k=1}^{N_\text{p}} \bigg(\frac{1}{2} \Big( j_{P_k}\big(\lambda_0, \lamb_c+\frac{\pi}{2}\eb_c\big)-j_{P_k}\big(\lambda_0, \lamb_c-\frac{\pi}{2}\eb_c\big) \Big)  + \nabla_{\lamb_c}j_{S_k} \bigg)\, \Delta x \, .
\end{split}
\label{eq:quantum_derivative_control}
\end{align}

\section{Numerical Results} \label{sec:Applications}
This section describes the application of the \textsc{VQA} to two prototypical \textsc{CFD} problems, i.e., the steady heat conduction (Sec. \ref{sec:Application:Poisson}) and the transient heat conduction (Sec. \ref{sec:Application:Heat}). Results are discussed for a range of Dirichlet, Neumann, and Robin boundary-condition combinations. The validation data is obtained from \textsc{FD}-solutions of the \textsc{PDE} problem. It employs the same discretization as the \textsc{VQA} to approximate the Laplace operator and the temporal derivative, cf. \Eq{eq:derivatives_space} and \Eq{eq:Generic_time}. 

The predictive deviation of the \textsc{VQA}-solutions $y^\text{VQ}$ from \textsc{FD}-solutions $y^\text{FD}$ is assessed using either the $l_2$-norm $\varepsilon_{l_2}$ or the frequently employed trace distance $\varepsilon_{\text{tr}}$, cf.  \cite{BravoPrieto2019,Sato2021,Liu2021}, viz.
\begin{align}
    \varepsilon_{l_2} = \sqrt{\sum_k (y_k^\text{FD}-y_k^\text{VQ})^2} \, , \qquad
    \varepsilon_{\text{tr}} = \sqrt{\Bigg(1- \sum_k \bigg|\frac{y_k^\text{FD}}{\sqrt{\sum_j (y_j^\text{FD})^2}}\frac{y_k^\text{VQ}}{\sqrt{\sum_j (y_j^\text{VQ})^2}}\bigg|^2 \Bigg)} \, .
\end{align}

Time-averaged measures are employed to assess the transient heat conduction and indicated by $ \Bar{\varepsilon}_{l_2}, \Bar{\varepsilon}_{\text{tr}}$. The number of control variables $c$  of  $\lamb_c$ is also stated in the results tables. A full dataset of the performed experiments is available via \cite{data}. 

\subsection{Steady Heat Conduction (Poisson Equation)}
\label{sec:Application:Poisson}
The first case refers to the steady heat conduction, where the time derivative in \Eq{eq:Generic_PDE-TR} vanishes, and $y$~[K] marks the temperature. The physical potential $p(x)$~[1/s] vanishes, and the diffusivity is assigned to a unit value $\nu=1$~[m$^2$/s]. In the case of pure Neumann boundary conditions, the source reads $f(x) =1$ for $x\leq 0.5$ and $f(x) =-1$ for $x>0.5$. In all other cases, the unit source responds to $f(x) \equiv 1$~[K/s]. Mind that singular solutions, e.g., caused by pure Neumann conditions, are fixed by defining a central zero crossing. The application of the \textsc{VQA} framework to this Poisson problem is presented for the following seven scenarios: in-/homogeneous Dirichlet, in-/homogeneous Neumann, mixed Neumann/Dirichlet conditions, Robin conditions, and periodic conditions. 
First and second-order derivative approximations are applied to Neumann, mixed Neumann/Dirichlet, and Robin boundaries.
The displayed results are obtained for a coarse grid, featuring $N_\text{p}=4$ interior points ($n=2$ qubits; dashed and dash-dotted lines), and a finer grid, featuring $N_\text{p}=16$ interior points ($n=4$ qubits; dotted and solid lines). The  \textsc{FD}-results are depicted by closed ($N_\text{p}=4$) and open ($N_\text{p}=16$) symbols. 

The periodic case serves as a verification reference and it is displayed in \Fig{fig:Results:Poisson:periodic}. \textsc{VQA}-results were obtained from shallow circuits $U(\lamb_c)$ using a depth of $d=1$, cf. \Fig{fig:VQA_Ansatz}.
Figure \ref{fig:Results:Poisson:periodic:cost} shows the convergence of the objective function with the dotted vertical lines indicating the transition from global (\textsc{PSO}) to local ({\textsc{GD}) optimization (cf. Alg. \ref{alg:VQA}). The \textsc{VQA}-results are compared with \textsc{FD}-solutions in \Fig{fig:Results:Poisson:periodic:res} and indicate an excellent agreement with the expected solution {$y_k \equiv 0$}. The \textsc{VQA}-result improves as the grid is refined. The latter is also indicated by a reduction of $\varepsilon_{l_2}$ from an already satisfactory value of $10^{-6}$ to $10^{-7}$, cf. \Tab{tab:Results:Poisson:Errors}. The reduction goes hand in hand with lower amplitude but higher frequency wiggles displayed in \Fig{fig:Results:Poisson:periodic:res}, which are reflected by an augmented trace distance $\varepsilon_{\text{tr}}$ and should be viewed with caution due to the vanishing mean value for this case.
 
\begin{figure}[htbp]
    \centering
    \begin{subfigure}[t]{0.48\textwidth}
        \input{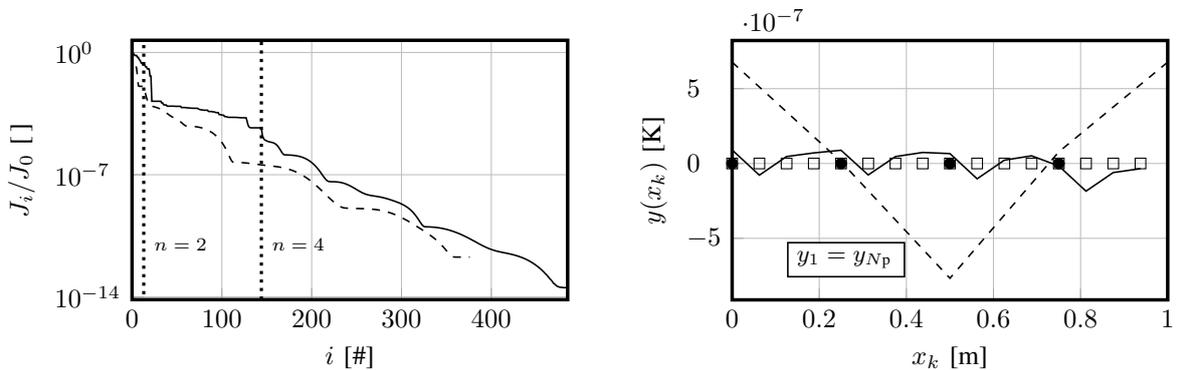}
        \caption{Evolution of the normalized objective function over the optimization count $i$. Dotted vertical lines indicate the transition from global (\textsc{PSO}) to local (GD) optimization. }
        \label{fig:Results:Poisson:periodic:cost}
    \end{subfigure}
    \hspace{0.3cm}
    \begin{subfigure}[t]{0.48\textwidth}
        \centering
        \tikzsetnextfilename{Results-PoissionPeriodic}
\begin{tikzpicture}
\begin{axis}[
tick pos=left,
xmajorgrids,
xmin=0, xmax=1,
ymajorgrids,
xlabel={$x_k$ [m]},
ylabel={$y(x_k)$ [K]},
y tick label style={
text width=1.3em
}
]

\addplot [semithick,dashed]
table [col sep = comma , row sep=crcr]{%
0, 0.677075705945995e-06 \\
0.25, 0.012310271024774e-06 \\
0.5, -0.766904719951000e-06 \\
0.75, 0.077519271002302e-06 \\
1.0, 0.677075705945995e-06 \\
};

\addplot [thin,black,mark={*},only marks]
table [col sep = comma , row sep=crcr]{%
0, 0.0 \\
0.25, 0.0 \\
0.5, 0.0 \\
0.75, 0.0 \\
};

\addplot [semithick]
table [col sep = comma , row sep=crcr]{%
0.0,  0.090309721034743e-06 \\
0.0625, -0.078152440075030e-06 \\
0.125,  0.046388298979849e-06 \\
0.1875,  0.070492960024815e-06 \\
0.25,  0.088381249985225e-06 \\
0.3125, -0.076744939958573e-06 \\
0.375,  0.046127813013719e-06 \\
0.4375,  0.073568444003413e-06 \\
0.5,  0.066002425969103e-06 \\
0.5625, -0.102459730033644e-06 \\
0.625, 0.022081002026830e-06 \\
0.6875, 0.051189618988090e-06 \\
0.75, -0.016663000090844e-06 \\
0.8125, -0.185125170082401e-06 \\
0.875, -0.060584429917299e-06 \\
0.9375, -0.034811779903166e-06 \\
};

\addplot [thin,black,mark={square},only marks]
table [col sep = comma , row sep=crcr]{%
0.0,  0.0 \\
0.0625, 0.0 \\
0.125,  0.0 \\
0.1875,  0.0 \\
0.25,  0.0 \\
0.3125, 0.0 \\
0.375,  0.0 \\
0.4375,  0.0 \\
0.5,  0.0 \\
0.5625, 0.0 \\
0.625, 0.0 \\
0.6875, 0.0 \\
0.75, 0.0 \\
0.8125, 0.0 \\
0.875, 0.0 \\
0.9375, 0.0 \\
};

\node[above,text width=1.3cm,draw, fill=white,semithick]  at (canvas cs:x=1.5cm,y=0.01cm) {\small{$y_1=y_{N\text{p}}$}}; 
\end{axis}
\end{tikzpicture}
        \caption{Comparisons of \textsc{VQA}- and \textsc{FD}-results, ($n=2$, closed circles; $n=4$, open squares). } 
        \label{fig:Results:Poisson:periodic:res}
    \end{subfigure}
    \caption{Verification results of the \textsc{VQA} application for a periodic steady heat conduction problem  using  $N_\text{p}=4$ ($n=2$, $d=1$; dashed lines) and $N_\text{p}=16$ ($n=4$, $d=1$; solid lines) interior points.}
    \label{fig:Results:Poisson:periodic}
\end{figure}

\begin{figure}[htbp]
\centering
    \hspace{0.5cm}\tikzsetnextfilename{Results-Legend}
 \begin{tikzpicture} 
    \begin{axis}[
        legend style={draw=black!15!black,legend cell align=center, fill opacity=0.8, draw opacity=1, text opacity=1},
        legend columns=4, 
        hide axis,
        width=0.005\textwidth,
        height=0.005\textwidth,
    ]

        \addlegendimage{thin,black,mark={*},only marks}
        \addlegendentry{$y^{\text{FD}}_{n=2} \quad $}
        \addlegendimage{thin,black,mark={square},only marks}
        \addlegendentry{$y^{\text{FD}}_{n=4} \quad $} 
        \addlegendimage{semithick,dashed}
        \addlegendentry{$y^{\text{VQ}}_{n=2} \quad $}
        \addlegendimage{semithick}
        \addlegendentry{$y^{\text{VQ}}_{n=4}$}
        \addplot [draw=none] coordinates {(0,0) (1,1)};
    \end{axis}
\end{tikzpicture}\\
    \vspace{0.2cm}
    \begin{subfigure}[t]{0.49\textwidth}
       \centering
        \tikzsetnextfilename{Results-PoissionDirichletHomo}
\begin{tikzpicture}
\begin{axis}[
tick pos=left,
xmajorgrids,
xmin=0, xmax=1,
ymajorgrids,
xlabel={$x_k$ [m]},
ylabel={$y(x_k)$ [K]},
y tick label style={
        /pgf/number format/.cd,
        fixed,
        fixed zerofill,
        precision=2,
        /tikz/.cd
    }
]

\addplot [semithick,dashed]
table [col sep = comma , row sep=crcr]{%
0,  0 \\  
0.2000,  7.999994135903538e-02 \\  
0.4000,  1.199999120385530e-01 \\  
0.6000,  1.199999120385530e-01 \\
0.8000,  7.999994135903538e-02 \\
1.0000, 0\\
};

\addplot [thin,black,mark={*},only marks]
table [col sep = comma , row sep=crcr]{%
0,  0 \\  
0.2000,  0.08 \\  
0.4000,  0.12  \\  
0.6000,  0.12  \\
0.8000,  0.08 \\
1.0000, 0\\
};

\addplot [semithick]
table [col sep = comma , row sep=crcr]{%
0,  0 \\  
0.0588, 2.772696154444362e-02 \\  
0.1176, 5.108725314849664e-02  \\  
0.1765, 7.170680259551429e-02  \\
0.2353, 9.004177295804902e-02  \\ 
0.2941, 1.038519970679225e-01  \\ 
0.3529, 1.141487869514948e-01  \\ 
0.4118, 1.214349758306554e-01  \\
0.4706, 1.245291314572673e-01  \\ 
0.5294, 1.245027854052222e-01\\ 
0.5882, 1.222219700424600e-01\\ 
0.6471, 1.149478250728057e-01\\
0.7059, 1.037729589117872e-01 \\  
0.7647, 8.998080066617320e-02 \\
0.8235, 7.139968290310275e-02\\  
0.8824, 5.074588358842645e-02  \\
0.9412, 2.771717586796972e-02 \\
1.0000, 0\\
};

\addplot [thin,black,mark={square},only marks]
table [col sep = comma , row sep=crcr]{%
0,  0 \\  
0.0588,  2.768166089965396e-02 \\  
0.1176,  5.190311418685116e-02 \\  
0.1765,   7.266435986159162e-02 \\
0.2353,  8.996539792387533e-02 \\ 
0.2941,  1.038062283737023e-01 \\ 
0.3529, 1.141868512110725e-01 \\ 
0.4118, 1.211072664359860e-01 \\
 0.4706, 1.245674740484427e-01 \\ 
0.5294,  1.245674740484427e-01 \\ 
0.5882, 1.211072664359859e-01 \\ 
0.6471, 1.141868512110724e-01 \\
0.7059,   1.038062283737023e-01 \\  
0.7647, 8.996539792387530e-02 \\
0.8235, 7.266435986159159e-02 \\  
0.8824, 5.190311418685114e-02 \\
0.9412 , 2.768166089965395e-02\\
1.0000, 0\\
};

\node[above,text width=1.5cm,draw, fill=white,semithick]  at (canvas cs:x=1.7cm,y=0.01cm) {\small{$y_0=0$ $y_{N\text{p}+1} = 0$}};
\end{axis}
\end{tikzpicture}
        \caption{Homogeneous Dirichlet.}
        \label{fig:Results:Poisson:homoDirichlet}
    \end{subfigure}
    \begin{subfigure}[t]{0.49\textwidth}
        \centering
\tikzsetnextfilename{Results-PoissionDirichletInHomo}
\begin{tikzpicture}
\begin{axis}[
tick pos=left,
xmajorgrids,
xmin=0, xmax=1,
ymajorgrids,
xlabel={$x_k$ [m]},
ylabel={},
y tick label style={
        /pgf/number format/.cd,
        fixed,
        fixed zerofill,
        precision=2,
        /tikz/.cd
    }
]
\addplot [semithick,dashed]
table [col sep = comma , row sep=crcr]{%
0, 1.000000000000000e-02 \\  
0.2000, 8.400002675319725e-02 \\  
0.4000, 1.179999588039245e-01 \\  
0.6000, 1.119999707949726e-01 \\
0.8000, 6.600006272634164e-02 \\
1.0000, -2.000000000000000e-02 \\
};

\addplot [thin,black,mark={*},only marks]
table [col sep = comma , row sep=crcr]{%
0, 0.01 \\  
0.2000, 0.08400000000000002  \\  
0.4000, 0.11800000000000002  \\  
0.6000, 0.11200000000000002  \\
0.8000, 0.066  \\
1.0000, -0.02  \\
};

\addplot [semithick]
table [col sep = comma , row sep=crcr]{%
0,  1.000000000000000e-02 \\  
0.0588, 3.598145388881607e-02  \\  
0.1176, 5.719107118041763e-02  \\  
0.1765, 7.633845775187212e-02  \\
0.2353, 9.296166817448953e-02  \\ 
0.2941, 1.050067335826017e-01  \\ 
0.3529, 1.138494323524782e-01  \\ 
0.4118, 1.188812609411945e-01  \\
0.4706, 1.204089855976630e-01  \\ 
0.5294, 1.186233473970270e-01  \\ 
0.5882, 1.145060701780188e-01  \\ 
0.6471, 1.058153176883413e-01  \\
0.7059, 9.266208614686322e-02  \\  
0.7647, 7.710860627373160e-02  \\
0.8235, 5.662279927669481e-02  \\  
0.8824, 3.420910649755920e-02  \\
0.9412, 9.493589218388308e-03  \\
1.0000, -2.000000000000000e-02  \\
};

\addplot [thin,black,mark={square},only marks]
table [col sep = comma , row sep=crcr]{%
0,  1.000000000000000e-02 \\  
0.0588, 3.591695501730102e-02   \\  
0.1176, 5.837370242214528e-02   \\  
0.1765, 7.737024221453279e-02    \\
0.2353, 9.290657439446355e-02   \\ 
0.2941, 1.049826989619376e-01   \\ 
0.3529, 1.135986159169548e-01  \\ 
0.4118, 1.187543252595154e-01  \\
0.4706, 1.204498269896192e-01  \\ 
0.5294, 1.186851211072662e-01   \\ 
0.5882, 1.134602076124566e-01  \\ 
0.6471, 1.047750865051901e-01  \\
0.7059, 9.262975778546699e-02   \\  
0.7647, 7.702422145328709e-02  \\
0.8235, 5.795847750865043e-02  \\  
0.8824, 3.543252595155703e-02   \\
0.9412, 9.446366782006886e-03 \\
1.0000, -2.000000000000000e-02 \\
};

\node[above,text width=2.3cm,draw, fill=white,semithick]  at (canvas cs:x=1.7cm,y=0.01cm) {\small{$y_0=0.01$ $y_{N\text{p}+1} = -0.02$}};

\end{axis}
\end{tikzpicture}
        \caption{Inhomogeneous Dirichlet.} 
        \label{fig:Results:Poisson:inhomoDirichlet}
    \end{subfigure}\\
    \smallskip
    \hspace{0.5cm}\tikzsetnextfilename{Results-Legend-SecondOrd}
\begin{tikzpicture} 
    \begin{axis}[
        legend style={draw=black!15!black,legend cell align=center, fill opacity=0.8, draw opacity=1, text opacity=1},
        legend columns=8, 
        hide axis,
        width=0.005\textwidth,
        height=0.005\textwidth,
    ]
        \addlegendimage{thin,black,mark={*},only marks}
        \addlegendentry{$y_{n=2}^{\text{FD } \mathcal{O}(1)} \ $}
        \addlegendimage{thin,black,mark=diamond*,mark options={fill=black},only marks}
        \addlegendentry{$y_{n=2}^{\text{FD } \mathcal{O}(2)} \ $}
        \addlegendimage{thin,black,mark={triangle},only marks}
        \addlegendentry{$y_{n=4}^{\text{FD } \mathcal{O}(1)} \ $}
        \addlegendimage{thin,black,mark={square},only marks}
        \addlegendentry{$y_{n=4}^{\text{FD } \mathcal{O}(2)} \ $}
        \addlegendimage{semithick,dashed}
        \addlegendentry{$y_{n=2}^{\text{VQ } \mathcal{O}(1)} \ $}
        \addlegendimage{semithick,dashdotted}
        \addlegendentry{$y_{n=2}^{\text{VQ } \mathcal{O}(2)} \ $}
        \addlegendimage{semithick,dotted}
        \addlegendentry{$y_{n=4}^{\text{VQ } \mathcal{O}(1)} \ $}
        \addlegendimage{semithick}
        \addlegendentry{$y_{n=4}^{\text{VQ } \mathcal{O}(2)}$}
        \addplot [draw=none] coordinates {(0,0) (1,1)};
    \end{axis}
\end{tikzpicture}\\
    \vspace{0.2cm}
    \begin{subfigure}[t]{0.49\textwidth}
        \centering
\tikzsetnextfilename{Results-PoissionNeumannHomo}
\begin{tikzpicture}
\begin{axis}[
tick pos=left,
xmajorgrids,
xmin=0, xmax=1,
ymajorgrids,
xlabel={$x_k$ [m]},
ylabel={$y(x_k)$ [K]},
yticklabel style={text width=2.4em},
y tick label style={
        /pgf/number format/.cd,
        fixed,
        fixed zerofill,
        precision=2,
        /tikz/.cd
    }
]
\addplot [semithick, dashed]
table [col sep = comma , row sep=crcr]{%
0, 8.032821598729457e-02 \\  
0.2000, 8.032821598729457e-02  \\  
0.4000, 4.032821386582131e-02  \\  
0.6000, -3.967179037712520e-02  \\
0.8000, -7.967179249859847e-02  \\
1.0000, -7.967179249859847e-02  \\
};

\addplot [thin,black,mark={*},only marks]
table [col sep = comma , row sep=crcr]{%
0, 0.080328215987295  \\  
0.2000, 0.080328215987295   \\  
0.4000, 0.040328215987295   \\  
0.6000, -0.039671784012705   \\
0.8000, -0.079671784012705   \\
1.0000, -0.079671784012705   \\
};

\addplot [semithick, dashdotted]
table [col sep = comma , row sep=crcr]{%
0.0   , 0.109094759131815   \\  
0.2000, 0.109094759131815   \\  
0.4000, 0.049094760505649   \\  
0.6000,-0.050905239818646   \\
0.8000,-0.110905244206992   \\
1.0000,-0.110905244206992   \\
};

\addplot [thin,black,mark=diamond*,mark options={fill=black},only marks]
table [col sep = comma , row sep=crcr]{%
0.000000000000000, 0.109094759131815  \\
0.200000000000000, 0.109094759131815  \\
0.400000000000000, 0.049094759131815  \\
0.600000000000000, -0.050905240868185  \\
0.800000000000000, -0.110905240868185  \\
1.000000000000000, -0.110905240868185  \\
};

\addplot [semithick, dotted]
table [col sep = comma , row sep=crcr]{%
0,  1.107000762841818e-01 \\  
0.0588, 1.107000762841818e-01   \\  
0.1176, 1.072648898255938e-01   \\  
0.1765, 1.008003575147008e-01   \\
0.2353, 8.992409517334442e-02   \\ 
0.2941, 7.609227618082548e-02   \\ 
0.3529, 5.858163341243307e-02   \\ 
0.4118, 3.691220467695272e-02   \\
0.4706, 1.389296537462902e-02   \\ 
0.5294,-1.382418640042653e-02   \\ 
0.5882,-3.778041560374296e-02   \\ 
0.6471,-5.774072689654697e-02   \\
0.7059,-7.627777080515419e-02   \\  
0.7647,-9.007535426741337e-02   \\
0.8235,-1.006538007338888e-01   \\  
0.8824,-1.078605460454617e-01   \\
0.9412,-1.106522374886819e-01   \\
1.0000,-1.106522374886819e-01   \\
};

\addplot [thin,black,only marks,mark={triangle}]
table [col sep = comma , row sep=crcr]{%
0,  1.107000762841818e-01 \\  
0.0588, 1.107000762841818e-01  \\  
0.1176, 1.072398686717251e-01  \\  
0.1765, 1.003194534468115e-01  \\
0.2353, 8.993883060944112e-02  \\ 
0.2941, 7.609800015961399e-02  \\ 
0.3529, 5.879696209733005e-02 \\ 
0.4118, 3.803571642258935e-02 \\
0.4706, 1.381426313539187e-02 \\ 
0.5294, -1.386739776426238e-02  \\ 
0.5882, -3.808885105145987e-02 \\ 
0.6471, -5.885009672620059e-02 \\
0.7059, -7.615113478848451e-02  \\  
0.7647, -8.999196523831163e-02 \\
0.8235, -1.003725880756820e-01 \\  
0.8824, -1.072930033005955e-01 \\
0.9412, -1.107532109130523e-01 \\
1.0000, -1.107532109130523e-01 \\
};

\addplot [semithick]
table [col sep = comma , row sep=crcr]{%
0.0, 0.123660972098363  \\
0.058823529411765, 0.123660972098363  \\
0.117647058823529, 0.119779820517933  \\
0.176470588235294, 0.110985388919829  \\
0.235294117647059, 0.097303174774497  \\
0.294117647058824, 0.081817436459678  \\
0.352941176470588, 0.060405986353675  \\
0.411764705882353, 0.038198339183315  \\
0.470588235294118, 0.015332040951928  \\
0.529411764705882, -0.014147168215224  \\
0.588235294117647, -0.039017403117995  \\
0.647058823529412, -0.061467280602198  \\
0.705882352941176, -0.082224943070627  \\
0.764705882352941, -0.097786642288185  \\
0.823529411764706, -0.107753336217749  \\
0.882352941176471, -0.117059950169051  \\
0.941176470588235, -0.124307284542221  \\
1.000000000000000, -0.124307284542221  \\
};

\addplot [thin,black,mark={square},only marks]
table [col sep = comma , row sep=crcr]{%
0.0,    0.123660972098363 \\
0.058823529411765, 0.123660972098363 \\
0.117647058823529, 0.118470660679678 \\
0.176470588235294, 0.109820141648536 \\
0.235294117647059, 0.097709415004938 \\
0.294117647058824, 0.082138480748882 \\
0.352941176470588, 0.063107338880370 \\
0.411764705882353, 0.040615989399401 \\
0.470588235294118, 0.014664432305975 \\
0.529411764705882, -0.014747332399908 \\
0.588235294117647, -0.040698889493334 \\
0.647058823529412, -0.063190238974303 \\
0.705882352941176, -0.082221380842815 \\
0.764705882352941, -0.097792315098871 \\
0.823529411764706, -0.109903041742469 \\
0.882352941176471, -0.118553560773611 \\
0.941176470588235, -0.123743872192296 \\
1.000000000000000, -0.123743872192296 \\
};

\node[above,text width=2.0cm,draw, fill=white,semithick]  at (canvas cs:x=1.7cm,y=0.01cm) {$\nicefrac{\partial y_0}{\partial x}$ \small{$= 0$} 
$\nicefrac{\partial y_{N_\text{p}+1}}{\partial x}$ \small{$= 0$} };

\end{axis}
\end{tikzpicture}
        \vspace{-0.4cm}
        \caption{Homogeneous Neumann.} 
        \label{fig:Results:Poisson:homoNeumann}
    \end{subfigure}
    \begin{subfigure}[t]{0.49\textwidth}
        \centering
\tikzsetnextfilename{Results-PoissionNeumannInHomo}
\begin{tikzpicture}
\begin{axis}[
tick pos=left,
xmajorgrids,
xmin=0, xmax=1,
ymajorgrids,
xlabel={$x_k$ [m]},
ylabel={},
yticklabel style={text width=2.4em},
y tick label style={
        /pgf/number format/.cd,
        fixed,
        fixed zerofill,
        precision=2,
        /tikz/.cd
    }
]
\addplot [semithick,dashed]
table [col sep = comma , row sep=crcr]{%
0, 5.777871889888819e-01 \\  
0.2000, 3.777871889888820e-01 \\  
0.4000, 1.377860564191096e-01 \\  
0.6000,-1.422139585277342e-01 \\
0.8000,-3.822124783283919e-01 \\
1.0000,-5.822124783283920e-01 \\
};

\addplot [thin,black,mark={*},only marks]
table [col sep = comma , row sep=crcr]{%
0, 0.577787188988882    \\  
0.2000, 0.377787188988882    \\  
0.4000, 0.137787188988882    \\  
0.6000, -0.142212811011118    \\
0.8000, -0.382212811011118    \\
1.0000, -0.582212811011118   \\
};

\addplot [semithick,dashdotted]
table [col sep = comma , row sep=crcr]{%
0     , 0.609994706193752  \\  
0.2000, 0.409994706193752  \\  
0.4000, 0.149994689227294  \\  
0.6000,-0.150005309266619  \\
0.8000,-0.410005312663044  \\
1.0000,-0.610005312663044  \\
};

\addplot [thin,black,mark=diamond*,mark options={fill=black},only marks]
table [col sep = comma , row sep=crcr]{%
0     ,  0.609994706193752    \\  
0.2000,  0.409994706193752    \\  
0.4000,  0.149994706193752    \\  
0.6000, -0.150005293806248    \\
0.8000, -0.410005293806248    \\
1.0000, -0.610005293806248    \\
};

\addplot [semithick,dotted]
table [col sep = comma , row sep=crcr]{%
0, 6.104808040248462e-01 \\  
0.0588, 5.516572746130814e-01 \\  
0.1176, 4.903343127947838e-01 \\  
0.1765, 4.245384510946750e-01 \\
0.2353, 3.543816496657263e-01 \\ 
0.2941, 2.817360290986010e-01 \\ 
0.3529, 2.034153135644836e-01 \\ 
0.4118, 1.239511361401863e-01 \\
0.4706, 4.312141193211083e-02 \\ 
0.5294,-4.340359677873851e-02 \\ 
0.5882,-1.247233760826053e-01 \\ 
0.6471,-2.043857780400317e-01 \\
0.7059,-2.822604883746087e-01 \\  
0.7647,-3.549207922404842e-01 \\
0.8235,-4.245214637292651e-01 \\  
0.8824,-4.902622630590603e-01 \\
0.9412,-5.521523672915885e-01 \\
1.0000,-6.109758967033532e-01 \\
};

\addplot [thin,black,mark={triangle},only marks]
table [col sep = comma , row sep=crcr]{%
0, 6.104808040248462e-01   \\  
0.0588, 5.516572746130814e-01  \\  
0.1176, 4.893735375888599e-01  \\  
0.1765, 4.236295929521814e-01  \\
0.2353, 3.544254407030459e-01  \\ 
0.2941, 2.817610808414536e-01  \\ 
0.3529, 2.056365133674043e-01  \\ 
0.4118, 1.260517382808982e-01  \\
0.4706, 4.300675558193523e-02  \\ 
0.5294, -4.349843472948449e-02  \\ 
0.5882, -1.265434174284474e-01  \\ 
0.6471, -2.061281925149535e-01  \\
0.7059, -2.822527599890028e-01  \\  
0.7647, -3.549171198505951e-01  \\
0.8235, -4.241212720997305e-01  \\ 
0.8824, -4.898652167364090e-01  \\
0.9412, -5.521489537606306e-01  \\
1.0000, -6.109724831723953e-01  \\
};

\addplot [semithick]
table [col sep = comma , row sep=crcr]{%
0.000000000000000, 0.623536309697084    \\
0.058823529411765, 0.564712780285320    \\
0.117647058823529, 0.498064272530470    \\
0.176470588235294, 0.430659054224650    \\
0.235294117647059, 0.362014594410399    \\
0.294117647058824, 0.287615791807549    \\
0.352941176470588, 0.211709092197560    \\
0.411764705882353, 0.130393938925820    \\
0.470588235294118, 0.043962628217989    \\
0.529411764705882, -0.044282320520101    \\
0.588235294117647, -0.128367751485341    \\
0.647058823529412, -0.209290392724581    \\
0.705882352941176, -0.287155086528781    \\
0.764705882352941, -0.361607495287021    \\
0.823529411764706, -0.431788593137331    \\
0.882352941176471, -0.499808625779601    \\
0.941176470588235, -0.565315210276391    \\
1.000000000000000, -0.624138739688155    \\
};

\addplot [thin,black,mark={square},only marks]
table [col sep = comma , row sep=crcr]{%
0.000000000000000, 0.623536309697084  \\
0.058823529411765, 0.564712780285319  \\
0.117647058823529, 0.500698939454869  \\
0.176470588235294, 0.433224891011962  \\
0.235294117647059, 0.362290634956599  \\
0.294117647058824, 0.287896171288778  \\
0.352941176470588, 0.210041500008500  \\
0.411764705882353, 0.128726621115766  \\
0.470588235294118, 0.043951534610574  \\
0.529411764705882, -0.044283759507074  \\
0.588235294117647, -0.129058846012265  \\
0.647058823529412, -0.210373724905000  \\
0.705882352941176, -0.288228396185277  \\
0.764705882352941, -0.362622859853098  \\
0.823529411764706, -0.433557115908462  \\
0.882352941176471, -0.501031164351369  \\
0.941176470588235, -0.565045005181819  \\
1.000000000000000, -0.623868534593583  \\
};

\node[above,text width=2.2cm,draw, fill=white,semithick]  at (canvas cs:x=1.6cm,y=0.01cm) {$\nicefrac{\partial y_0}{\partial x}$ \small{$= -1$} 
$\nicefrac{\partial y_{N_\text{p}+1}}{\partial x}$ \small{$ = -1$} };

\end{axis}
\end{tikzpicture}
        \caption{Inhomogeneous Neumann.}
    \label{fig:Results:Poisson:inhomoNeumann}
    \end{subfigure}\\
    \smallskip
    \begin{subfigure}[t]{0.49\textwidth}
        \centering       
        \tikzsetnextfilename{Results-PoissionMixed}
\begin{tikzpicture}
\begin{axis}[
tick pos=left,
xmajorgrids,
xmin=0, xmax=1,
ymajorgrids,
xlabel={$x_k$ [m]},
ylabel={$y(x_k)$ [K]},
y tick label style={
        /pgf/number format/.cd,
        fixed,
        fixed zerofill,
        precision=2,
        /tikz/.cd
    }
]

\addplot [semithick,dashed]
table [col sep = comma , row sep=crcr]{%
0, 0.399999968619869 \\ 
0.2000, 0.399999968619869 \\  
0.4000, 0.359999971060695 \\  
0.6000, 0.279999975474656 \\
0.8000, 0.159999985667602 \\
1.0000, 0 \\
};

\addplot [thin,black,mark={*},only marks]
table [col sep = comma , row sep=crcr]{%
0, 0.400000000000000   \\  
0.2000, 0.400000000000000    \\  
0.4000, 0.360000000000000    \\  
0.6000, 0.280000000000000    \\
0.8000, 0.160000000000000    \\
1.0000, 0    \\
};

\addplot [semithick, dashdotted]
table [col sep = comma , row sep=crcr]{%
0.0   , 0.510000007737554 \\  
0.2000, 0.480000007737554 \\  
0.4000, 0.420000005586033 \\  
0.6000, 0.320000002759379 \\
0.8000, 0.180000000348863 \\
1.0000, 0.0 \\
};

\addplot [thin,black,mark=diamond*,mark options={fill=black},only marks]
table [col sep = comma , row sep=crcr]{%
0.000000000000000, 0.510000000000000 \\
0.200000000000000, 0.480000000000000 \\
0.400000000000000, 0.420000000000000 \\
0.600000000000000, 0.320000000000000 \\
0.800000000000000, 0.180000000000000 \\
1.000000000000000, 0.0  \\
};

\addplot [semithick,dotted]
table [col sep = comma , row sep=crcr]{%
0.0, 0.470588235294118 \\   
0.0588, 0.470588235294118 \\    
0.1176, 0.467128027681661 \\     
0.1765, 0.460207612456747 \\    
0.2353, 0.449826989619377 \\     
0.2941, 0.435986159169550 \\    
0.3529, 0.418685121107266 \\   
0.4118, 0.397923875432526 \\  
0.4706, 0.373702422145329 \\  
0.5294, 0.346020761245675 \\  
0.5882, 0.314878892733564 \\  
0.6471, 0.280276816608997 \\ 
0.7059, 0.242214532871972 \\   
0.7647, 0.200692041522491 \\  
0.8235, 0.155709342560554 \\    
0.8824, 0.107266435986159 \\ 
0.9412, 0.055363321799308 \\  
1.0000, 0 \\
};

\addplot [thin,black,mark={triangle},only marks]
table [col sep = comma , row sep=crcr]{%
0.0, 0.470613389567865 \\    
0.0588, 0.470613389567865 \\    
0.1176, 0.466885180637723 \\    
0.1765, 0.459867552176784 \\ 
0.2353, 0.449580873544664 \\  
0.2941, 0.435721757012530 \\  
0.3529, 0.418691433915085 \\  
0.4118, 0.398039800866966 \\ 
0.4706, 0.373690555308196 \\  
0.5294, 0.346058724463163 \\  
0.5882, 0.315076486221058 \\  
0.6471, 0.280545551090251 \\
0.7059, 0.242586753804909 \\   
0.7647, 0.201026961167694 \\ 
0.8235, 0.155841563544137 \\   
0.8824, 0.107294638575343 \\ 
0.9412, 0.055361674535804 \\
1.0000, 0.0 \\
};

\addplot [semithick]
table [col sep = comma , row sep=crcr]{%
0.000000000000000, 0.500874984140014   \\
0.058823529411765, 0.498279828430671   \\
0.117647058823529, 0.492873870785460   \\
0.176470588235294, 0.484233839094504   \\
0.235294117647059, 0.472304784379440   \\
0.294117647058824, 0.456730511121352   \\
0.352941176470588, 0.437843929341765   \\
0.411764705882353, 0.415356214826708   \\
0.470588235294118, 0.389262373612820   \\
0.529411764705882, 0.359849400634862   \\
0.588235294117647, 0.327171666095242   \\
0.647058823529412, 0.290807802343409   \\
0.705882352941176, 0.250872668766537   \\
0.764705882352941, 0.207620984156392   \\
0.823529411764706, 0.160774753007692   \\
0.882352941176471, 0.110621785773604   \\
0.941176470588235, 0.057094347567786   \\
1.000000000000000, 0.0   \\
};

\addplot [thin,black,mark={square},only marks]
table [col sep = comma , row sep=crcr]{%
0.000000000000000, 0.500865051903114 \\
0.058823529411765, 0.498269896193772 \\
0.117647058823529, 0.493079584775086 \\
0.176470588235294, 0.484429065743945 \\
0.235294117647059, 0.472318339100346 \\
0.294117647058824, 0.456747404844291 \\
0.352941176470588, 0.437716262975779 \\
0.411764705882353, 0.415224913494810 \\
0.470588235294118, 0.389273356401384 \\
0.529411764705882, 0.359861591695502 \\
0.588235294117647, 0.326989619377163 \\
0.647058823529412, 0.290657439446367 \\
0.705882352941176, 0.250865051903114 \\
0.764705882352941, 0.207612456747405 \\
0.823529411764706, 0.160899653979239 \\
0.882352941176471, 0.110726643598616 \\
0.941176470588235, 0.057093425605536 \\
1.000000000000000, 0.0 \\
};

\node[above,text width=1.7cm,draw, fill=white,semithick]  at (canvas cs:x=1.7cm,y=0.01cm) {$\nicefrac{\partial y_0}{\partial x}$ \small{$= 0$} 
\small{$y_{N_\text{p}+1} = 0$}};
\end{axis}
\end{tikzpicture}
        \caption{Mixed Neumann/Dirichlet.}
        \label{fig:Results:Poisson:Mixed}    
        \end{subfigure}
    \begin{subfigure}[t]{0.49\textwidth}
        \centering
        \tikzsetnextfilename{Results-PoissionRobin}
\begin{tikzpicture}
\begin{axis}[
tick pos=left,
xmajorgrids,
xmin=0, xmax=1,
ymajorgrids,
xlabel={$x_k$ [m]},
ylabel={},
yticklabel style={text width=2.5em},
y tick label style={
        /pgf/number format/.cd,
        fixed,
        fixed zerofill,
        precision=2,
        /tikz/.cd
    }
]

\addplot [semithick,dashed]
table [col sep = comma , row sep=crcr]{%
0.0, -0.062857144068556 \\
0.20, 0.074285711862889 \\
0.40, 0.171428570993395 \\
0.60, 0.228571429176039 \\  
0.80, 0.245714284135476 \\ 
1.00, 0.222857142067738 \\
};

\addplot [thin,black,mark={*},only marks]
table [col sep = comma , row sep=crcr]{%
0.0,  -0.062857142857143 \\
0.20,   0.074285714285714 \\
0.40,   0.171428571428572 \\
0.60,   0.228571428571429 \\
0.80,   0.245714285714286 \\
1.00,   0.222857142857143 \\
};

\addplot [semithick,dashdotted]
table [col sep = comma , row sep=crcr]{%
0.0,  -0.052857132707344  \\
0.20,  0.094285734585313 \\
0.40,  0.191428588642811 \\
0.60,  0.248571436893727 \\  
0.80,  0.265714289272738 \\ 
1.00,  0.232857144636369 \\
};

\addplot [thin,black,mark=diamond*,mark options={fill=black},only marks]
table [col sep = comma , row sep=crcr]{%
0.0,  -0.052857142857143  \\
0.20,  0.094285714285714  \\
0.40,  0.191428571428572  \\
0.60,  0.248571428571429  \\
0.80,  0.265714285714286  \\
1.00,  0.232857142857143  \\
};

\addplot [semithick, dotted]
table [col sep = comma , row sep=crcr]{%
0.0, -0.024941370471424  \\
0.058823529411765, 0.00894078846891669  \\
0.117647058823529, 0.0385359489125765  \\ 
0.176470588235294, 0.0657783981060733  \\
0.235294117647059, 0.0898392904257255  \\
0.294117647058824, 0.109844287901088  \\
0.352941176470588, 0.126974039146231  \\
0.411764705882353, 0.139856523203727  \\
0.470588235294118, 0.149085864480187  \\
0.529411764705882, 0.155276773904836  \\
0.588235294117647, 0.158619278295906  \\
0.647058823529412, 0.157683557945546  \\
0.705882352941176, 0.153073721469675  \\
0.764705882352941, 0.145451373241442  \\
0.823529411764706, 0.133322390858337  \\
0.882352941176471, 0.119133984817197  \\
0.941176470588235, 0.101840453467265  \\
1.000000000000000, 0.080331991439515  \\
};

\addplot [thin,black,mark={triangle},only marks]
table [col sep = comma , row sep=crcr]{%
0.0, -0.024949918047714   \\
0.058823529411765, 0.008923693316336  \\
0.117647058823529, 0.039337097067929  \\ 
0.176470588235294, 0.066290293207066  \\
0.235294117647059, 0.089783281733746  \\
0.294117647058824, 0.109816062647969  \\
0.352941176470588, 0.126388635949736  \\
0.411764705882353, 0.139501001639045  \\
0.470588235294118, 0.149153159715898  \\
0.529411764705882, 0.155345110180295  \\
0.588235294117647, 0.158076853032234  \\
0.647058823529412, 0.157348388271717  \\
0.705882352941176, 0.153159715898743  \\
0.764705882352941, 0.145510835913312  \\
0.823529411764706, 0.134401748315425  \\
0.882352941176471, 0.119832453105081  \\
0.941176470588235, 0.101802950282280  \\
1.000000000000000, 0.080313239847022  \\
};

\addplot [semithick]
table [col sep = comma , row sep=crcr]{%
0.000000000000000, -0.024019284914696   \\
0.058823529411765, 0.010784959582372    \\
0.117647058823529, 0.040265538828842    \\
0.176470588235294, 0.066988399601247    \\
0.235294117647059, 0.091076928667884    \\
0.294117647058824, 0.111117738883410    \\
0.352941176470588, 0.128278678031549    \\
0.411764705882353, 0.141667686576969    \\
0.470588235294118, 0.150892833295396    \\
0.529411764705882, 0.157069727857551    \\
0.588235294117647, 0.160562836579699    \\
0.647058823529412, 0.159825876523897    \\
0.705882352941176, 0.155209381947258    \\
0.764705882352941, 0.147680227322350    \\
0.823529411764706, 0.135436467218476    \\
0.882352941176471, 0.120757186977753    \\
0.941176470588235, 0.103506441768324    \\
1.000000000000000, 0.081164985590044    \\
};

\addplot [thin,black,mark={square},only marks]
table [col sep = comma , row sep=crcr]{%
0.000000000000000, -0.024084866144600  \\
0.058823529411765, 0.010653797122564  \\
0.117647058823529, 0.041067200874158  \\
0.176470588235294, 0.068020397013294  \\
0.235294117647059, 0.091513385539974  \\
0.294117647058824, 0.111546166454198  \\
0.352941176470588, 0.128118739755964  \\
0.411764705882353, 0.141231105445274  \\
0.470588235294118, 0.150883263522127  \\
0.529411764705882, 0.157075213986523  \\
0.588235294117647, 0.159806956838463  \\
0.647058823529412, 0.159078492077945  \\
0.705882352941176, 0.154889819704971  \\
0.764705882352941, 0.147240939719541  \\
0.823529411764706, 0.136131852121653  \\
0.882352941176471, 0.121562556911309  \\
0.941176470588235, 0.103533054088508  \\
1.000000000000000, 0.081178291750137  \\
};

\node[above,text width=3.7cm,draw, fill=white,semithick]  at (canvas cs:x=3.7cm,y=0.01cm){
$\nicefrac{-y_0     }{\Delta x}+\nicefrac{\partial y_0        }{\partial x}$ \small{$= 1$} 
$\nicefrac{y_{N_p+1}}{\Delta x}+\nicefrac{\partial y_{N_p + 1}}{\partial x}$ \small{$= 1$}};
\end{axis}
\end{tikzpicture}
        \caption{Inhomogeneous Robin.}
        \label{fig:Results:Poisson:Robin}
    \end{subfigure}\\
    
    \caption{Comparisons of \textsc{VQA}-result (lines) and \textsc{FD}-results (symbols) for the steady heat conduction in combination with different boundary conditions on a grid with $N_\text{p}=4$ ($n=2$; dashed or dash-dotted lines \& closed symbols) and $N_\text{p}=16$ ($n=4$; dotted or solid lines \& open symbols) interior points. The depth of the ansatz circuit $U(\lamb_c)$ is $d=1$ for $n=2$ qubits and $d=3$ for $n=4$ qubits, cf. \Fig{fig:VQA_Ansatz}.}
    \label{fig:Results:Poisson}
\end{figure}

As indicated in \Fig{fig:Results:Poisson}, the \textsc{VQA} predictions perfectly agree with the \textsc{FD} reference for all Dirichlet, Neumann, Robin, or mixed boundary conditions and grid resolutions. Furthermore, the first-order accurate  approximation of Neumann conditions is obviously improved by using second-order techniques in both the \textsc{VQA} and the \textsc{FD}-results, cf. for example \Fig{fig:Results:Poisson:homoNeumann}. Note that, the increase in the order of accuracy does not imply higher computational costs}, since the quantum circuits described in Sec.~\ref{sec:Compilation} can be employed without additional modifications and only the source term $\Tilde{\fb}$ is adjusted. It is noteworthy to remark that using second-order Neumann boundary conditions does not induce a substantial change in the Robin boundary condition experiment, which is confirmed by the \textsc{FD} results. The required depth $d$ increases for non-periodic conditions on the finer grid. To this end, a depth $d=3$ is sufficient to mimic all investigated boundary conditions in conjunction with the finer grid. 

\Tab{tab:Results:Poisson:Errors} quantifies the deviation of the \textsc{VQA}-results from the \textsc{FD}-results for all cases depicted in \Fig{fig:Results:Poisson}. The table displays a deviation of $\varepsilon_{l_2} \approx 10^{-6}$--$10^{-7}$ for two qubits ($N_\text{p}=4$), disregarding the imposed boundary conditions. The good agreement of $\varepsilon_{l_2}$ for the periodic case, where no specific boundary condition circuits are required, and the non-periodic cases reveal that no substantial sensitivity of the attainable accuracy is observed when applying the current strategy for a quantum-based implementation of boundary conditions. For the finer discretization associated with $n=4$ qubits, \Tab{tab:Results:Poisson:Errors} shows higher values of $\varepsilon_{l_2} \approx 10^{-3}$ for the non-periodic boundary conditions and thereby a less satisfactory agreement with the periodic arrangement. The latter is attributed to difficulties of finding the global optimum in non-convex optimization and highlights a bottleneck of the \textsc{VQA} in combination with versatile ansatz functions $U(\lamb_c)$, cf. \Fig{fig:VQA}. Moreover, it also outlines that sufficient accuracy of the ancilla measurements is crucial for detecting the (global) minimum. 

The trace distance $\varepsilon_{\text{tr}}$ is given in the third column of \Tab{tab:Results:Poisson:Errors} and can be interpreted as point-wise correlation between the quantum- and the \textsc{FD}-result. The trace distance is related to the convergence to the global minimum, while the $\varepsilon_{l_2}$ measures the expressibility of the ansatz. Since the two measures address different aspects, considering just one of them seems insufficient. Moreover, the trace distance sometimes displays numerical zero values, cf. \Tab{tab:Results:Poisson:Errors}, which does not necessarily guarantee a perfect agreement and also advocates assessing both measures. In addition to the error measures used to assess the \textsc{VQA}-results, the origin of the observed differences in predictive accuracy is also of importance. 
Interestingly, the results reported by Sato et al. \cite{Sato2021} -for example- show a much better trace distance compared to the present results for periodic boundaries, although the predictive agreement of the current results appears to be slightly better.  A key difference between the current and similar recent efforts, e.g., \cite{Sato2021}, refers to the employed \textsc{\it SO4} ansatz illustrated in \Fig{fig:quantum_ansatz_SO4}, which seems to feature an improved ansatz expressibility but is also harder to train. 
Furthermore, the number of control variables $c$ given in \Tab{tab:Results:Poisson:Errors} roughly scales as $\mathcal{O}(n \, d)$, which suggests room for future improvements in the design of the quantum ansatz circuits towards more problem-specific strategies \cite{Holmes2022}. 

\begin{table}[htbp]
   \centering
    \begin{tabular}{@{}lcccccc@{}}
\toprule
Boundary Setting & \multicolumn{2}{c}{$l_2$-norm $\varepsilon_{l_2}$}  & \multicolumn{2}{c}{trace distance $\varepsilon_{\text{tr}}$}&\multicolumn{2}{c}{\textsc{DOF} $c$} \\ 
\midrule
\multicolumn{1}{r}{interior points}&\multicolumn{1}{l}{$N_\text{p}=4$}& $N_\text{p}=16$& \multicolumn{1}{|l}{$N_\text{p}=4$}& $N_\text{p}=16$& \multicolumn{1}{|l}{$N_\text{p}=4$}& $N_\text{p}=16$   \\
\midrule
Periodic          & \multicolumn{1}{c}{$1.23 \times 10^{-6}$}&\multicolumn{1}{c}{$3.26 \times 10^{-7}$}&\multicolumn{1}{|c}{$9.70 \times 10^{-1}$}&\multicolumn{1}{c}{$1.0 \times 10^{0}$}&\multicolumn{1}{|c}{$3$}&$6$\\
Homog. Dirichlet   & \multicolumn{1}{c}{$1.5 \times 10^{-7}$}&\multicolumn{1}{c}{$2.54\times 10^{-3}$}&\multicolumn{1}{|c}{$0.00$}&\multicolumn{1}{c}{$6.76 \times 10^{-3}$}&\multicolumn{1}{|c}{$3$}&$24$\\
Homog. Neumann $\mathcal{O}(1)$  & \multicolumn{1}{c}{$1.37 \times 10^{-8}$
}&\multicolumn{1}{c|}{$1.82 \times 10^{-3}$}&\multicolumn{1}{c}{$6.32 \times 10^{-8}$}&\multicolumn{1}{c}{$5.05 \times 10^{-3}$}&\multicolumn{1}{|c}{$3$}&$24$\\
Homog. Neumann $\mathcal{O}(2)$  & \multicolumn{1}{c}{$5.03 \times 10^{-9}$
}&\multicolumn{1}{c|}{$5.53 \times 10^{-3}$}&\multicolumn{1}{c}{$1.49 \times 10^{-8}$}&\multicolumn{1}{c}{$1.36 \times 10^{-2}$}&\multicolumn{1}{|c}{$3$}&$24$\\
Inhomog. Dirichlet & \multicolumn{1}{c}{$8.49 \times 10^{-8}$}&\multicolumn{1}{c}{$2.83 \times 10^{-3}$}&\multicolumn{1}{|c}{$4.31 \times 10^{-7}$}&\multicolumn{1}{c}{$7.87 \times 10^{-3}$}&\multicolumn{1}{|c}{$3$}&$24$\\
Inhomog. Neumann  $\mathcal{O}(1)$ & \multicolumn{1}{c}{$1.68 \times 10^{-6}$}&\multicolumn{1}{c}{$4.22 \times 10^{-3}$}&\multicolumn{1}{|c}{$1.65 \times 10^{-6}$
}&\multicolumn{1}{c}{$2.55 \times 10^{-3}$}&\multicolumn{1}{|c}{$3$}&$24$\\
Inhomog. Neumann  $\mathcal{O}(2)$ & \multicolumn{1}{c}{$3.52 \times 10^{-8}$}&\multicolumn{1}{c}{$5.28 \times 10^{-3}$}&\multicolumn{1}{|c}{$2.58 \times 10^{-8}$
}&\multicolumn{1}{c}{$2.78 \times 10^{-3}$}&\multicolumn{1}{|c}{$3$}&$24$\\
Mixed Neumann/Dirichlet  $\mathcal{O}(1)$ & \multicolumn{1}{c}{$6.01 \times 10^{-8}$
}&\multicolumn{1}{c}{$8.38 \times 10^{-4}$
}&\multicolumn{1}{|c}{$0.00$}&\multicolumn{1}{c}{{$5.66
 \times 10^{-4}$}}&\multicolumn{1}{|c}{$3$}&$24$\\ 
Mixed Neumann/Dirichlet  $\mathcal{O}(2)$ & \multicolumn{1}{c}{$1.26 \times 10^{-8}$
}&\multicolumn{1}{c}{$4.44 \times 10^{-4}$ }&\multicolumn{1}{|c}{$0.00$}&\multicolumn{1}{c}{{$2.87 \times 10^{-4}$}}&\multicolumn{1}{|c}{$3$} & $24$\\ 
Inhomog. Robin   $\mathcal{O}(1)$   & \multicolumn{1}{c}{$3.32 \times 10^{-9}$
}&\multicolumn{1}{c}{$1.86 \times 10^{-3}$
}&\multicolumn{1}{|c}{{$0.00$}}&\multicolumn{1}{c}{$3.7 \times 10^{-3}$}&\multicolumn{1}{|c}{$3$}&{$24$}\\ 
Inhomog. Robin $\mathcal{O}(2)$ & \multicolumn{1}{c}{$2.99 \times 10^{-8}$
}&\multicolumn{1}{c}{$2.21 \times 10^{-3}$
}&\multicolumn{1}{|c}{{$5.16 \times 10^{-8}$}}&\multicolumn{1}{c}{$4.34 \times 10^{-3}$}&\multicolumn{1}{|c}{$3$}&{$24$}\\ 
\bottomrule
\end{tabular}
\caption{Deviations of \textsc{VQA} and \textsc{FD}-results in the $l_2$-norm ($\varepsilon_{l_2}$) supplemented by their trace distance ($\varepsilon_{\text{tr}}$) for results depicted in \Fig{fig:Results:Poisson}. The degrees of freedom (\textsc{DOF}) are indicated by the number of control variables~$c$.}
\label{tab:Results:Poisson:Errors}
\end{table}

\subsection{Transient Heat Conduction (Heat Equation)}
\label{sec:Application:Heat}
The second case deals with the transient heat conduction where the time derivative in \Eq{eq:Generic_PDE-TR-2} is treated as discussed in Sec. \ref{sec:Mathematics}. The contributions of the potential term $p(x,t)$~[1/s] as well as the physical source term $f(x)$~[K/s] in \Eq{eq:Generic_PDE-TR} are neglected. The diffusivity is again set to a unit value of $\nu=1$~[m$^2$/s]. Four different boundary condition settings are investigated using a constant time step of $\Delta t \nu/\Delta x^2 =1/(2^{\abs{n-3}}(N_p-n))$. The discretizations are the same two equidistant grids used in Sec. \ref{sec:Application:Poisson}, i.e., a coarse ($N_\text{p}=4$) and a fine grid ($N_\text{p}=16$). All cases studied should reach a steady state, and due to the different convergence speeds to the steady state, different time step numbers are shown for different boundary and initial conditions. The convergence of the optimizer is controlled by \Eq{alg:convergence} using a threshold level of $10^{-7}$ for two qubits and $10^{-4}$ for four qubits, respectively. 

\begin{figure}[htbp]
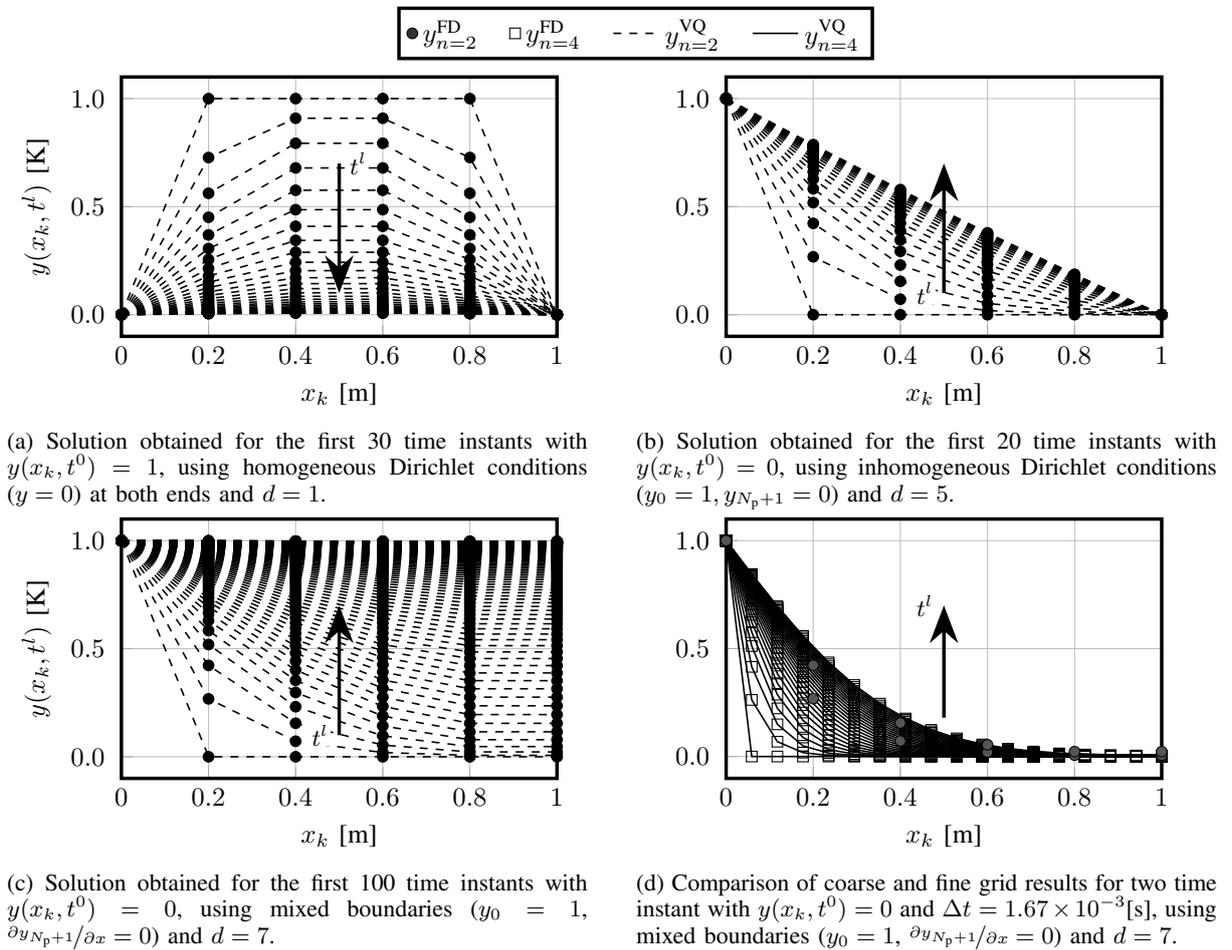

\centering
\hspace{0.5cm}\tikzsetnextfilename{Results-Legend-Heat}
\begin{tikzpicture} 
    \begin{axis}[
        legend style={draw=black!15!black,legend cell align=center, fill opacity=0.8, draw opacity=1, text opacity=1},
        legend columns=4, 
        hide axis,
        width=0.005\textwidth,
        height=0.005\textwidth,
    ]

        \addlegendimage{thin,black,mark={*},only marks}
        \addlegendentry{$y^{\text{FD}}_{n=2} \quad $}
        \addlegendimage{thin,black,mark={square},only marks}
        \addlegendentry{$y^{\text{FD}}_{n=4} \quad $} 
        \addlegendimage{semithick,dashed}
        \addlegendentry{$y^{\text{VQ}}_{n=2} \quad $}
        \addlegendimage{semithick}
        \addlegendentry{$y^{\text{VQ}}_{n=4}$}
        \addplot [draw=none] coordinates {(0,0) (1,1)};
    \end{axis}
\end{tikzpicture}\\
\vspace{0.2cm}
    \begin{subfigure}[t]{0.48\textwidth}
       \centering
        \input{Figures/Results/R_Heat_Dirichlet_f1_homogeneous.tex}
        \caption{Solution obtained for the first 30 time instants with $y(x_k,t^0)=1$, using homogeneous Dirichlet conditions ($y = 0$) at both ends and $d=1$.}
        \label{fig:Results:Heat:homoDirichlet}
        \end{subfigure}
        \hfill
        \begin{subfigure}[t]{0.48\textwidth}
        \centering
        \tikzsetnextfilename{Results-HeatDirichletInhomoF0}
\begin{tikzpicture}
\begin{axis}[
tick pos=left,
xmajorgrids,
xmin=0, xmax=1,
ymajorgrids,
xlabel={$x_k$ [m]},
ylabel={},
y tick label style={
        text width=1.5em,
        /pgf/number format/.cd,
        fixed,
        fixed zerofill,
        precision=1,
        /tikz/.cd
}
]

\addplot [semithick,dashed]
table [col sep = comma , row sep=crcr]{%
0.0, 1.0 \\  
0.2, 0.0 \\
0.4, 0.0 \\
0.6, 0.0 \\
0.8, 0.0 \\
1.0, 0.0 \\
};

\addplot [thin,mark={*},mark options={solid},only marks]
table [col sep = comma , row sep=crcr]{%
0.0, 1.0 \\  
0.2, 0.0 \\
0.4, 0.0 \\
0.6, 0.0 \\
0.8, 0.0 \\
1.0, 0.0 \\
};

\addplot [semithick,dashed]
table [col sep = comma , row sep=crcr]{%
0.0, 1.0 \\  
0.2, 0.267942585868218  \\
0.4, 0.0717703195363581 \\
0.6, 0.0191387333116379 \\
0.8, 0.00478465949320228 \\
1.0, 0.0 \\
};

\addplot [thin,mark={*},mark options={solid},only marks]
table [col sep = comma , row sep=crcr]{%
0.0, 1.0 \\  
0.2, 0.26794258373205737 \\
0.4, 0.07177033492822965 \\
0.6, 0.019138755980861236 \\
0.8, 0.004784688995215308 \\
1.0, 0.0  \\
};

\addplot [semithick,dashed]
table [col sep = comma , row sep=crcr]{%
0,  1.0 \\  
0.2, 0.422609368242536 \\
0.4, 0.154552297384516 \\
0.6, 0.0520592179550561 \\
0.8, 0.0154071282780572 \\
1.0, 0.0\\
};

\addplot [thin,mark={*},mark options={solid},only marks]
table [col sep = comma , row sep=crcr]{%
0.0, 1.0 \\  
0.2, 0.4226093724960508 \\
0.4, 0.15455232252008877 \\
0.6, 0.05205924772784504 \\
0.8, 0.01540715642956891 \\
1.0, 0.0 \\
};

\addplot [semithick,dashed]
table [col sep = comma , row sep=crcr]{%
0, 1.0 \\  
0.2, 0.518737353303533 \\
0.4, 0.229730665379588 \\
0.6, 0.0910807106171441 \\
0.8, 0.0304737441273506 \\
1.0, 0.0 \\
};

\addplot [thin,mark={*},mark options={solid},only marks]
table [col sep = comma , row sep=crcr]{%
0.0, 1.0 \\  
0.2, 0.5187373573676662 \\
0.4, 0.22973068447856348 \\
0.6, 0.09108073550641012 \\
0.8, 0.030473762091386983 \\
1.0, 0\\
};

\addplot [semithick,dashed]
table [col sep = comma , row sep=crcr]{%
0,  1.0 \\ 
0.2, 0.5826798856546 \\
0.4, 0.293244847331632 \\
0.6, 0.130838172105505 \\
0.8, 0.047946413941419 \\
1.0, 0.0 \\
};

\addplot [thin,mark={*},mark options={solid},only marks]
table [col sep = comma , row sep=crcr]{%
0.0, 1.0 \\  
0.2, 0.5826798945826295 \\
0.4, 0.29324486359518576 \\
0.6, 0.1308381908409865 \\
0.8, 0.04794642875594011 \\
1.0, 0.0 \\
};

\addplot [semithick,dashed]
table [col sep = comma , row sep=crcr]{%
0,  1.0 \\  
0.2, 0.627751629018659 \\ 
0.4, 0.345646755846778 \\ 
0.6, 0.168345686264924 \\ 
0.8, 0.0660596290817647 \\ 
1.0, 0.0 \\
};

\addplot [thin,mark={*},mark options={solid},only marks]
table [col sep = comma , row sep=crcr]{%
0.0, 1.0 \\  
0.2, 0.6277516386858845 \\
0.4, 0.3456467655782787 \\
0.6, 0.16834569643685915 \\
0.8, 0.06605963848718484 \\
1.0, 0\\
};

\addplot [semithick,dashed]
table [col sep = comma , row sep=crcr]{%
0.0, 1.0 \\  
0.2, 0.661035743219225 \\ 
0.4, 0.388639707669206 \\ 
0.6, 0.202229573977647 \\ 
0.8, 0.0835872083007691 \\ 
1.0, 0.0 \\
};

\addplot [thin,mark={*},mark options={solid},only marks]
table [col sep = comma , row sep=crcr]{%
0.0, 1.0 \\  
0.2, 0.6610357480796012 \\
0.4, 0.38863971494663613 \\
0.6, 0.20222958055038578 \\
0.8, 0.08358721438118887 \\
1.0, 0.0 \\
};

\addplot [semithick,dashed]
table [col sep = comma , row sep=crcr]{%
0.0, 1.0 \\  
0.2, 0.686508183379355 \\ 
0.4, 0.423961163170295 \\ 
0.6, 0.232057046408136 \\ 
0.8, 0.0998078675090238 \\ 
1.0, 0.0 \\
};

\addplot [thin,mark={*},mark options={solid},only marks]
table [col sep = comma , row sep=crcr]{%
0.0, 1.0 \\  
0.2, 0.6865081641458367 \\
0.4, 0.4239611604241446 \\
0.6, 0.23205704765746946 \\
0.8, 0.09980786910496178 \\
1.0, 0.0 \\
};

\addplot [semithick,dashed]
table [col sep = comma , row sep=crcr]{%
0.0, 1.0 \\  
0.2, 0.70652547227065 \\
0.4, 0.453085516719719 \\
0.6, 0.257894278594008 \\
0.8, 0.114377502385194 \\
1.0, 0.0 \\
};

\addplot [thin,mark={*},mark options={solid},only marks]
table [col sep = comma , row sep=crcr]{%
0.0, 1.0 \\  
0.2, 0.7065254608446302  \\
0.4, 0.4530855150868477  \\
0.6, 0.2578942786544713  \\
0.8, 0.1143775042160987  \\
1.0, 0.0 \\
};

\addplot [semithick,dashed]
table [col sep = comma , row sep=crcr]{%
0.0, 1.0 \\  
0.2, 0.722561363138816 \\
0.4, 0.477194539200957 \\
0.6, 0.280045830369763 \\
0.8, 0.127200202851091 \\
1.0, 0.0 \\
};

\addplot [thin,mark={*},mark options={solid},only marks]
table [col sep = comma , row sep=crcr]{%
0.0, 1.0 \\  
0.2, 0.7225613698797702 \\
0.4, 0.4771945578298205 \\
0.6, 0.2800458312658166 \\
0.8, 0.12720020992450348 \\
1.0, 0.0 \\
};

\addplot [semithick,dashed]
table [col sep = comma , row sep=crcr]{%
0.0, 1.0 \\  
0.2, 0.73558600788731 \\
0.4, 0.497221308295737 \\
0.6, 0.29891015402017 \\
0.8, 0.138327637622034 \\
1.0, 0.0 \\
};

\addplot [thin,mark={*},mark options={solid},only marks]
table [col sep = comma , row sep=crcr]{%
0.0, 1.0 \\  
0.2, 0.7355860154612651 \\
0.4, 0.4972213220855201 \\
0.6, 0.29891015722117464 \\
0.8, 0.13832764426754537 \\
1.0, 0.0 \\
};

\addplot [semithick,dashed]
table [col sep = comma , row sep=crcr]{%
0.0, 1.0 \\
0.2, 0.746268943490133 \\
0.4, 0.513903756604512 \\
0.6, 0.314903438608766 \\
0.8, 0.147889671433962 \\
1.0, 0.0 \\
};

\addplot [thin,mark={*},mark options={solid},only marks]
table [col sep = comma , row sep=crcr]{%
0.0, 1.0 \\  
0.2, 0.746268947083884  \\
0.4, 0.5139037574130063  \\
0.6, 0.31490343839710083  \\
0.8, 0.14788968173304787  \\
1.0, 0.0 \\
};

\addplot [semithick,dashed]
table [col sep = comma , row sep=crcr]{%
0.0, 1.0 \\
0.2, 0.755092064715287 \\
0.4, 0.52783036645087 \\
0.6, 0.328421890214315 \\
0.8, 0.156050309554279 \\
1.0, 0.0 \\
};

\addplot [thin,mark={*},mark options={solid},only marks]
table [col sep = comma , row sep=crcr]{%
0.0, 1.0 \\  
0.2, 0.755092065389608 \\
0.4, 0.5278303673906646 \\
0.6, 0.32842188934703737 \\
0.8, 0.15605031320328325 \\
1.0, 0.0 \\
};
 
\addplot [semithick,dashed]
table [col sep = comma , row sep=crcr]{%
0.0, 1.0 \\
0.2, 0.762414817751949 \\
0.4, 0.539475158985577  \\
0.6, 0.339825092245722  \\
0.8, 0.162981425289659  \\
1.0, 0.0 \\
};

\addplot [thin,mark={*},mark options={solid},only marks]
table [col sep = comma , row sep=crcr]{%
0.0, 1.0 \\  
0.2, 0.7624148233805689 \\
0.4, 0.5394751627430595 \\
0.6, 0.3398250928103402 \\
0.8, 0.16298142980422667 \\
1.0, 0.0 \\
};
 
\addplot [semithick,dashed]
table [col sep = comma , row sep=crcr]{%
0.0, 1.0 \\
0.2, 0.768513292254645 \\
0.4, 0.549223528646855 \\
0.6, 0.34943051203766  \\
0.8, 0.168848343537185 \\
1.0, 0.0 \\
};

\addplot [thin,mark={*},mark options={solid},only marks]
table [col sep = comma , row sep=crcr]{%
0.0, 1.0 \\  
0.2, 0.7685132952296969 \\
0.4, 0.5492235341576502 \\
0.6, 0.349430515914785 \\
0.8, 0.16884834388080958 \\
1.0, 0.0 \\
};

\addplot [semithick,dashed]
table [col sep = comma , row sep=crcr]{%
0.0, 1.0 \\
0.2, 0.773604479203481 \\
0.4, 0.557391314526195  \\
0.6, 0.357513741099182  \\
0.8, 0.173802604184609  \\
1.0, 0.0 \\
};26

\addplot [thin,mark={*},mark options={solid},only marks]
table [col sep = comma , row sep=crcr]{%
0.0, 1.0 \\  
0.2, 0.773604478016916 \\
0.4, 0.5573913216082704 \\
0.6, 0.35751374010086534 \\
0.8, 0.1738026069656211 \\
1.0, 0.0 \\
};

\addplot [semithick,dashed]
table [col sep = comma , row sep=crcr]{%
0.0, 1.0 \\
0.2, 0.777861976977681 \\
0.4, 0.564239005723056 \\
0.6, 0.364311412753909 \\
0.8, 0.177979155658468 \\
1.0, 0.0 \\
};

\addplot [thin,mark={*},mark options={solid},only marks]
table [col sep = comma , row sep=crcr]{%
0.0, 1.0 \\  
0.2, 0.7778619919659372 \\
0.4, 0.5642390118299168 \\
0.6, 0.3643114121371887 \\
0.8, 0.17797915651710772 \\
1.0, 0.0 \\
};
 
\addplot [semithick,dashed]
table [col sep = comma , row sep=crcr]{%
0.0, 1.0 \\
0.2, 0.78142660216431 \\
0.4, 0.569982481504838 \\
0.6, 0.37002530686335 \\
0.8, 0.181495908713189 \\
1.0, 0.0 \\
};

\addplot [thin,mark={*},mark options={solid},only marks]
table [col sep = comma , row sep=crcr]{%
0.0, 1.0 \\  
0.2, 0.7814266175369099 \\
0.4, 0.5699824862157656 \\
0.6, 0.37002530366631914 \\
0.8, 0.18149590417513362 \\
1.0, 0.0 \\
};
 
\addplot [semithick,dashed]
table [col sep = comma , row sep=crcr]{%
0.0, 1.0 \\
0.2, 0.784413632904398 \\
0.4, 0.574801303953981 \\
0.6, 0.374826634424463 \\
0.8, 0.184454613851449 \\
1.0, 0.0 \\
};

\addplot [thin,mark={*},mark options={solid},only marks]
table [col sep = comma , row sep=crcr]{%
0.0, 1.0 \\  
0.2, 0.7844136363075965 \\
0.4, 0.5748013101565667 \\
0.6, 0.3748266318871391 \\
0.8, 0.18445461005935157 \\
1.0, 0.0 \\
};

\addplot [semithick,dashed]
table [col sep = comma , row sep=crcr]{%
0.0, 1.0 \\
0.2, 0.7869181407213 \\
0.4, 0.578845243474482 \\
0.6, 0.378860218243567 \\
0.8, 0.186942355706748 \\
1.0, 0.0 \\
};

\addplot [thin,mark={*},mark options={solid},only marks]
table [col sep = comma , row sep=crcr]{%
0.0, 1.0 \\  
0.2, 0.7869181284515006 \\
0.4, 0.5788452411908093 \\
0.6, 0.37886021599860353 \\
0.8, 0.18694235902932663 \\
1.0, 0.0 \\
};

\draw[-{Stealth[length=5mm]}] (0.5, 0.1) -- (0.5, 0.7) ;
\draw[fill=white,draw=none] (0.42,0.05) rectangle (0.48,0.15);
\node[below, left] at (0.5,0.1) {\small{$t^l$}};

\end{axis}
\end{tikzpicture}
        \caption{Solution obtained for the first 20 time instants with $y(x_k,t^0)=0$, using inhomogeneous Dirichlet conditions ($y_{0} = 1, y_{N_\text{p}+1}=0$) and $d=5$.}
        \label{fig:Results:Heat:inhomoDirichlet}
    \end{subfigure}\\
    \smallskip
    \begin{subfigure}[t]{0.48\textwidth}
        \centering
        \input{Figures/Results/R_Heat_Mixed_f0_inhomogeneous_nq2.tex}
        \caption{Solution obtained for the first 100 time instants with $y(x_k,t^0)=0$, using mixed boundaries ($y_{0}=1$, $\nicefrac{\partial y_{N_\text{p}+1}}{\partial x} = 0$) and $d=7$.}
        \label{fig:Results:Heat:mixed_nq2}
    \end{subfigure}
    \hfill
    \begin{subfigure}[t]{0.48\textwidth}
        \centering
        \input{Figures/Results/R_Heat_Mixed_f0_inhomogeneous_nq4.tex}
        \caption{Comparison of coarse and fine grid results for two time instant with $y(x_k,t^0)=0$ and $\Delta t = 1.67\times10^{-3}$[s], using mixed boundaries ($y_{0}=1$, $\nicefrac{\partial y_{N_\text{p}+1}}{\partial x} = 0$) and $d=7$.} 
       \label{fig:Results:Heat:mixed_nq4}
    \end{subfigure}
    \caption{
    Comparisons of \textsc{VQA}-result (lines) and \textsc{FD}-results (symbols) for the transient heat conduction in combination with four different boundary conditions on grids with $N_\text{p}=4$ ($n=2$; dashed lines) and $N_\text{p}=16$ ($n=4$; solid lines) interior points.} 
    \label{fig:Results:Heat}
\end{figure}

The predicted temperature distributions are depicted in \Fig{fig:Results:Heat}, where arrows indicate the temporal evolution from the initial to the final time step, and  \textsc{FD}-results are again illustrated by closed (coarse grid) and open (fine grid) symbols. Mind that a larger scatter of depth $d$, cf. \Fig{fig:VQA_Ansatz}, is seen for the employed ansatz $U(\lamb_c)$  
which is motivated by the complexity of the initial solution as explained below. 

Figure \ref{fig:Results:Heat:homoDirichlet} shows the results obtained for the 
homogeneous Dirichlet problem over the first 30 time instants. Since the results for the two grids are very similar, we mostly confine the discussion to coarse grid results for the sake of brevity. The initial solution was assigned to $y(x_k, t^0)=1$ in the interior, which yields negative heat fluxes at the boundaries in combination with the imposed homogeneous Dirichlet boundaries. The figure indicates an excellent agreement between the \textsc{VQA}-results and the \textsc{FD}-results. This also applies to the results obtained for the inhomogeneous Dirichlet problem depicted in  \Fig{fig:Results:Heat:inhomoDirichlet}, where the initial solution refers to $y(x_k, t^0)=0$ in the interior domain, and a constant heat flux from the left to the right is predicted at the twentieth time step. An interesting detail refers to the increased depth $d$ of the ansatz circuit $U(\lamb_c)$, cf. \Fig{fig:VQA_Ansatz}. The increase is explained by the different initial conditions and the related steep solution in the initial phase that requires a deeper ansatz circuit, even when using the fairly rich \textsc{\it SO4} approach for $\tilde U(\lamb_i)$. On the other hand, deeper circuits are afflicted with more parameters to be optimized, and the optimization effort could be reduced for lowered depth circuits as the solution advances to the simpler steady state.

Results obtained for a mixed Dirichlet/Neumann problem are outlined in \Fig{fig:Results:Heat:mixed_nq2} and \Fig{fig:Results:Heat:mixed_nq4}. The initial temperature distribution is assigned to $y(x_k,t^0)=0$ in the interior as well as to the Neumann condition on the right side. The final solution should agree with the left side Dirichlet value $y(0,t^l)=1$ in the whole domain. For the coarse grid, $N_\text{t} = 100$ time instants are displayed in \Fig{fig:Results:Heat:mixed_nq2}. They indicate a remarkable visual agreement between the \textsc{VQA}-results and the \textsc{FD}-results over the entire simulation period. In line with the \textsc{FD}-results, the \textsc{VQA}-results unveil that the temperature on the right increases for the adiabatic condition until the global heat exchange finally vanishes. 
Figure~\ref{fig:Results:Heat:mixed_nq4} compares the evolution of the solution on the coarse and the fine grid for the initial $N_\text{t} = 2$ time instants. Results indicate that the predictive accuracy of the \textsc{VQA} is similar on both grids. 

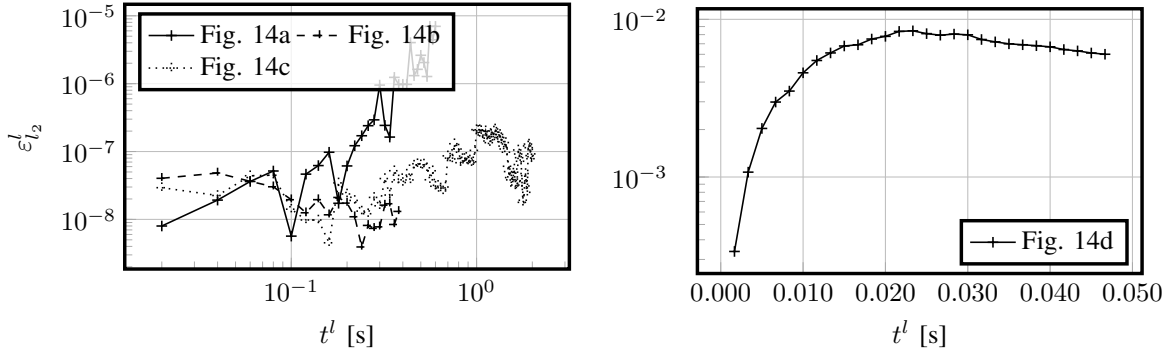
\begin{figure}[htbp]
\centering    
    \begin{subfigure}[t]{0.49\textwidth}
        \centering
        \tikzexternaldisable
\begin{tikzpicture}
\begin{axis}[
tick pos=left,
xmajorgrids,
ymajorgrids,
log basis y={10},
legend pos=north west,
legend style={draw=black!15!black,legend cell align=left, fill opacity=0.8, draw opacity=1, text opacity=1},
legend columns=2, 
ymode=log,
xmode=log,
xlabel={$t^l$ [s]},
ylabel={$\varepsilon_{l_2}^l$}
]

\addplot [semithick, mark=+]
table[col sep = comma, row sep=crcr] {%
0.02,7.9937684770942845e-09\\
0.04,1.9242657886596816e-08\\
0.06, 3.6218898932440903e-08\\
0.08, 5.1660053731094659e-08\\
0.1,5.6686276032443576e-09\\
0.12,4.6442476419440557e-08\\
0.14,6.2100280424507281e-08\\
0.16,9.7636240255460675e-08\\
0.18,1.7280524006074975e-08\\
0.2,6.1337537857750195e-08 \\
0.22,1.2186465030331132e-07\\
0.24,1.7088049687593401e-07\\
0.26,2.4020838875007354e-07\\
0.28,2.9290621598700665e-07\\
0.3,9.5235644936346244e-07\\
0.32,2.4269131489898345e-07\\
0.34,1.6331702834691359e-07\\
0.36,1.2378999481416310e-06\\
0.38,9.7289656798652760e-07\\
0.4,9.7987218248498399e-07\\
0.42,9.7657555016249430e-07\\
0.44,4.0298303614569466e-06\\
0.46,1.3130151102246358e-06\\
0.48, 1.6258362459068678e-06\\
0.5,2.6196430020525633e-06\\
0.52,2.0477856843144930e-06\\
0.54,1.2749979672652012e-06\\
0.56,6.9924579104539211e-06\\
0.58,4.4303297900199218e-06\\
0.6,6.992457910453921e-06\\
};
\addlegendentry{\Fig{fig:Results:Heat:homoDirichlet}}

\addplot [semithick, mark=+, dashed]
table[col sep = comma, row sep=crcr] {%
0.02, 4.0320408623393105e-08 \\
0.04, 4.8257807440515838e-08 \\
0.06, 3.6379530095993001e-08 \\
0.08, 3.0244027406821383e-08 \\
0.10, 1.9495834096949913e-08 \\ 
0.12, 1.2520249424505043e-08 \\ 
0.14, 1.9534006794650537e-08 \\ 
0.16, 1.1686576268455702e-08 \\
0.18, 2.1054954156760666e-08 \\ 
0.20, 1.7376195382605138e-08 \\
0.22, 1.0940049429253216e-08 \\
0.24, 3.9249618014233990e-09 \\
0.26, 8.1547625349327948e-09 \\
0.28, 7.3735998180101036e-09 \\
0.30, 7.7649464934523161e-09 \\
0.32, 1.6219099189146940e-08 \\
0.34, 1.7009544928690025e-08 \\
0.36, 8.4185417151278201e-09 \\
0.38, 1.3108872113730400e-08 \\
};
\addlegendentry{\Fig{fig:Results:Heat:inhomoDirichlet}}

\addplot [semithick, mark=+, dotted]
table[col sep = comma, row sep=crcr] {%
0.02,  2.9328583757121922e-08 \\ 
0.04,  2.2503882645211664e-08 \\
0.06,  4.3984030107920000e-08 \\
0.08,  4.4755319971126264e-08 \\ 
0.1,   1.4142600751215915e-08 \\
0.12,  9.8083978597893131e-09 \\
0.14,  9.7884929014551195e-09 \\
0.16,  4.5216791354467599e-09 \\
0.18,  3.4502558200895945e-08 \\
0.2,   2.3702432783951331e-08 \\
0.22,  1.9040663646251850e-08 \\
0.24,  1.3251610326252077e-08 \\
0.26,  1.2127955377794734e-08 \\
0.28,  2.1088694770507444e-08 \\
0.3,   1.5874061405436622e-08 \\
0.32,  3.9986825802509540e-08 \\
0.34,  2.2860354789100878e-08 \\
0.36,  5.2139375948612465e-08 \\
0.38,  3.8275983878809113e-08 \\
0.4,   3.4959624396976492e-08 \\
0.42,  4.6420802617676475e-08 \\
0.44,  3.8161354251724726e-08 \\
0.46,  6.2941170476501788e-08 \\
0.48,  6.6087168685254818e-08 \\
0.5,   6.3737941384191218e-08 \\
0.52,  6.5454669443873987e-08 \\
0.54,  5.7631579742005161e-08 \\
0.56,  3.8387838408262986e-08 \\
0.58,  5.2059185906907740e-08 \\
0.6,   3.0791806207935689e-08 \\
0.62,  3.1765995812965199e-08 \\
0.64,  2.9007914337995749e-08 \\
0.66,  2.6272349729068019e-08 \\
0.68,  3.0004342807684083e-08 \\
0.7,   8.6339715886367559e-08 \\
0.72,  7.9985199762374074e-08 \\
0.74,  7.7702948697258386e-08 \\
0.76,  1.3030360352228469e-07 \\
0.78,  1.0590476520094089e-07 \\
0.8,   1.0891024595384498e-07 \\
0.82,  6.8475920111104860e-08 \\
0.84,  7.8694402662020694e-08 \\
0.86,  7.8959092231457446e-08 \\
0.88,  8.2093435793713754e-08 \\
0.9,   7.2183663404795154e-08 \\
0.92,  7.5046219147680680e-08 \\
0.94,  6.8393578682037191e-08 \\
0.96,  7.0607503862800164e-08 \\
0.98,  1.0398875910735464e-07 \\
1.,    2.0996326117787629e-07 \\
1.02,  2.0521825838622115e-07 \\
1.04,  2.0923065173518130e-07 \\
1.06,  2.0269629222550778e-07 \\
1.08,  1.7555097081492056e-07 \\
1.1,   1.7060105708704391e-07 \\
1.12,  2.0186008663886867e-07 \\
1.14,  1.6789087451247547e-07 \\
1.16,  1.4290507986156924e-07 \\
1.18,  1.3511748745440588e-07 \\
1.2,   1.7893526877750727e-07 \\
1.22,  1.6124710128452425e-07 \\
1.24,  2.0059039820635291e-07 \\
1.26,  2.1500085181263438e-07 \\
1.28,  1.9274062359421217e-07 \\
1.3,   1.6608723990625687e-07 \\
1.32,  1.5075463603816281e-07 \\
1.34,  1.7862021629133567e-07 \\
1.36,  1.6030838303678290e-07 \\
1.38,  1.4009584319996571e-07 \\
1.4,   1.2019829694822696e-07 \\
1.42,  1.0897979883370637e-07 \\
1.44,  1.3562212087550838e-07 \\
1.46,  9.7672454762002111e-08 \\
1.48,  6.0652756730724255e-08 \\
1.5,   5.0628745414072147e-08 \\
1.52,  4.1328235906414639e-08 \\
1.54,  3.8241707668048574e-08 \\
1.56,  4.2318917110026126e-08 \\
1.58,  4.8050010086629494e-08 \\
1.6,   5.5230004576629319e-08 \\
1.62,  3.1101713046022398e-08 \\
1.64,  3.2088389158797281e-08 \\
1.66,  3.6211997961441273e-08 \\
1.68,  1.0542315504012040e-07 \\
1.7,   9.1972136339424024e-08 \\
1.72,  8.1115172216274415e-08 \\
1.74,  7.1915402505094940e-08 \\
1.76,  3.4145414071056143e-08 \\
1.78,  1.7839360957527023e-08 \\
1.8,   1.9177026767115510e-08 \\
1.82,  3.7151868399073258e-08 \\
1.84,  4.7403392486356965e-08 \\
1.86,  3.1113398937977574e-08 \\
1.88,  2.5068369899661422e-08 \\
1.9,   1.2091626767702596e-07 \\
1.92,  1.2874216249320741e-07 \\
1.94,  1.1392034871853851e-07 \\
1.96,  8.7160441055939332e-08 \\
1.98,  7.8411177159980840e-08 \\
2.,    9.2316206714000319e-08 \\
};
\addlegendentry{\Fig{fig:Results:Heat:mixed_nq2}}
\end{axis}
\end{tikzpicture}
\tikzexternalenable
        \caption{Evolution for coarse-grid results displayed in \Fig{fig:Results:Heat:homoDirichlet}-\subref{fig:Results:Heat:mixed_nq2}.} 
        \label{fig:Results:Heat:Error:nq2}
    \end{subfigure} 
    \begin{subfigure}[t]{0.49\textwidth}
        \centering
        \tikzexternaldisable
\begin{tikzpicture}
\begin{axis}[
tick pos=left,
xmajorgrids,
ymajorgrids,
log basis y={10},
legend pos=south east,
ymode=log,
xlabel={$t^l$ [s]},
ylabel={},
scaled x ticks=false,
y tick label style={text width=2.8em},
 x tick label style={
          /pgf/number format/.cd,
          fixed,
          fixed zerofill,
          precision=3,
          /tikz/.cd
 }
]

\addplot [semithick, mark=+]
table[col sep = comma, row sep=crcr] {%
0.001666666666667,  0.0003371484114523485 \\
0.003333333333334,  0.001073689360097617  \\
0.005000000000001,  0.0020326137421642593 \\
0.006666666666668,  0.0029886206893710385 \\
0.008333333333335,  0.0035076321644051237 \\
0.010000000000002,  0.004579714904971679 \\
0.011666666666669,  0.00548381747476837 \\
0.013333333333336,  0.006122613710801446 \\
0.015000000000003,  0.006776989282555962 \\
0.01666666666667,   0.006897860523917461 \\
0.018333333333337,  0.007500511872991561 \\ 
0.020000000000004,  0.007801762849049892 \\
0.021666666666671,  0.008384804603151665 \\
0.023333333333338,  0.008454953353354545 \\
0.025000000000005,  0.008114678805533251 \\
0.026666666666672,  0.007934017534551118 \\
0.028333333333339,  0.008062166382034597 \\
0.030000000000006,  0.00797837892449917 \\
0.031666666666673,  0.0074510706041031076 \\
0.03333333333334 ,   0.007197147564470886  \\
0.03500000000000701, 0.006994639208560086 \\
0.03666666666667401, 0.006899930116695015 \\ 
0.038333333333341005,  0.006803199068269109 \\
0.040000000000008,     0.006694399762532808 \\
0.04166666666667501, 0.00643430454478981 \\ 
0.04333333333334201, 0.006306883742593956 \\ 
0.045000000000009005, 0.006118867514920258 \\
0.046666666666676, 0.006007139696595618 \\
};
\addlegendentry{\Fig{fig:Results:Heat:mixed_nq4}}

\end{axis}
\end{tikzpicture}
\tikzexternalenable

        \caption{Evolution of fine-grid results displayed in \Fig{fig:Results:Heat:mixed_nq4}.}
    \label{fig:Results:Heat:Errors:mixed_nq4}
    \end{subfigure}
    \caption{Temporal evolution of the deviation measure ${\varepsilon^l_{\text{l}_2}}$ for the results displayed in \Fig{fig:Results:Heat}.}
    \label{fig:Results:Heat:Errors}
\end{figure}

The time evolution of the $\varepsilon_{l_2}$ measure is displayed, using a \textit{log-log} representation in \Fig{fig:Results:Heat:Error:nq2} and a \textit{semi-log} representation in \Fig{fig:Results:Heat:Errors:mixed_nq4}. For the inhomogeneous Dirichlet and the mixed boundary condition, the maximum error occurs during the initial time steps and decays as the solution advances in time until approximately $t=0.1$~[s] (5 time steps). 
Subsequently, the measure remains at a low level for these cases. For the homogeneous Dirichlet case with $d=1$, displayed in \Fig{fig:Results:Heat:homoDirichlet}, the error increases significantly after approximately 10 time steps but remains small, i.e., under $10^{-5}$. 
In contrast, the ansatz with $d=7$ that was used to compute the inhomogeneous steady state solution depicted in \Fig{fig:Results:Heat:mixed_nq2} outperforms the trainability in the homogeneous Dirichlet setting ($d=1$).  

Despite a general temporal error transport, the implicit numerical scheme remains stable and does not indicate any amplification of the error, which is in the range of the error of the steady-state solution, cf. Sec. \ref{sec:Application:Poisson}. 

As indicated by \Fig{fig:Results:Heat:Error:nq2}, the 
maximum values agree with the results of the steady heat conduction case provided by  \Tab{tab:Results:Poisson:Errors}. The corresponding 
\Tab{tab:Results:Heat:Errors} lists the time-averaged values of the $l_2$-norm $\Bar{\varepsilon}_{\text{l}_2}$ and the trace distance measure $\Bar{\varepsilon}_{\text{tr}}$.   Since the changes of the solution over time are small, maximum values of $\Bar{\varepsilon}_{\text{l}_2}$ and $\Bar{\varepsilon}_{\text{tr}}$ are bounded, and the averaged values remain within the range of the steady state results.

\begin{table}[htbp]
   \centering
    \begin{tabular}{@{}lcccccc@{}}
    \toprule
Boundary Setting & \multicolumn{2}{c}{$l_2$-norm $\Bar{\varepsilon}_{\text{l}_2}$}  & \multicolumn{2}{c}{trace distance $\Bar{\varepsilon}_{\text{tr}}$}&\multicolumn{2}{c}{DOF $c$} \\ 
\midrule
\multicolumn{1}{r}{interior points}&\multicolumn{1}{l}{$N_\text{p}=4$}& $N_\text{p}=16$& \multicolumn{1}{|l}{$N_\text{p}=4$}& $N_\text{p}=16$& \multicolumn{1}{|l}{$N_\text{p}=4$}& $N_\text{p}=16$  
\\
\midrule
Homog. Dirichlet  & \multicolumn{1}{c}{{$6.70 \times 10^{-6}$}}&\multicolumn{1}{c|}{-}&\multicolumn{1}{c}{{$4.79
 \times 10^{-5}$ }
}&\multicolumn{1}{c}{-}&\multicolumn{1}{|c}{$3$}&-\\
Inhomog. Dirichlet & \multicolumn{1}{c}{$1.84\times 10^{-8}$
}&\multicolumn{1}{c|}{-}&\multicolumn{1}{c}{$2.77\times 10^{-7}$ }&\multicolumn{1}{c}{-}&\multicolumn{1}{|c}{$15$}&-\\
Mixed Dirichlet/Neumann & \multicolumn{1}{c}{$8.08 \times 10^{-8}$
}&\multicolumn{1}{c|}{{$6.10 \times 10^{-3}$}}&\multicolumn{1}{c}{$2.12 \times 10^{-8}$}&\multicolumn{1}{c}{{$4.07 \times 10^{-3}$}}&\multicolumn{1}{|c}{$21$}&$60$\\
    \bottomrule
    \end{tabular}
   \caption{Time-averaged deviation measures $\Bar{\varepsilon}_{\text{l}_2}$ and $\Bar{\varepsilon}_{\text{tr}}$ for the results depicted in \Fig{fig:Results:Heat}. The degrees of freedom (\textsc{DOF}) are indicated by the number of control variables $c$.}
   \label{tab:Results:Heat:Errors}
\end{table}

\section{Conclusion} \label{sec:Conclusion}

The study presents a flexible and general boundary condition treatment for the numerical solution of \textsc{PDE}s on a \textsc{QC}. The strategy is based on the \textsc{VQA} approach by Lubasch et al. \cite{Lubasch2020}. To this end, initial-boundary value problems are reformulated as optimal control problems where all terms, including the boundary conditions, are cast into case-independent \textsc{QNPU}s. The boundary treatment combines ghost points with boundary cost function contributions that complement the objective function of the control problem. The method relaxes the requirements for the ansatz functions $U(\lamb_c)$, which must only be valid for the interior points that remain decoupled from the boundary points and avoids the use of penalty terms. Verification studies are performed for steady and unsteady heat conduction problems, covering a variety of boundary conditions. 

\smallskip
The obtained \textsc{VQA}-results reveal an excellent agreement with a classical \textsc{FD}-results. 
The increase of the circuit depth could attenuate the optimization efficiency due to multi-controlled gates, limiting the implementation on \textsc{QC} hardware \cite{Cerezo2021,Sato2021,Sarma2023}. Analogously to the implementation of Laplacian and potential terms, improvements in this regard are achieved by replacing the multi-controlled gates with additional carry qubits. The remarkable flexibility and accuracy of the proposed approach are thereby complemented by a favorable scaling of the quantum circuits with the number of qubits.

\smallskip 
As for all \textsc{VQA} approaches, the choice of the ansatz is crucial since it defines the underlying function space for the \textsc{PDE} solution. At the same time, the approach can introduce a non-convex property into the optimization. The hardware-efficient bricklayer topology of the employed ansatz shows favorable expressibility in terms of small deviations from \textsc{FD} solutions in noise-less \textsc{QC} simulations. The richer expressibility of the \textsc{\it SO4} ansatz in combination with shallow depths $d$ is advantageous, for example, to express general source terms. However, the strategy also unveils unfavorable with regard to the case-dependent increase of control parameters $\lamb_c$ and depth $d$, and the related poorer trainability. The combination of challenging trainability and the non-convex nature of the optimization can imply serious convergence issues in tracking the global optimum for larger numbers of qubits. 

Future research will be devoted to encoding first derivatives as a \textsc{QNPU} operator to represent convective fluxes. In view of extending the present algorithm to \textsc{2D} and \textsc{3D} spatial problems, the non-convex optimization problem will be addressed by selecting more appropriate (trainable, expressible, shallow) dynamic ansatz strategies. For instance, strategies for avoiding barren plateaus in the cost function \cite{Ragone2023} include utilizing classical shadows \cite{Sack2022} and convolutional neural networks \cite{Pesah2021}. For the time evolution, the use of adaptive ansatz depths and circuit re-compilation techniques \cite{Jaderberg2020} seem promising ways to save computational cost and increase the range of applications towards industrial \textsc{QCFD} in the current \textsc{NISQ} era. 

\section*{Acknowledgment}
\addcontentsline{toc}{section}{Acknowledgment}
This publication and the current work have received funding from the European Union's Horizon Europe research and innovation program (HORIZON-CL4-2021-DIGITAL-EMERGING-02-10) under grant agreement No. 101080085 QCFD. The authors thank N.-L. van H{\"u}lst, G.~S. Reese,  E.~L.~Fesefeldt, D.~Schmeckpeper, P.~M.~M{\"u}ller, and M.~Kiffner for fruitful discussions. 

\section*{Author Declarations}
\addcontentsline{toc}{section}{Author Declarations}

\subsection*{Declaration of Interest}
\addcontentsline{toc}{subsection}{Declaration of Interest}
The authors have no interest to disclose. 

\subsection*{Author Contributions}
\addcontentsline{toc}{subsection}{Author Contributions}
\textbf{Paul~Over:} Conceptualization, Methodology, Software, Validation, Formal analysis, Investigation, Writing - original draft, Writing - review \& editing, Visualization. 
\textbf{Sergio~Bengoechea:} Conceptualization, Methodology, Software, Validation, Formal analysis, Investigation, Writing - original draft, Writing - review \& editing, Visualization. 
\textbf{Thomas~Rung:} Project administration, Funding acquisition, Supervision, Conceptualization, Methodology, Resources, Writing - original draft, Writing - review \& editing. 
\textbf{Francesco~Clerici:} Writing - review \& editing.  
\textbf{Leonardo~Scandurra:} Writing - review \& editing. 
\textbf{Eugene~de~Villiers:} Project administration, Funding acquisition, Supervision, Resources. 
\textbf{Dieter~Jaksch:} Project administration, Funding acquisition, Supervision, Resources, Writing - review \& editing.

\section*{Data availability}
\addcontentsline{toc}{section}{Data availability}
The data is available via \href{https://doi.org/10.25592/uhhfdm.14123}{https://doi.org/10.25592/uhhfdm.14123} \cite{data}.

\bibliographystyle{IEEEtran} %
\bibliography{lib}

\appendix
\def\theequation{\thesubsection.\arabic{equation}}
\def\thefigure{\thesubsection.\arabic{figure}}    
\renewcommand{\thesubsection}{\Alph{subsection}}
In this appendix, the fundamental quantum gates and the gate sequence of the quantum half-adder are summarized together with their corresponding matrix representation. In addition, the matrix gate conversion for the four qubit version of the circuits in \Fig{fig:matrix2gate} is also given.

\subsection{Fundamentals}
\label{app:Fundamentals}
\setcounter{equation}{0}
\setcounter{figure}{0}
The Dirac notation defines $\psib = \psi^i \eb_i= \ket{\psi} \; \text{(called "ket psi") and} \; \psib^\intercal = \psi_j  \eb^j= \bra{\psi} \; \text{(called "bra psi")}$. Quantum circuits are accordingly assembled by inner (dot/scalar product) and outer products (tensor/dyadic product) using the computational basis of a qubit, viz. $\ket{0} = \left(1,0\right)^\intercal$ and $\ket{1}=\left(0,1\right)^\intercal$. The inner product contracts two identical coordinates, e.g., in a scalar $z = \phib^\intercal \psib = \bra{\phi} \ket{\psi}= \sum \limits_k \phi^k \psi_k$. The outer product instead conserves the sum of the ranks (orders) of the multiplied factors, e.g., in the matrix $\phib\otimes\psib^\text{T} = \ket{\phi} \otimes \bra{\psi} = \ket{\phi} \bra{\psi}   \to  \sum \limits_k \sum \limits_l \phi^k \psi_l \eb_k \otimes \eb^l$. The application of these basic rules allows to derive various operations, e.g., the fundamental set of one-qubit operations 
\begin{align}
\begin{split}
    &\sigma_0 = I = \ket{0}\bra{0}+ \ket{1} \bra{1} =\begin{bmatrix}
        1 &0 \\
        0 & 1\\
    \end{bmatrix}, \quad
   \sigma_x = X = \ket{0} \bra{1} + \ket{1} \bra{0} = \begin{bmatrix}
        0 &1 \\
        1 & 0\\
    \end{bmatrix},
    \end{split}
    \end{align}

    \begin{align}
    \begin{split}
     \hspace{1cm} \sigma_y = Y = i(\ket{1}\bra{0}- \ket{0} \bra{1}) = \begin{bmatrix}
        0 &-i \\
        i & 0\\
    \end{bmatrix}, \quad 
    \sigma_z = Z = \ket{0} \bra{0}- \ket{1} \bra{1} = \begin{bmatrix}
        1 & 0\\
        0 & -1\\
    \end{bmatrix}, 
    \end{split}\\
    \begin{split}
    R_y(\theta) = \exp \Bigg( i\frac{-\theta}{2}\sigma_y \Bigg) = 
    \begin{bmatrix}
        cos\left(\frac{\theta}{2}\right) & -sin\left(\frac{\theta}{2}\right) \\
        sin\left(\frac{\theta}{2}\right) & cos\left(\frac{\theta}{2}\right)
    \end{bmatrix}, \quad
    R_z(\theta) = \exp \Bigg( i\frac{-\theta}{2}\sigma_z \Bigg) = 
    \begin{bmatrix}
      e^{-i\nicefrac{\theta}{2}} & 0 \\
     0 & e^{i \nicefrac{\theta}{2}}
 \end{bmatrix}
    \label{eq:principlegates}
    \end{split}
\end{align}
\noindent
and the two-qubit operations 
\begin{align}
    &\textsc{CNOT} = I \otimes \ket{0} \bra{0} + \sigma_x \otimes \ket{1} \bra{1}   =\begin{bmatrix}
        1 & 0 & 0 & 0 \\
        0 & 0 & 0 & 1 \\
        0 & 0 & 1 & 0 \\
        0 & 1 & 0 & 0 \\
    \end{bmatrix}, \textsc{CZ} = I \otimes \ket{0} \bra{0} + \sigma_z \otimes \ket{1} \bra{1} =
    \begin{bmatrix}
         1 & 0 & 0 & 0\\
         0 & 1 & 0 & 0\\
         0 & 0 & 1 & 0\\
         0 & 0 & 0 &-1\\
          \end{bmatrix} .
\end{align}
Mind that the shape of the \textsc{CNOT} matrix results from the \textit{little-endian} convention.

\medskip

\subsection{Matrix Gate Conversion}
\label{app:MatrixGateConversion}
\setcounter{equation}{0}
\setcounter{figure}{0}

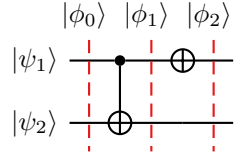
\begin{figure}[h]
    \centering
    \tikzsetnextfilename{ExpAdder}
    \begin{quantikz}[slice style=blue]
         \lstick{$\ket{\psi_1}$} \slice{$\ket{\phi_0}$} &\ctrl{1}  \slice{$\ket{\phi_1}$}   & \targ{}\slice{$\ket{\phi_2}$}  & \qw  \\
          \lstick{$\ket{\psi_2}$}  & \targ{}& \qw  &\qw 
    \end{quantikz}
    \caption{Quantum half-adder circuit. }
    \label{fig:App:examplecirc}
\end{figure}

We present two approaches to derive the corresponding matrix representation of the two-qubit quantum half-adder in \Fig{fig:App:examplecirc}. The first technique, referred to as the ``formal way'', details each intermediate state $\ket{\phi_r}; r\in\{0,1,2\}$ of the circuit while the second scans all the possible outputs to construct the matrix representation of the network by inspection. The latter is restricted to real-valued states and, we referred to as the ``fast way''. The intermediate break-points, indicated by red dashed lines, serve as orientation in the following computations. 

\textbf{The formal way} 
\begin{align}
    \begin{split}
            \ket{\phi_0} &= \ket{\psi_2} \otimes \ket{\psi_1} = \ket{\psi_2\psi_1}
    \end{split}\\
    \begin{split}
    \ket{\phi_1} &= \textsc{CNOT}\ket{\psi_2\psi_1} = (I\otimes \ket{0} \bra{0}+ \sigma_x \otimes \ket{1}\bra{1}) \ket{\psi_2\psi_1} \\
     &= (I \otimes \ket{0} \bra{0}) \ket{\psi_2\psi_1} + (\sigma_x \otimes \ket{1}\bra{1}) \ket{\psi_2\psi_1}
     = I \ket{\psi_2} \otimes \ket{0} \bra{0} \ket{\psi_1} + \sigma_x \ket{\psi_2} \otimes \ket{1} \bra{1} \ket{\psi_1}
    \end{split}\\
    \begin{split}
    \ket{\phi_2} &= I \ket{\psi_2} \otimes (\sigma_x \ket{0} \bra{0} \ket{\psi_1}) + \sigma_x \ket{\psi_2} \otimes \sigma_x \ket{1} \bra{1}\ket{\psi_1}    \\
    &= (I \otimes \sigma_x \ket{0}\bra{0}) \ket{\psi_2 \psi_1} + (\sigma_x \otimes \sigma_x \ket{1} \bra{1})\ket{\psi_2 \psi_1}=  \underbrace{(I \otimes \sigma_x \ket{0} \bra{0} + \sigma_x \otimes \sigma_x \ket{1}\bra{1})}_{A} \ket{\psi_2 \psi_1}
    \end{split}
    \end{align}

    \begin{align}
    \begin{split}
   \quad A = \begin{bmatrix}
       1& 0\\
       0& 1 \\
   \end{bmatrix} \otimes \begin{bmatrix}
       0 &1 \\
       1& 0\\
   \end{bmatrix} \cdot \begin{bmatrix}
       1 & 0 \\
       0 & 0 \\
   \end{bmatrix} + \begin{bmatrix}
       0 &1 \\
       1& 0\\
   \end{bmatrix} \otimes \begin{bmatrix}
       0 &1 \\
       1& 0\\
   \end{bmatrix} \cdot \begin{bmatrix}
       0 & 0\\
       0 & 1\\
   \end{bmatrix}= 
   \begin{bmatrix}
       0&0&0&1\\
       1&0&0&0\\
       0&1&0&0\\
       0&0&1&0\\
   \end{bmatrix}
   \label{eq:transform_Matrix_HalfAdder}
   \end{split}\\
   \begin{split}
       \ket{\phi_2} &= A \ket{\psi_2 \psi_1} = A\ket{\phi_0}.
   \end{split}
\end{align}

\textbf{The fast way} 
\begin{align}
1) \quad
\begin{split}
    \ket{\phi_0}&= \ket{\psi_2\psi_1} = \ket{00} = (1,0,0,0)^\intercal\\ 
    \ket{\phi_1}&= \ket{00}\\
    \ket{\phi_2}&= \ket{01} \qquad \underline{\ket{00} \rightarrow \ket{01}},
\end{split}
 \qquad \qquad  2) \hspace{-0.8cm}
\begin{split}
    \ket{\phi_0}&= \ket{\psi_2\psi_1} = \ket{01}=(0,1,0,0)^\intercal\\ 
    \ket{\phi_1}&= \ket{11}\\
    \ket{\phi_2}&= \ket{10} \qquad \underline{\ket{01} \rightarrow \ket{10}},
\end{split}
\end{align}
\begin{align}
3) \quad
\begin{split}
    \ket{\phi_0}&= \ket{\psi_2\psi_1} = \ket{10}=(0,0,1,0)^\intercal\\ 
    \ket{\phi_1}&= \ket{10}\\
    \ket{\phi_2}&= \ket{11} \qquad \underline{\ket{10} \rightarrow \ket{11}},
\end{split}
 \qquad \qquad  4) \hspace{-0.8cm}
\begin{split}
    \ket{\phi_0}&= \ket{\psi_2 \psi_1} = \ket{11}=(0,0,0,1)^\intercal\\ 
    \ket{\phi_1}&= \ket{01}\\
    \ket{\phi_2}&= \ket{00} \qquad \underline{\ket{11} \rightarrow \ket{00}}.
\end{split}
\end{align}
The underlined statements deliver the information to reconstruct the rows of matrix $A$. By inspection, it is straightforward to extract the matrix representation in \Eq{eq:transform_Matrix_HalfAdder}. 

\medskip
\subsection{Boundary Circuit: Matrix and Gate Representation for 4-qubits}
\label{app:BoundaryMatrixGate}
\setcounter{equation}{0}
\setcounter{figure}{0}

\begin{figure}[t]
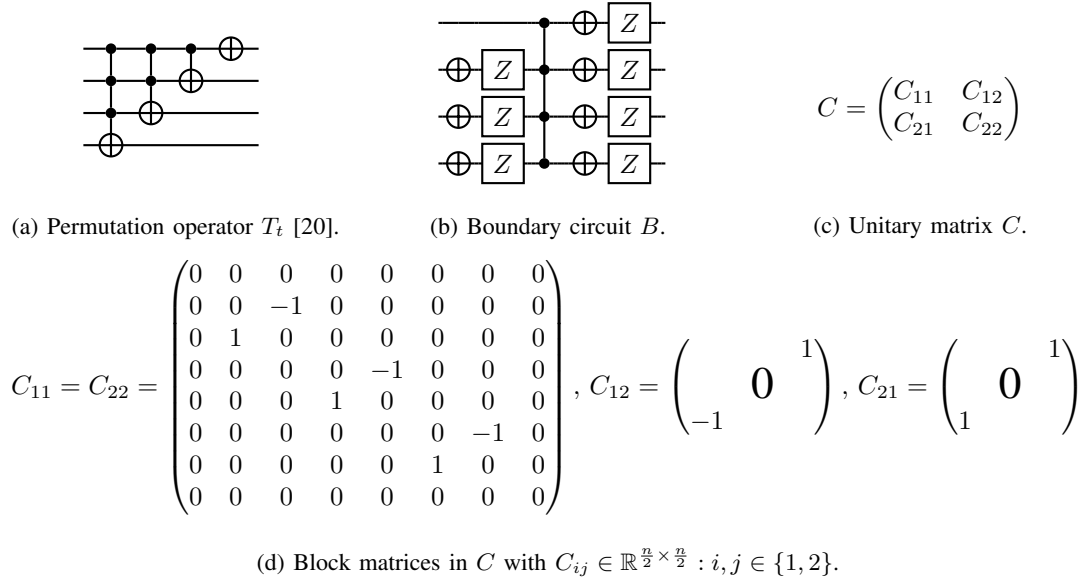

\centering
    \begin{subfigure}[t]{0.3\textwidth}
        \centering
        \input{Figures/F_Transformation_nq4}
        \vspace{0.45cm}
        \caption{Permutation operator $T_t$ \cite{Sato2021}.}
        \label{fig:Shift_nq4}
    \end{subfigure}
    \begin{subfigure}[t]{0.3\textwidth}
        \centering
        \input{Figures/F_Dirichlet_QNPU_nq4}
        \caption{Boundary circuit $B$.}
       \label{fig:BC_nq4}
    \end{subfigure}
    \begin{subfigure}[t]{0.3\textwidth}
        \centering
            \vspace{-0.5cm}
            \begin{equation*}
            C = \begin{pmatrix} C_{11} & C_{12} \\ C_{21} & C_{22} \end{pmatrix}
            \vspace{0.53cm}
        \end{equation*}
        \caption{Unitary matrix $C$.}
       \label{fig:Matrix_nq4}
    \end{subfigure}
    
    \begin{subfigure}[t]{\textwidth}
        \centering
        \begin{equation*}
        C_{11} = C_{22} = \begin{pmatrix}
            0 & 0 & 0 & 0 & 0 & 0 & 0 & 0 \\
            0 & 0 &-1 & 0 & 0 & 0 & 0 & 0 \\
            0 & 1 & 0 & 0 & 0 & 0 & 0 & 0 \\
            0 & 0 & 0 & 0 &-1 & 0 & 0 & 0 \\
            0 & 0 & 0 & 1 & 0 & 0 & 0 & 0 \\
            0 & 0 & 0 & 0 & 0 & 0 &-1 & 0 \\
            0 & 0 & 0 & 0 & 0 & 1 & 0 & 0 \\
            0 & 0 & 0 & 0 & 0 & 0 & 0 & 0
            \end{pmatrix} , \,
        C_{12} = \begin{pmatrix}
                &            & 1    \\
                & \text{\LARGE{0}} & \\
            -1  &            & 
            \end{pmatrix} , \, 
        C_{21} = \begin{pmatrix}
                &            & 1     \\
                & \text{\LARGE{0}} &  \\
            1   &            & 
            \end{pmatrix}
        \end{equation*}
    \caption{Block matrices in $C$ with $C_{ij} \in \mathbb{R}^{\frac{n}{2}\times \frac{n}{2}}: i,j \in  \{1,2\}$.}
    \label{fig:Block_matrix_nq4}
    \end{subfigure}
    \caption{Boundary circuits and matrix $C$ for 4 qubits.}
    \label{fig:Permu_BC_nq4}
\end{figure}

The boundary circuits for $n = 4$, and the corresponding transformation matrix $C$ are depicted in \Fig{fig:Permu_BC_nq4}. In the case of more grid points (or qubits), the different negative signed terms in the interior of $C$ do not fall only on the antidiagonal, cf. \Fig{fig:Block_matrix_nq4}. Consequently, at least two matrix configurations work, and thus, the design of circuits in \Fig{fig:Shift_nq4} and \Fig{fig:BC_nq4} is not unique with respect to the number of qubits. However, there are three important properties to be respected. First, the matrix entries are required to be symmetric with respect to the antidiagonal and the main diagonal. Second, the interior entries need to have a different sign with respect to their main diagonal-mirrored counterparts in order to cancel when taking the expectation value $\langle \sigma_z \rangle$. Last and most importantly, the corners of the antidiagonal have to be of the same sign to match the periodic entries of the derivative matrix. 

\end{document}